%Paper: hep-th/9505080
%From: "Gerald B. Cleaver" <gcleaver@pacific.mps.ohio-state.edu>
%Date: Fri, 12 May 1995 21:28:29 -0400
%Date (revised): Sat, 13 May 1995 18:52:04 -0400
%Date (revised): Sat, 13 May 1995 19:24:13 -0400
%Date (revised): Mon, 15 May 1995 00:47:18 -0400
%Date (revised): Thu, 18 May 1995 20:38:43 -0400
%Date (revised): Sat, 3 Jun 1995 19:41:43 -0400
%Date (revised): Wed, 9 Aug 1995 18:32:09 -0400
%Date (revised): Mon, 20 Nov 1995 11:00:47 -0500

%%%%%%%%%%%%%%%%%%%%%%%%%%%%%%%%%%%%%%%%%%%%%%%%%%%%%%%%%%%%%%%%%%%
% Supersymmetries in Free Fermionic Strings             %
%%%%%%%%%%%%%%%%%%%%%%%%%%%%%%%%%%%%%%%%%%%%%%%%%%%%%%%%%%%%%%%%%%%
% By Gerald B. Cleaver (gcleaver@ohstpy.bitnet)                   %
%%%%%%%%%%%%%%%%%%%%%%%%%%%%%%%%%%%%%%%%%%%%%%%%%%%%%%%%%%%%%%%%%%%
% Use plain TeX to compile                                        %
%%%%%%%%%%%%%%%%%%%%%%%%%%%%%%%%%%%%%%%%%%%%%%%%%%%%%%%%%%%%%%%%%%%
% paper starts at ``start here''
% NOTE: integer symbol defined by \Z (at line 2168)
% using Jytex fonts not always available

\magnification=\magstep1

%******************************************************************************
\input jytex.tex

%******************************************************************************
%\input defs.tex

% DEFINITIONS:
%
% Command Abbreviations
%
% temperary
%\def\ast{\cdot}

%%typo "lifejackets"

%% horizontal spaces
\def\h{\hbox to .5cm{\hfill}}
\def\hof{\hbox to .15cm{\hfill}}
\def\hi{\hbox to .2cm{\hfill}}
\def\htwo{\hbox to .2cm{\hfill}}
\def\htf{\hbox to .35cm{\hfill}}

\def\ha#1{\n\hbox to .6cm{\n {#1}\hfill}}
\def\hb#1{\indent\hbox to .7cm{\n {#1}\hfill}}
\def\hc#1{\indent\hbox to .7cm{\hfill}\hbox to 1.1cm{\n {#1}\hfill}}
\def\hd#1{\indent\hbox to 1.8cm{\hfill}\hbox to 1.4cm{\n {#1}\hfill}}

\def\hj{{\hbox to 0.06truecm{\hfill}}}
%%
%% vertical spaces

\def\vskipa{\vskip .2cm}
\def\eq#1{\eqno\eqnlabel{#1}}

\def\mpr#1{\markup{[\putref{#1}]}}
\def\pr#1{[\putref{#1}]}

\def\n{\noindent}
\def\no{\noindent}
\def\ref{\reference}

\def\undbib{\underbar {\hbox to 2cm{\hfill}}, }
%
% SUSY:
%
\def\st{ ST-SUSY}

\def\nall{ $N= 4$, 2, 1, 0 ST-SUSY }
\def\nall1{  $N= 4$, 2,    0 ST-SUSY }

%
% Math Symbols:
%

\def\balpha{{\bmit \alpha}}

\def\bbeta{{\bmit \beta}}

\def\bgamma{{\bmit \gamma}}

\def\bF{ {\bmit F}}
\def\bQ{ {\bmit Q}}
\def\bS{ {\bmit S}}
\def\bV{ {\bmit V}}
\def\b#1{ {\bmit{#1}} }

\def\ra{\rightarrow  }

\def\mod#1{ ( {\rm mod}\, {#1} )}
\def\phm{{\phantom{-}}}
%
%% fractions
%
\def\fract#1#2{{#1\over #2}}

\def\half{{1\over 2}}
\def\onehalf{{1\over 2}}

\def\third{{1\over 3}}
\def\onethird{{1\over 3}}
\def\twothird{{2\over 3}}

\def\fourth{{1\over 4}}
\def\onefourth{{1\over 4}}

\def\threefourth{{3\over 4}}

\def\fifth{{1\over 5}}
\def\onefifth{{1\over 5}}
\def\twofifth{{2\over 5}}
\def\threefifth{{3\over 5}}
\def\fourfifth{{4\over 5}}

\def\sixth{{1\over 6}}
\def\onesixth{{1\over 6}}

\def\fivesixth{{5\over 6}}

\def\onetenth{{1\over 10}}

\def\threetenth{{3\over 10}}

\def\fivetenth{{5\over 10}}

\def\sevententh{{7\over 10}}

\def\ninetenth{{9\over 10}}

\def\twelf{{1\over 12}}
\def\onetwelf{{1\over 12}}

\def\threetwelf{{3\over 12}}

\def\fivetwelf{{5\over 12}}

\def\seventwelf{{7\over 12}}

\def\ninetwelf{{9\over 12}}

\def\eleventwelf{{11\over 12}}

\def\thetaa{{\theta\over 2}}
\def\thetab{{\theta\over 3}}
\def\thetac{{\theta\over 4}}
\def\thetad{{\theta\over 5}}
\def\thetae{{\theta\over 6}}

%
%General
%
\def\eg{  {\it e.g.} }

%
%Publications
%

% Math Fonts
\begin{ignore}

\font\twelveBbb=msym10 scaled \magstep1
\font\nineBbb=msym9
\font\sevenBbb=msym7
\newfam\Bbbfam
\textfont\Bbbfam=\twelveBbb
\scriptfont\Bbbfam=\nineBbb
\scriptscriptfont\Bbbfam=\sevenBbb

 \def\Z{{\Bbb Z}}
\end{ignore}
\def\Z{Z\!\!\!Z}

%******************************************************************************

% start here
%\draft

\topmargin= 1truein\vsize=9truein
\leftmargin=1truein\hsize=6.5truein

\baselinestretch=1200
%\baselinestretch=1500

\foot={\hfil\normalfonts\numstyle\pagenum\hfil}
\head={\hfil}

%\begin{ignore}

\begin{title}
\pagenumstyle{blank}

{\rightline{\hbox to 4.5cm{\vtop{\hsize= 4.5cm
\baselinestretch=960\footnotefonts
\hfill\\
OHSTPY-HEP-T-95-004\\
DOE/ER/01545-643\\
hep-th/9505080\\
May 1995}}{\hbox to .35cm{\hfill}}}}

\vskip 1.5 truecm
\footnotenumstyle{symbols}
\footnotenum= 0

\centertext{\bf Supersymmetries in Free Fermionic Strings}
\vskip 1.2 truecm

\centertext{{Gerald B.~Cleaver}\footnote{gcleaver@ohstpy.bitnet}}

\vskip .3 truecm
\centertext{{\it The Ohio State University, Columbus,} OH 43210}
\vskip .1 truecm

\vskip 1.2 truecm
\centerline{\bf Abstract}
\vskip .3 truecm
Consistent heterotic free fermionic string models are classified in terms
of their number of spacetime supersymmetries, $N$.
For each of the six distinct choices of gravitino sector,
we determine what number of supersymmetries can survive additional
GSO projections.
We prove by exhaustive search that only
three of the six can yield N = 1, in addition to the N = 4, 2, or 0 that five
of the six can yield. One choice of gravitino sector can only produce N = 4
or 0.
Relatedly, we find that only $\Z_2$, $\Z_4$, and $\Z_8$ twists
of the internal fermions with worldsheet supersymmetry
are consistent with $N=1$ in free fermionic models.
Any other twists obviate $N=1$.
\vskip 1.0truecm

\centertext{\it To appear in Nucl.~Phys.~B456 (1995) 219-256.}

\end{title}
\newpage

\footnotenumstyle{symbols}
\footnotenum= 0
\pagenum=0\pagenumstyle{arabic}

\sectionnum=0\equationnum=0
\sectionnumstyle{arabic}
%\subsectionnum=0
%\subsectionnumstyle{alphabetic}
\equationnumstyle{arabic}

\def\ss#1{\vskip .1truecm
\no\hbox to .4truecm{\hfill}{\bf #1}
\vskip .1truecm}

{\bigfonts\bf\section{Review of Free Fermionic Strings}}

Over the last decade, string model building has grown into a field with
sometimes parallel, sometimes intersecting avenues.
Essentially these avenues can be viewed as the various
approaches to ``compactification'' from ten-dimensions to four.
Some of which lead to the geometrical interpretation of actual
compactified dimensions, while others do not.
The primary avenues correspond to
bosonic lattices,\mpr{narain86,lerche87}
free fermions,\mpr{antoniadis87,kawai87a}
$\Z_n$-orbifolds,\mpr{dixon85,dixon87,narain87}
Calabi-Yau manifolds,\mpr{candelas85a,candelas88a} and
$N=2$ minimal models.\mpr{gepner88a}
Some of these avenues intersect with others only at a few locations;
some overlap with others significantly.
Some turn out to be sections of a larger avenue; some
are eventually found to be identical to another, covering the
same path, as a single road given two different names.
Along varying avenues of string models and varying the locations on a specific
avenue,
the number, $N$, of spacetime supersymmetries in the early stringy
universe can be as large as 8 and as small as 0, depending on location.
Recent LEP results, however, point strongly towards the sections of these
various avenues that yield exactly $N=1$ spacetime supersymmetry
(ST-SUSY).\mpr{langacker92}
In this paper we traverse one specific avenue, that of free fermions
(and in particular the sub-avenue of heterotic free fermions),
and search for the sections along this path that do, indeed, produce
$N=1$ spacetime supersymmetry.

Free fermionic model building was developed simultaneously
by
Antoniadis, Bachas, Kounas, and Windey in \pr{antoniadis87}
and by
Kawai, Lewellen, and Tye in \pr{kawai87a}.
In light-cone gauge, a heterotic free fermionic string model contains
64 real worldsheet (WS) fermions $\psi^n$
($1\leq n \leq 20$ for left-moving (LM) WS fermions, $21\leq n \leq 64$ for
right-moving (RM) WS fermions),
in addition to the LM/RM WS scalars
($X_{\mu =1,2}\, ,\bar X_{\bar\mu =1,2}$) embedding transverse coordinates of
four-dimensional spacetime. $\psi^1$ and $\psi^2$ are the WS
superpartners of the two LM transverse scalars; the
remaining 62 fermions, $\{ \psi^{n= 3\, {\rm to}\, 64} \}$
are internal degrees of freedom.
Some or all of these internal fermions may be paired to form complex
fermions $\psi^{n,m}\equiv \psi^n + i\psi^m$.
 If $n$ and $m$ both denote
to either left-movers or right-movers, then $\psi^{n,m}$ is a Weyl fermion.
If $n$ denotes a left-mover and $m$ a right-mover, then
$\psi^{n,m}$ is a Mayorana fermion.
A specific model is
defined by (1) sets of 64-component boundary vectors
(with components for complex fermions counted twice) describing how the WS
fermions transform around non-contractible loops on the worldsheet,
and (2) sets of coefficients weighting contributions to
the one-loop partition function from
fermions with specific boundary conditions.

Modular invariance is a requirement for a sensible model and exists
if: (1) the one-loop partition function is invariant under
$S:\, \tau\rightarrow -1/\tau$ and $T:\, \tau\rightarrow \tau+1$
transformations of the complex WS parameter $\tau$ defining the
one-loop worldsheet (a torus);
and (2) either a specific additional two-loop constraint is
satisfied\mpr{antoniadis87}
or, equivalently, the states surviving the one-loop GSO projection
(GSOP)
``are sensible''\mpr{kawai87a}.
Under transportation around either of the non-contractable loops
(which we denote respectively as $l_{\alpha}$ and $l_{\beta}$)
of the one-loop worldsheet,
a WS fermion $\psi^{n\, (n,m)}$ can undergo transformations.
Consistency demands diagonalizable transformations, which are
expressible purely as phase changes:
$${\psi^{j}
  {\buildrel l_{\alpha}\over \longrightarrow}
- \exp\{\pi \, i\,\alpha^{j} \}
   \psi^{j}}\,\, ,
  \eqno\eqnlabel{trans1}
$$
where $\psi^j$ represents either a real fermion $\psi^{n}$ or a complex
fermion $\psi^{n,m}$ and $-1<\alpha^{j}\leq 1$.
(For the $l_\beta$ loop, $\beta^{j}$ replaces $\alpha^{j}$.)
The boundary conditions for an unpaired real fermion $\psi^{n}$ must be either
periodic (Ramond) or antiperiodic (Neveu-Schwarz),
{\it i.e.}
$\alpha^n,\, \beta^n \in \{ 1 , 0 \}$;
whereas, $\alpha^{n,m}$ and $\beta^{n,m}$ can both be rational
for Weyl fermions $\psi^{n,m}$.
For each loop,
the set of phases $\alpha^{j}$ and $\beta^{j}$, respectively,
define the 64-dimensional boundary vectors ${\balpha}$ and ${\bbeta}$.

The contribution to the one-loop partition function, ${\cal Z}_{\rm fermion}$,
from the WS fermions with their chosen sets of boundary vectors,
$\{\balpha\}$ and $\{\bbeta\}$, can be expressed as a weighted summation
over the individual partition functions, ${\cal Z} ({\balpha\atop\bbeta})$,
 for specific pairs of boundary vectors,
$${\cal Z}_{\rm fermion}= \sum_{{\balpha}\in\{ {\balpha}\}\atop
                         {\bbeta}\in\{ {\bbeta}\} }
C\left({\balpha\atop\bbeta}\right)
Z\left({\balpha\atop\bbeta}\right)\,\, .
\eq{partfunc}$$
The weights $C({\balpha\atop\bbeta})$ can be either complex or real ($\pm 1$)
phases when either $\balpha$ or $\bbeta$ have rational, non-integer components,
but only real phases when $\balpha$ and $\bbeta$
are both integer vectors. One-loop modular invariance requires that
$\{ {\balpha}\}$ and $\{ {\bbeta}\}$ be identical sets and that if
$\balpha$ and $\bgamma$ are in $\{ {\balpha}\}$ then
${\balpha} + {\bgamma}$ must be also.
Thus,
$\{ {\balpha}\}$ and $\{ {\bbeta}\}$ can be defined by
a set of basis vectors (BVs) $\{ \bV_i\}$:
$$\balpha = \sum_{i=1}^{D} a_i \bV_i\,\, \mod{2}\, ,\,\,\,\,
  \bbeta  = \sum_{i=1}^{D} b_i \bV_i\,\, \mod{2}\, ,
  \eqno\eqnlabel{expand}
$$
where $a_i$ and $b_i$ are integers in the range 0 to $N_i -1$,
with $N_i$ (the order of $\bV_i$)
the smallest positive integer such that $N_i \bV_i= 0\,\,\,\, \mod{2}$
for all 64 components of $\bV_i$.
The vector $\bV_0$, with 1
for all components, must be present in the basis set, as
required by modular invariance.

Modular invariance dictates the allowed form of the
phase weights:
%\footnote{From hereon we generally
%express the modular invariance constraints
%in the $k_{i,j}$ language of \pr{kawai87a}
%rather than in the equivalent
%$C({\balpha\atop\bbeta})$ language
%of \pr{antoniadis87},
%to correlate with upcoming papers of ours\mpr{cleaver95b, cleaver95d}
%in which the $k_{i,j}$ form has some advantages.}
$$ C( { \balpha \atop \bbeta } )
    = (-1)^{s_{\balpha}+ s_{\bbeta}}
        \exp \{
             \pi i \, \sum_{i,j} b_i
                      \,( k_{i,j} - \half\, \bV_i\cdot \bV_j  )\,
                                 a_j \}  \, ,
\eqno\eqnlabel{cab}
$$
where $s_{\balpha}$ is the spacetime component of $\balpha$,
{\it etc.} ($s_i$ is similarly used for $\bV_i$ below.)
The rational numbers $k_{i,j}$, where $-1< k_{i,j}\leq 1$,
and BVs $\bV_i$ are constrained by,
\subequationnumstyle{alphabetic}
$$\eqalignno{
k_{i,j} + k_{j,i} & = \half\, \bV_i\cdot \bV_j\,\,\,\, \mod{2} ,
                 &\eqnlabel{const-a}\cr
     N_j k_{i,j} & = 0 \,\, \mod{2}\,  ,
                 &\eqnlabel{const-b}\cr
k_{i,i} + k_{i,0} & = - s_i + \fourth\, {\bV_i}\cdot {\bV_i}\,\,\,\,
\mod{2}\, .
                 &\eqnlabel{const-c}}
$$
\subequationnumstyle{alphabetic}
%$$\eqalignno{
%C({{\balpha}\atop{\bbeta}})
%    & = (-1)^{\balpha_{1,2}}\exp\left{2\pi i\, r/N_{\balpha}\right}\,
%      & \eqnlabel{coef-a}\cr
%    & = (-1)^{\bbeta_{1,2}} \exp\left{2\pi i\, s/N_{\bbeta}\right}
%                            \exp\left{\pi i\, \balpha*\bbeta/2\right}
%      & \eqnlabel{coef-b}\cr}
%$$
\subequationnumstyle{blank}
After appropriate integer multiplication, the dependence upon the $k_{i,j}$
can be removed to yield three direct constraints on the $\bV_i$:
\subequationnumstyle{alphabetic}
$$\eqalignno{
N_{i,j} \bV_i\cdot \bV_j & = 0 \,\,\,\, \mod{4}\, &\eqnlabel{bv-a}\cr
N_{i}   \bV_i\cdot \bV_i & = 0 \,\,\,\, \mod{8}\, &\eqnlabel{bv-b}\cr
{\rm The~number~of~real~fermio}&{\rm ns~simultaneously~periodic}\cr
       {\rm for~any~three~basi}&{\rm s~vectors~is~even.}&\eqnlabel{bv-c}}
$$
$N_{i,j}$ is the lowest common multiple of $N_i$ and $N_j$.
\subequationnumstyle{blank}

One of the non-contractible
loops, $l_{\alpha}$, may be regarded as space-like,
and the other loop, $l_{\beta}$, as time-like.
Each  ${\balpha}$ corresponds to a sector of excitations of the vacuum by
fermion modes at frequencies $\mu_{\psi^j}$ proportional to $\alpha^{j}$:
$$\mu_{\psi^j} = \left\{ {1 \pm \alpha^j\over 2} + \Z \right\}\,\, ,
  \eq{freq}$$
($+$ for $\psi^j$ and $-$ for $\psi^{j\, *}$).
In sector $\balpha$, each complex Weyl fermion $\psi^{n,m}$
carries a U(1) charge, $Q(\psi^{n,m})$, proportional
to its boundary condition:
$$Q(\psi^{n,m})= \alpha^{n,m}/2 + F(\psi^{n,m}) \,\, .\eq{qcharge}$$
$F$ is the fermion number operator with eigenvalues
%(that are dependent upon the specific state $F$ is acting on)
%of
$\{0,\pm 1 \}$ for non-periodic fermions and $\{0,-1\}$ for periodic.
Hence, the charge $Q(\psi^{n,m})$ has possible values of $\{0,\pm 1\}$
for antiperiodic fermions, and $\{\pm\half\}$ for periodic.

%The gauge group in a model
%depends on the states in the sectors.

%Together, the charges of all states in all sectors form a
%lattice upon which the roots and weights of an algebra
%can be embedded.

The boundary vectors $\bbeta$
contribute a set of GSOPs that act on the
$\balpha$-sector states,
projecting some of them out of the model.
Which states survive is a function
of the phase coefficients
$\{ C({  {\balpha} \atop {\bbeta} } ) \}$
(or equivalently of the
$\{C({{\balpha}\atop \bV_i })\}$).
In a given $\balpha $-sector, a state is removed from the model unless it
satisfies the GSOP eq. imposed by each $\bV_j$:
\subequationnumstyle{alphabetic}
$$ \bV_j\cdot {\bmit F}_{\balpha} = \left(\sum_i k_{j,i} a_i\right) + s_j
    - \half\, \bV_j\cdot {\balpha}\,\,\,\, \mod{2}\, ,
  \eqno\eqnlabel{gso1-a}$$
or, equivalently,
$$ \bV_j\cdot {\bmit Q}_{\balpha} = \left(\sum_i k_{j,i} a_i\right) + s_j
\,\,\,\, \mod{2}\, .
  \eqno\eqnlabel{gso1-b}$$
\subequationnumstyle{blank}
Eqs.~(\puteqn{gso1-a}-b) imply
that if a state with charge vector $\bQ$ in sector $m\balpha$
survives a given set of GSOPs,
then a state with charge vector $-\bQ$ in sector $(N_{\balpha}-m)\balpha$
also survives this same set.
This is guaranteed because $s_j = 0$, 1 and
\subequationnumstyle{alphabetic}
$$\eqalignno{
(N_i -m)\bS_i       & = -m\bS_i\,\,\,\, \mod{2}
                    & \eqnlabel{identn-a}\cr
(N_i- m)k_{j,\bS_i} &= - mk_{j,\bS_i}\,\,\,\, \mod{2}.
                    & \eqnlabel{identn-b}}
$$
\subequationnumstyle{blank}
Since $\bQ_{\alpha}(\psi^{1,2})$ gives the chirality of a spacetime
fermion, a model with a surviving spacetime fermion of a given chirality
in sector $m\balpha$ must also have a
surviving spacetime fermion of opposite chirality and opposite internal
charges in sector $(N_{\balpha}-m)\balpha$.

{\bigfonts\bf\section{Number of Supersymmetries in Free Fermionic Strings}}

%\subsectionnumstyle{alphabetic}
\ss{2.a Classes of Basis Vectors}
%\subsectionnumstyle{blank}

In $D$-dimensional heterotic free fermionic models,
the $3(10-D)$ real internal LM fermions (henceforth denoted
by the $\chi^I$)
non-linearly realize a worldsheet
supersymmetry using a supercurrent of the form\mpr{antoniadis87,kawai87a}
$$ T_{\rm F} = \psi^{\mu}\partial X_{\mu} + f_{IJK}\chi^I\chi^J\chi^K\, .
\eqno\eqnlabel{scharge}$$
The $f_{IJK}$ are the structure constants of a semi-simple Lie algebra
$\cal L$ of dimension $3(10-D)$.  Four dimensional models can involve
any one of the three 18-dimensional Lie algebras: SU(2)$^6$,
SU(2)$\otimes$SU(4), and SU(3)$\otimes$SO(5).
When $T_{\rm F}$ is transported around the non-contractible loops on the
worldsheet, it must transform the same as $\psi^{\mu}$ does,
periodically for spacetime fermions and antiperiodically for
spacetime bosons. This requirement severely constrains the BVs in consistent
models. Each BV must represent an automorphism (up to a minus sign) of the
chosen algebra, since $f_{IJK}\chi^I\chi^J\chi^K$ must also transform
as $\psi^{\mu}$ does.

The simplest modular invariant heterotic four-dimensional string
model built from free fermions is non-supersymmetric.
This model contains a single BV, the all-periodic $\bV_0$, which means
it has only two sectors: $\bV_0$ and the
all antiperiodic, $\vec 0 \equiv \bV_0 + \bV_0$.
%$$\eqalignno{
%\bV_0   & = \left\{ \hbox to .5cm {$1^{2}$\hfill} \htwo 1^{18} \mid 1^44
%\right\}\cr
%\vec 0  & = \left\{ \hbox to .5cm {$0^{2}$\hfill} \htwo 0^{18} \mid 0^44
%\right\}
%        & \eqnlabel{so44} }
%$$
The graviton, dilaton, antisymmetric tensor, and spin-1 gauge particles
all originate in the $\vec 0$ sector.
Each of the three choices for Lie algebra
allow various possibilities for an additional BV, $\bmit S_i$,
that satisfies the automorphism constraint and can also contribute
massless gravitinos.
Every $\{ \bV,\, \bS_i \}$ set generates an $N=4$ supergravity model.
Additional BVs (with their corresponding GSOPs)
must be added to reduce the number of spacetime supersymmetries below four.
Ref.~\pr{dreiner89b} indicates that neither
SU(2)$\otimes$SU(4) nor SU(3)$\otimes$SO(5) algebras can be used to
obtain $N=1$ ST-SUSY.
This work also presents two examples of different BV combinations
(one being the NAHE set\mpr{faraggix}
 that can yield $N=1$ for SU(2)$^6$, while also
revealing a situation where presence of a specific BV forbids $N=1$.

Our objective is to continue the work begun in \pr{dreiner89b}.
That is, we {\it completely} classify the sets of LM
BVs that can produce  exactly $N=1$ ST-SUSY
(and $N=4$, 2, and 0 ST-SUSY solutions in the process).
Therefore, we select SU(2)$^6$ for the supercurrent's Lie algebra
which gives (\puteqn{scharge}) the form of,
$$ T_{\rm F} = \psi^{\mu}\partial X_{\mu}
             + i\sum_{J=1}^6 \chi^{3J}\chi^{3J+1}\chi^{3J+2}\, .
\eqno\eqnlabel{scharge2}$$
The fermion triplet $(\chi^{3J},\, \chi^{3J+1},\, \chi^{3J+2})$
yields the three generators of the $J^{th}$ SU(2).
The generators for each SU(2)
can be written using either the Cartan-Weyl (CW) basis
($J_3$, $J_+$, and $J_-$) or the non-Cartan-Weyl basis
($J_3$, $J_1$, and $J_2$).

An automorphism of SU(2)$^6$ is
the product of inner automorphisms for the individual SU(2) algebras and
an outer automorphism of the whole SU(2)$^6$ product
algebra.\mpr{antoniadis88,dreiner89b}
An outer automorphism can be expressed as an element of the permutation group
$P_6$ that mixes the SU(2) algebras.\mpr{dreiner89b}
The elements of $P_6$ can be resolved into disjoint commuting
cycles, and fit into eleven classes defined by the different
possible lengths, $n_k$, of the cycles in the permutation
(with a set of lengths written as $n_1\cdot n_2\cdots n_i$)
such that $\sum_k n_k = 6$. The set of these eleven classes is
$$\eqalignno{{\bmit n} \in \{
           & 1\cdot1\cdot1\cdot1\cdot1\cdot1,\,\,\,\,
             2\cdot1\cdot1\cdot1\cdot1,\,\,\,\, 2\cdot2\cdot1\cdot1,\,\,\,\,
             2\cdot2\cdot2,\cr
           & 3\cdot1\cdot1\cdot1,\,\,\,\, 3\cdot2\cdot1,\,\,\,\,
             3\cdot3,\,\,\,\,   4\cdot1\cdot1,\,\,\,\, 4\cdot2,\,\,\,\,
             5\cdot1,\,\,\,\, 6 \}\, .
           &\eqnlabel{perms}}$$
The first element in this set,
$1\cdot1\cdot1\cdot1\cdot1\cdot1$, is the $P_6$ identity element,
which does not permute any of the six individual SU(2)$_J$ algebras.
\subequationnumstyle{alphabetic}
$$ 1\cdot1\cdot1\cdot1\cdot1\cdot1
   :\quad (\chi^{3J},\, \chi^{3J+1},\, \chi^{3J+2})\leftrightarrow
   (\chi^{3J},\, \chi^{3J+1},\, \chi^{3J+2})\,\,
{\rm ~for~} J= 1 {\rm ~to~} 6\, .
\eqno\eqnlabel{perm-a}$$
$2\cdot1\cdot1\cdot1\cdot1$ denotes a cyclic permutation between two
SU(2) algebras, with $J$ subscripts indicating which two, \eg
$$2_{1,2}\cdot1\cdot1\cdot1\cdot1:\quad
(\chi^3,\, \chi^4,\, \chi^5),\, \leftrightarrow
(\chi^6,\, \chi^7,\, \chi^8)\, .
\eqno\eqnlabel{perm-b}$$
Similarly, $2\cdot2\cdot1\cdot1$ denotes two separate
cyclic permutations between two pairs of
SU(2) algebras, \eg
$$\eqalignno{
2_{1,2}\cdot2_{3,4}\cdot1\cdot1:\quad(\chi^3,\, \chi^4,\, \chi^5)
         & \leftrightarrow  (\chi^6,\, \chi^7,\, \chi^8)\, ,\cr
(\chi^9,\, \chi^{10},\, \chi^{11})
         & \leftrightarrow  (\chi^{12},\, \chi^{13},\, \chi^{14})\, .
  & \eqnlabel{perm-c}}
$$
while, as our last example,
$3\cdot1\cdot1\cdot1$ performs a cyclic permutation between three
SU(2) algebras, \eg
$$\eqalignno{3_{1,2,3}\cdot1\cdot1\cdot1:\quad
(\chi^3,\, \chi^4,\, \chi^5)
  & \rightarrow (\chi^6,\, \chi^7,\, \chi^8)\, ,\cr
(\chi^6,\, \chi^{7},\, \chi^{8})
  & \rightarrow (\chi^{9},\, \chi^{10},\, \chi^{11})\, ,
  & \eqnlabel{perm-d}\cr
(\chi^9,\, \chi^{10},\, \chi^{11})
  & \rightarrow  (\chi^{3},\, \chi^{4},\, \chi^{5})\, .}
$$
\subequationnumstyle{blank}
All eleven classes of cyclic permutations
can be diagonalized in a new fermion
basis by similarity transformations. The eigenvalue contributions from
a non-disjoint cyclic permutation of length $n$ is the set of
$n^{\rm th}$-roots of 1.

For spacetime fermions, diagonalized inner automorphisms of a
SU(2) triplet gives phases
$(1,\, \exp\{i\pi\theta\},$ $ \exp\{-i\pi\theta\})$,
with $0\leq\theta<2$.
For generic $\theta$
the triplet eigenstates can only be written in the CW basis.
However, for the particular values of
$\theta = 0$, $1$ the eigenstates can
correspond to non-CW generators.
Besides $(1,\, -1,\, -1)$ and $(1,\, 1,\, 1)$
for $\theta = 0,1$ in the CW basis,
the non-CW basis can include the additional eigenvalue sets of
$(-1,\, 1,\, -1)$ and $(-1,\, -1,\, 1)$.

When inner and outer automorphisms are combined, the number of independent
$\theta$ is reduced from one for each
$(\chi^{3J},\chi^{3J+1},\chi^{3J+2})$ triplet
to one for each non-disjoint cycle of length $n_k$ in the
outer automorphism. Table 1 shows the eigenvalues for a given cycle and
$\theta$ combination. For odd $n_k$, the
contribution to the GSOP from the fermions in the cyclic permutation
is of order $2n_k$ when $\theta = 0,1$; whereas, for even $n_k$,
it is of order $2n_k$ when $\theta= 1$, and of order $n_k$ when
$\theta = 0$.

Massless sectors correspond to $\theta = 1$
for all non-disjoint cycles in a permutation.
This means that massless sectors can be written in either the
CW or the non-CW bases,
depending on the basis required by massive sectors.
Any other value (including 0) for any of the $\theta$'s
increases the vacuum mass of a sector.
Thus, consistent models {\it cannot}
contain specetime fermion sectors with tachyonic vacuums, since such
sectors would not correspond to automorphisms of the WS supercurrent.

Table 2 gives the $\alpha^j$ boundary vector components associated with the
$-\exp \{\pi i \alpha^j\}$ eigenvalues
in the eleven massless LM sectors $\bS_i$. These sectors
are formed by combining $\theta= 1$ in the inner automorphisms
with a representative element from each the eleven classes
of $P_6$ cyclic permutations listed above.
Note that $\theta = 1$ for a cycle gives exactly one real periodic fermion
(in that cycle). Thus, massless sectors with an even number
of LM real periodic fermions must be built from
permutation classes having an even number of disjoint cycles.
Since the RM part of a gravitino sector is completely antiperioidic
and the number of real LM periodic fermions in a gravitino-generating
sector must be even by rule (\puteqn{bv-c}),
the six $\bS_i$ qualifying as gravitino-sources are:
$\bS_1$ ($1\cdot1\cdot1\cdot1\cdot1\cdot1$ class), $\bS_3$
($2\cdot2\cdot1\cdot1$
class),
$\bS_5$ ($3\cdot1\cdot1\cdot1$ class),     $\bS_7$ ($3\cdot3$ class),
$\bS_9$ ($4\cdot2$ class), and $\bS_{10}$ ($5\cdot1$ class).\mpr{dreiner89b}
These sectors are of orders $N_1= 2$,
$N_3= 4$, $N_5= 6$, $N_7= 6$, $N_1= 8$, and $N_5= 10$, respectively.
Models formed simply from the periodic sector $\bV_0$ and
any one of these six $\bS_i$ yield $N=4$ ST-SUSY after GSOPs.
The four gravitinos are dispersed among the sectors $m\bS_i$, where
$m= 1,$ $3,$ $5,\dots , N_i - 1$.
(Therefore, only for $\bS_1$ do all four gravitinos
appear in the same sector.)

%\subsectionnumstyle{alphabetic}
\ss{2.b General Requirements of ${\bmit N = }$ 1 Solutions}
%\subsectionnumstyle{blank}

If any one of the six gravitino-generating $\bS_i$ is to
produce $N=1$ ST-SUSY,
additional BVs giving satisfactory GSOPs
must be added to the initial $\{ \bV_0,\, \bS_i\}$ set.
(We often refer to these additional BVs as reduction vectors.)
Focusing on ($N=1$) supersymmetric free fermionic models offers
one obvious simplification for model classification:
whether a basis vector is a spacetime fermion or boson is of no
physical consequence.
Therefore, we can choose that all BV be spacetime fermions, which
is of use in analysis of GSOPs.
Another non-physical degree of freedom in $N=1$ models
is the choice of which sector $m\bS_i$ (of order $N_{\bS_i}$)
should produce the one surviving gravitino.
We choose the LM gravitino to always appear in $\bS_i$.
Our discussion following eq.~(\puteqn{gso1-a}-b)
indicates that if a gravitino of given chirality and
charge vector $\bQ$ in sector $m\bS_i$ survives a set of GSOPs,
a gravitino of opposite chirality and charge vector $-\bQ$ in sector
$(N_{S_i}-m)\bS_i$ survives also.
Hence,
$N=1$ ST-SUSY implies that the set of GSOPs from the reduction vectors
must act together to form a chiral operator that
projects out all gravitinos from $\bS_i$ (for $i\ne 1$)
except one left-handed. Then only a single right-handed gravitino
remains in $(N_{S_i}-1)\bS_1$.

For a consistent model, the additional BVs
must be derived from automorphisms that commute among themselves
and with $\bS_i$.
This requires the various SU(2)$^6$ permutations
to commute, necessitating
a disjoint cycle or product of disjoint cycles in one BV to be
a power or ``root" of disjoint cycles in all other BVs.\mpr{dreiner89b}
For example, $\bS_5$ uses the permutation
$3_{1,2,3}\cdot1_{4}\cdot1_{5}\cdot1_{6}$.
Together the identity transformations,
$1_{4}\cdot1_{5}\cdot1_{6}$,
for the last three SU(2) algebras,
%$$ 1_{4}\cdot1_{5}\cdot1_{6}:\,\,\,\, (\chi^{3J},\, \chi^{3J+1},\,
%%\chi^{3J+2})
%   \rightarrow (\chi^{3J},\, \chi^{3J+1},\, \chi^{3J+2})\,\,\,\,
%{\rm ~for~} J= 4,\, 5,\, 6\, ,
%\eqno\eqnlabel{perm3111}
%$$
correspond to the cube of the $3_{4,5,6}$ permutation
%$$\eqalignno{
% 3_{4,5,6}:\,\,\,\, (\chi^{12},\, \chi^{13},\, \chi^{14}) & \rightarrow
%     (\chi^{15},\, \chi^{16},\, \chi^{17})\, ,\cr
%     (\chi^{15},\, \chi^{16},\, \chi^{17}) & \rightarrow
%     (\chi^{18},\, \chi^{19},\, \chi^{20})\, ,& \cr
%     (\chi^{18},\, \chi^{19},\, \chi^{20}) & \rightarrow
%     (\chi^{12},\, \chi^{13},\, \chi^{14})\, .&
%\eqnlabel{perm33}}
%$$
and to the square of the $2_{4,5}\cdot1_{6}$ permutation,
%$$
% 2_{4,5}\cdot1{6}:\,\,\,\,
%     (\chi^{12},\, \chi^{13},\, \chi^{14}) \leftrightarrow
%     (\chi^{15},\, \chi^{16},\, \chi^{17})\, .
%\eqno\eqnlabel{perm21}$$
Similarly, $3_{1,2,3}$ commutes only with itself, with its square
$3_{1,3,2}$, and with its cube
$1_{1}\cdot1_{2}\cdot1_{3}$.
Since  $3_{4,5,6}$ and $2_{4,5}\cdot1_{6}$ do not commute, there are two
(unique classes of) permutation sets we can use to form the reduction vectors
for $\bS_5$:
$$\eqalignno{
\left\{ 3_{1,2,3}\cdot1_{4}\cdot1_{5}\cdot1_{6},\,
           3_{1,2,3}\cdot3_{4,5,6},\,
           1_{1}\cdot1_{2}\cdot1_{3}\right.
           &\left. \cdot3_{4,5,6},\,
           1_{1}\cdot1_{2}\cdot1_{3}\cdot1_{4}\cdot1_{5}\cdot1_{6}\right\}
\, ,
&\eqnlabel{commute3111b}\cr
{\rm and}&\cr
\left\{ 3_{1,2,3}\cdot1_{4}\cdot1_{5}\cdot1_{6},\,
           3_{1,2,3}\cdot2_{4,5}\cdot1_{6},\,
           1_{1}\cdot1_{2}\cdot1_{3} \right.
           &\left. \cdot2_{4,5}\cdot1_{6},\,
           1_{1}\cdot1_{2}\cdot1_{3}\cdot1_{4}\cdot1_{5}\cdot1_{6}\right\}
           \, .
&\eqnlabel{commute3111c}}$$
The potential set of reduction vectors is formed by combining
the allowed cyclic permutations with inner automorphisms.

Generically, $N=1$ ST-SUSY requires the non-CW basis
with $\theta= 0,1$ for inner automorphisms.\mpr{dreiner89b}
Therefore, the inner automorphisms that may be combined with a given
set of permutations are $(1,-1,-1)$, $(-1,1,-1)$, $(-1,-1,1)$,
and $(1,1,1)$ for spacetime fermions.
One constraint on inner automorphisms
occurs when a product of cycles of equal length
is a power of a single larger cycle also appearing,
for example, when $1_{4}\cdot1_{5}\cdot1_{6}$ and $3_{4,5,6}$
are both used. In this event, all cycles in the product must be
tensored with the same inner automorphism.
A second constraint is that LM BVs must satisfy
(\puteqn{bv-a}-c) when $\bS_i$ is one of the $\bV_i$.

In Table 3,
we find all boundary vectors formed from powers of
a non-disjoint cyclic permutation with length $n_k$ in the range
$1\leq n_k\leq 6$ with $\theta=1,0$.
These form maximal sets of commuting boundary vectors of length $n_k$.
Boundary vectors that could be used to construct the unique
massless gravitino generators are found.
The independent classes of GSOPs are then determined for these
specific boundary vectors.
We use these results in the next subsection to form all possible
sets of reduction vectors and their related GSOPs on each of
the six gravitino generators.
For a given gravitino generator we classify the reduction sets by
specifying the longest length independent cycles appearing in a reduction set.
If we include the gravitino generator in its own reduction set, then
there are 11 classes of models for $\bS_1$,
four classes for $\bS_3$, two classes for $\bS_5$, two for $\bS_7$, and
one each for $\bS_9$ and $\bS_{10}$.
We analyze the sets of GSOPs to determine which (if any)
are sufficient to produce {\it exactly} $N=1$ ST-SUSY from
a given gravitino generator.

%\subsectionnumstyle{alphabetic}
\ss{2.c Classification of Supersymmetry Solutions}
%\subsectionnumstyle{blank}

%\input table4.tex

\def\noc{\no}

\def\poss{POSSIBLE GSO PROJECTIONS ON  }
\def\secondz{The second GSOP reduces the initial $N=4$ ST-SUSY to $N=2$
or $N=0$.  }

\def\second{The second GSOP reduces the initial $N=4$ ST-SUSY to $N=2$.  }

\def\NAHE{The first four GSOPs give the standard NAHE set of GSOPs,
although the basis vectors generating these GSOPs
do not correspond to the NAHE basis vectors.
The combination of any three of these four GSOPs produce $N=1$ ST-SUSY,
independent of their RHS values.
The remaining GSOP either keeps or breaks $N=1$,
depending upon its RHS value.  }

\def\FNAHE{The first four GSOPs imitate the NAHE set.
The combination of any three of these four GSOPs produce $N=1$ ST-SUSY
independent of their RHS values.
The remaining GSOP either keeps or breaks $N=1$,
depending upon its RHS value.  }

\def\vfa{\vskip .3truecm }
\def\hbp{\no\hbox to 16.5truecm}
\def\haf{\hbox to .4truecm}
\def\hbf{\hbox to .7truecm}

\def\headf{\vskip .3truecm
\hrule height0.1truecm
\vskip .3truecm}

\def\dotf{
\no.\dotfill.

\vskip .1truecm}

\def\st{ {\rm st}}

\def\clss{\hbox to 1.0truecm{\hfill{class }}}

\def\hast{\hbox to 0.4truecm{\hfill$\ast$\hfill}}
\def\phast{\hbox to 0.4truecm{\hfill$\phantom{\ast}$\hfill}}

\def\ran{\hbox to 1.8truecm{\hfill$\rightarrow N=0$\hfill}}
\def\phn{\hbox to 1.8truecm{\hfill$\phantom{\rightarrow N=0}$\hfill}}

\def\cy{\hbox to 1.00truecm{\hfill\clss}}
\def\cn{\hbox to 1.45truecm{\hfill\hast\clss}}
\def\cyz{\hbox to 3.25truecm{\hfill\ran\phast\clss}}
\def\cnz{\hbox to 3.25truecm{\hfill\ran\hast\clss}}

%**********************************************************
%1
\def\gayo#1#2#3#4#5{
\hbp{\haf{\hfill $1$} $F_0$\hbf{\hfill $#1$} $F_1$\hbf{\hfill $#2$}
$F_2$\hbf{\hfill $#3$} $F_3
\in \{ #4\}$\hfill\cy\hbox to 2.2truecm{$#5$\hfill}}}

\def\gano#1#2#3#4#5{
\hbp{\haf{\hfill $1$} $F_0$\hbf{\hfill $#1$} $F_1$\hbf{\hfill $#2$}
$F_2$\hbf{\hfill $#3$} $F_3
\in \{ #4\}$\hfill\cn\hbox to 2.2truecm{$#5$\hfill}}}

\def\gbyo#1#2#3#4{
\hbp{\haf{\hfill $1$} $F_0$\hbf{\hfill $#1$} $F_1$\hbf{\hfill $#2$}
$F_2
\in \{ #3\}$\hfill\cy\hbox to 2.2truecm{$#4$\hfill}}}

\def\gbno#1#2#3#4{
\hbp{\haf{\hfill $1$} $F_0$\hbf{\hfill $#1$} $F_1$\hbf{\hfill $#2$} $F_2
\in \{ #3\}$\hfill\cn\hbox to 2.2truecm{$#4$\hfill}}}

\def\gcyo#1#2#3{
\hbp{\haf{\hfill $1$} $F_0$\hbf{\hfill $#1$} $F_1
\in \{ #2\}$\hfill\cy\hbox to 2.2truecm{$#3$\hfill}}}

\def\gcno#1#2#3{
\hbp{\haf{\hfill $1$} $F_0$\hbf{\hfill $#1$} $F_1
\in \{ #2\}$\hfill\cn\hbox to 2.2truecm{$#3$\hfill}}}

\def\gayzo#1#2#3#4#5{
\hbp{\haf{\hfill $1$} $F_0$\hbf{\hfill $#1$} $F_1$\hbf{\hfill $#2$}
$F_2$\hbf{\hfill $#3$} $F_3
\in \{ #4\}$\hfill\cyz\hbox to 2.2truecm{$#5$\hfill}}}

\def\ganzo#1#2#3#4#5{
\hbp{\haf{\hfill $1$} $F_0$\hbf{\hfill $#1$} $F_1$\hbf{\hfill $#2$}
$F_2$\hbf{\hfill $#3$} $F_3
\in \{ #4\}$\hfill\cnz\hbox to 2.2truecm{$#5$\hfill}}}

\def\gbyzo#1#2#3#4{
\hbp{\haf{\hfill $1$} $F_0$\hbf{\hfill $#1$} $F_1$\hbf{\hfill $#2$}
$F_2 \in \{ #3\}$\hfill\cyz\hbox to 2.2truecm{$#4$\hfill}}}

\def\gbnzo#1#2#3#4{
\hbp{\haf{\hfill $1$} $F_0$\hbf{\hfill $#1$} $F_1$\hbf{\hfill $#2$} $F_2
\in \{ #3\}$\hfill\cnz\hbox to 2.2truecm{$#4$\hfill}}}

\def\gcyzo#1#2#3{
\hbp{\haf{\hfill $1$} $F_0$\hbf{\hfill $#1$} $F_1
\in \{ #2\}$\hfill\cyz\hbox to 2.2truecm{$#3$\hfill}}}

\def\gcnzo#1#2#3{
\hbp{\haf{\hfill $1$} $F_0$\hbf{\hfill $#1$} $F_1
\in \{ #2\}$\hfill\cnz\hbox to 2.2truecm{$#3$\hfill}}}

%**********************************************************
%**********************************************************
%2
\def\gayw#1#2#3#4#5{
\hbp{\haf{\hfill $1$} $F_0$\hbf{\hfill $#1$} $F_1$\hbf{\hfill $#2$}
$F_2$\hbf{\hfill $#3$} $F_3
\in \{ #4\}$\hfill\cy\hbox to 3.65truecm{$#5$\hfill}}}

\def\ganw#1#2#3#4#5{
\hbp{\haf{\hfill $1$} $F_0$
\hbf{\hfill $#1$} $F_1$\hbf{\hfill $#2$} $F_2$\hbf{\hfill $#3$} $F_3
\in \{ #4\}$\hfill\cn\hbox to 3.65truecm{$#5$\hfill}}}

\def\gbyw#1#2#3#4{
\hbp{\haf{\hfill $1$} $F_0$\hbf{\hfill $#1$} $F_1$\hbf{\hfill $#2$} $F_2
\in \{ #3\}$\hfill\cy\hbox to 3.65truecm{$#4$\hfill}}}

\def\gbnw#1#2#3#4{
\hbp{\haf{\hfill $1$} $F_0$\hbf{\hfill $#1$} $F_1$\hbf{\hfill $#2$} $F_2
\in \{ #3\}$\hfill\cn\hbox to 3.65truecm{$#4$\hfill}}}

\def\gcyw#1#2#3{
\hbp{\haf{\hfill $1$} $F_0$\hbf{\hfill $#1$} $F_1
\in \{ #2\}$\hfill\cy\hbox to 3.65truecm{$#3$\hfill}}}

\def\gcnw#1#2#3{
\hbp{\haf{\hfill $1$} $F_0$\hbf{\hfill $#1$} $F_1
\in \{ #2\}$\hfill\cn\hbox to 3.65truecm{$#3$\hfill}}}

\def\gayzw#1#2#3#4#5{
\hbp{\haf{\hfill $1$} $F_0$\hbf{\hfill $#1$} $F_1$\hbf{\hfill $#2$}
$F_2$\hbf{\hfill $#3$} $F_3
\in \{ #4\}$\hfill\cyz\hbox to 3.65truecm{$#5$\hfill}}}

\def\ganzw#1#2#3#4#5{
\hbp{\haf{\hfill $1$} $F_0$\hbf{\hfill $#1$} $F_1$\hbf{\hfill $#2$}
$F_2$\hbf{\hfill $#3$} $F_3
\in \{ #4\}$\hfill\cnz\hbox to 3.65truecm{$#5$\hfill}}}

\def\gbyzw#1#2#3#4{
\hbp{\haf{\hfill $1$} $F_0$\hbf{\hfill $#1$} $F_1$\hbf{\hfill $#2$} $F_2
\in \{ #3\}$\hfill\cyz\hbox to 3.65truecm{$#4$\hfill}}}

\def\gbnzw#1#2#3#4{
\hbp{\haf{\hfill $1$} $F_0$\hbf{\hfill $#1$} $F_1$\hbf{\hfill $#2$} $F_2
\in \{ #3\}$\hfill\cnz\hbox to 3.65truecm{$#4$\hfill}}}

\def\gcyzw#1#2#3{
\hbp{\haf{\hfill $1$} $F_0$\hbf{\hfill $#1$} $F_1
\in \{ #2\}$\hfill\cyz\hbox to 3.65truecm{$#3$\hfill}}}

\def\gcnzw#1#2#3{
\hbp{\haf{\hfill $1$} $F_0$\hbf{\hfill $#1$} $F_1
\in \{ #2\}$\hfill\cnz\hbox to 3.65truecm{$#3$\hfill}}}

%**********************************************************
%**********************************************************
%2b
\def\gaywb#1#2#3#4#5{
\hbp{\haf{\hfill $1$} $F_0$\hbf{\hfill $#1$} $F_1$\hbf{\hfill $#2$}
$F_2$\hbf{\hfill $#3$} $F_3
\in \{ #4\}$\hfill\cy\hbox to 4.05truecm{$#5$\hfill}}}

\def\ganwb#1#2#3#4#5{
\hbp{\haf{\hfill $1$} $F_0$
\hbf{\hfill $#1$} $F_1$\hbf{\hfill $#2$} $F_2$\hbf{\hfill $#3$} $F_3
\in \{ #4\}$\hfill\cn\hbox to 4.05truecm{$#5$\hfill}}}

\def\gbywb#1#2#3#4{
\hbp{\haf{\hfill $1$} $F_0$\hbf{\hfill $#1$} $F_1$\hbf{\hfill $#2$} $F_2
\in \{ #3\}$\hfill\cy\hbox to 4.05truecm{$#4$\hfill}}}

\def\gbnwb#1#2#3#4{
\hbp{\haf{\hfill $1$} $F_0$\hbf{\hfill $#1$} $F_1$\hbf{\hfill $#2$} $F_2
\in \{ #3\}$\hfill\cn\hbox to 4.05truecm{$#4$\hfill}}}

\def\gcywb#1#2#3{
\hbp{\haf{\hfill $1$} $F_0$\hbf{\hfill $#1$} $F_1
\in \{ #2\}$\hfill\cy\hbox to 4.05truecm{$#3$\hfill}}}

\def\gcnwb#1#2#3{
\hbp{\haf{\hfill $1$} $F_0$\hbf{\hfill $#1$} $F_1
\in \{ #2\}$\hfill\cn\hbox to 4.05truecm{$#3$\hfill}}}

\def\gayzwb#1#2#3#4#5{
\hbp{\haf{\hfill $1$} $F_0$\hbf{\hfill $#1$} $F_1$\hbf{\hfill $#2$}
$F_2$\hbf{\hfill $#3$} $F_3
\in \{ #4\}$\hfill\cyz\hbox to 4.05truecm{$#5$\hfill}}}

\def\ganzwb#1#2#3#4#5{
\hbp{\haf{\hfill $1$} $F_0$\hbf{\hfill $#1$} $F_1$\hbf{\hfill $#2$}
$F_2$\hbf{\hfill $#3$} $F_3
\in \{ #4\}$\hfill\cnz\hbox to 4.05truecm{$#5$\hfill}}}

\def\gbyzwb#1#2#3#4{
\hbp{\haf{\hfill $1$} $F_0$\hbf{\hfill $#1$} $F_1$\hbf{\hfill $#2$} $F_2
\in \{ #3\}$\hfill\cyz\hbox to 4.05truecm{$#4$\hfill}}}

\def\gbnzwb#1#2#3#4{
\hbp{\haf{\hfill $1$} $F_0$\hbf{\hfill $#1$} $F_1$\hbf{\hfill $#2$} $F_2
\in \{ #3\}$\hfill\cnz\hbox to 4.05truecm{$#4$\hfill}}}

\def\gcyzwb#1#2#3{
\hbp{\haf{\hfill $1$} $F_0$\hbf{\hfill $#1$} $F_1
\in \{ #2\}$\hfill\cyz\hbox to 4.05truecm{$#3$\hfill}}}

\def\gcnzwb#1#2#3{
\hbp{\haf{\hfill $1$} $F_0$\hbf{\hfill $#1$} $F_1
\in \{ #2\}$\hfill\cnz\hbox to 4.05truecm{$#3$\hfill}}}

%**********************************************************
%**********************************************************
%3
\def\gayt#1#2#3#4#5{
\hbp{\haf{\hfill $1$} $F_0$\hbf{\hfill $#1$} $F_1$\hbf{\hfill $#2$}
$F_2$\hbf{\hfill $#3$} $F_3
\in \{ #4\}$\hfill\cy\hbox to 5.8truecm{$#5$\hfill}}}

\def\gant#1#2#3#4#5{
\hbp{\haf{\hfill $1$} $F_0$\hbf{\hfill $#1$} $F_1$\hbf{\hfill $#2$}
$F_2$\hbf{\hfill $#3$} $F_3
\in \{ #4\}$\hfill\cn\hbox to 5.8truecm{$#5$\hfill}}}

\def\gbyt#1#2#3#4{
\hbp{\haf{\hfill $1$} $F_0$\hbf{\hfill $#1$} $F_1$\hbf{\hfill $#2$} $F_2
\in \{ #3\}$\hfill\cy\hbox to 5.8truecm{$#4$\hfill}}}

\def\gbnt#1#2#3#4{
\hbp{\haf{\hfill $1$} $F_0$\hbf{\hfill $#1$} $F_1$\hbf{\hfill $#2$} $F_2
\in \{ #3\}$\hfill\cn\hbox to 5.8truecm{$#4$\hfill}}}

\def\gcyt#1#2#3{
\hbp{\haf{\hfill $1$} $F_0$\hbf{\hfill $#1$} $F_1
\in \{ #2\}$\hfill\cy\hbox to 5.8truecm{$#3$\hfill}}}

\def\gcnt#1#2#3{
\hbp{\haf{\hfill $1$} $F_0$\hbf{\hfill $#1$} $F_1
\in \{ #2\}$\hfill\cn\hbox to 5.8truecm{$#3$\hfill}}}

\def\gayzt#1#2#3#4#5{
\hbp{\haf{\hfill $1$} $F_0$\hbf{\hfill $#1$} $F_1$\hbf{\hfill $#2$}
$F_2$\hbf{\hfill $#3$} $F_3
\in \{ #4\}$\hfill\cyz\hbox to 5.8truecm{$#5$\hfill}}}

\def\ganzt#1#2#3#4#5{
\hbp{\haf{\hfill $1$} $F_0$\hbf{\hfill $#1$} $F_1$\hbf{\hfill $#2$}
$F_2$\hbf{\hfill $#3$} $F_3
\in \{ #4\}$\hfill\cnz\hbox to 5.8truecm{$#5$\hfill}}}

\def\gbyzt#1#2#3#4{
\hbp{\haf{\hfill $1$} $F_0$\hbf{\hfill $#1$} $F_1$\hbf{\hfill $#2$} $F_2
\in \{ #3\}$\hfill\cyz\hbox to 5.8truecm{$#4$\hfill}}}

\def\gbnzt#1#2#3#4{
\hbp{\haf{\hfill $1$} $F_0$\hbf{\hfill $#1$} $F_1$\hbf{\hfill $#2$} $F_2
\in \{ #3\}$\hfill\cnz\hbox to 5.8truecm{$#4$\hfill}}}

\def\gbnztb#1#2#3#4{
\hbp{\haf{\hfill $1$} $F_0$\hbf{\hfill $#1$} $F_1$\hbf{\hfill $#2$} $F_2
\in \{ #3\}$\hfill}

\hbp{\haf{\hfill}\hbf{\hfill}\hbf{\hfill}\hfill\cnz\hbox to
5.8truecm{$#4$\hfill}}}

\def\gcyzt#1#2#3{
\hbp{\haf{\hfill $1$} $F_0$\hbf{\hfill $#1$} $F_1
\in \{ #2\}$\hfill\cyz\hbox to 5.8truecm{$#3$\hfill}}}

\def\gcnzt#1#2#3{
\hbp{\haf{\hfill $1$} $F_0$\hbf{\hfill $#1$} $F_1
\in \{ #2\}$\hfill\cnz\hbox to 5.8truecm{$#3$\hfill}}}

%**********************************************************
%**********************************************************
%4
\def\gayr#1#2#3#4#5{
\hbp{\haf{\hfill $1$} $F_0$\hbf{\hfill $#1$} $F_1$\hbf{\hfill $#2$}
$F_2$\hbf{\hfill $#3$} $F_3
\in \{ #4\}$\hfill\cy\hbox to 6.45truecm{$#5$\hfill}}}

\def\ganr#1#2#3#4#5{
\hbp{\haf{\hfill $1$} $F_0$\hbf{\hfill $#1$} $F_1$\hbf{\hfill $#2$}
$F_2$\hbf{\hfill $#3$} $F_3
\in \{ #4\}$\hfill\cn\hbox to 6.45truecm{$#5$\hfill}}}

\def\gbyr#1#2#3#4{
\hbp{\haf{\hfill $1$} $F_0$\hbf{\hfill $#1$} $F_1$\hbf{\hfill $#2$} $F_2
\in \{ #3\}$\hfill\cy\hbox to 6.45truecm{$#4$\hfill}}}

\def\gbnr#1#2#3#4{
\hbp{\haf{\hfill $1$} $F_0$\hbf{\hfill $#1$} $F_1$\hbf{\hfill $#2$} $F_2
\in \{ #3\}$\hfill\cn\hbox to 6.45truecm{$#4$\hfill}}}

\def\gcyr#1#2#3{
\hbp{\haf{\hfill $1$} $F_0$\hbf{\hfill $#1$} $F_1
\in \{ #2\}$\hfill\cy\hbox to 6.45truecm{$#3$\hfill}}}

\def\gcnr#1#2#3{
\hbp{\haf{\hfill $1$} $F_0$\hbf{\hfill $#1$} $F_1
\in \{ #2\}$\hfill\cn\hbox to 6.45truecm{$#3$\hfill}}}

\def\gayzr#1#2#3#4#5{
\hbp{\haf{\hfill $1$} $F_0$\hbf{\hfill $#1$} $F_1$\hbf{\hfill $#2$}
$F_2$\hbf{\hfill $#3$} $F_3
\in \{ #4\}$\hfill\cyz\hbox to 6.45truecm{$#5$\hfill}}}

\def\ganzr#1#2#3#4#5{
\hbp{\haf{\hfill $1$} $F_0$\hbf{\hfill $#1$} $F_1$\hbf{\hfill $#2$}
$F_2$\hbf{\hfill $#3$} $F_3
\in \{ #4\}$\hfill\cnz\hbox to 6.45truecm{$#5$\hfill}}}

\def\gbyzr#1#2#3#4{
\hbp{\haf{\hfill $1$} $F_0$\hbf{\hfill $#1$} $F_1$\hbf{\hfill $#2$} $F_2
\in \{ #3\}$\hfill\cyz\hbox to 6.45truecm{$#4$\hfill}}}

\def\gbnzr#1#2#3#4{
\hbp{\haf{\hfill $1$} $F_0$\hbf{\hfill $#1$} $F_1$\hbf{\hfill $#2$} $F_2
\in \{ #3\}$\hfill\cnz\hbox to 6.45truecm{$#4$\hfill}}}

\def\gcyzr#1#2#3{
\hbp{\haf{\hfill $1$} $F_0$\hbf{\hfill $#1$} $F_1
\in \{ #2\}$\hfill\cyz\hbox to 6.45truecm{$#3$\hfill}}}

\def\gcnzr#1#2#3{
\hbp{\haf{\hfill $1$} $F_0$\hbf{\hfill $#1$} $F_1
\in \{ #2\}$\hfill\cnz\hbox to 6.45truecm{$#3$\hfill}}}
%**********************************************************
%**********************************************************

\def\extra#1{\no\righttext{\hbox to 4.61truecm{w/ $#1$\hfill}}}

\def\title{LONGEST CYCLIC PERMUTATIONS: }
%length non-disjoint cyclic

\def\gen{ALLOWED GRAVITINO GENERATORS: }
\def\phan#1{$\phantom{\rm #1}$}

%\no
%Sets of reduction vectors are classification as defined above.
%All subgroups of GSOPs leading to $N=1$ ST-SUSY are found.
%(of order $N_{S_i}$)

In this subsection we present the details of our research.
We investigate all possible combinations of unique GSOPs
imposed by the BVs in reduction sets
for each of the six distinct gravitino generators.
The BVs are formed by tensoring various
$S(n_k,y,z,w)$ boundary vectors together with
a periodic spacetime component (denoted by ``$(\st)$"),
$$\bV_i = (\st) \prod_k S(n_k,y,z,w)\,\, ,\eqno\eqnlabel{videfs}$$
such that $\sum_k n_k= 6$.
(By convention we use the
ordering $n_{k_1}\geq n_{k_2}$ if ${k_1}<{k_2}$.)
Thus, only the periodic spacetime component can be combined with
an $S(6,y,z,w)$, whereas
the various $S(5,y,z,w)$ must also be combined with the $S(1,1,1,w')$.
In the latter case, we require $w,w'\in \{1,4\}$ or $w,w'\in \{2,3\}$.
This enables the two real perioidic $J_3$ fermions in
$S(5,y,z,w)$ and $S(1,1,1,w')$ to form a complex fermion, as is required
by (\puteqn{bv-c}) for the related choices of $\bS_1$ and $\bS_{10}$ in
class X below.
There are similar constraints for tensor products of the $S(n_k,y,z,w)$
in the other classes also.

The GSOPs from two generic BVs, $\bV_1$ and $\bV_2$,
in gravitino generator $\bS_i$'s reduction set
are in the same projection class if
(1) their $\bF$ coefficients all match or
differ only by an overall sign factor,
and (2) the difference between the dot products of $\bV_1$ and $\bV_2$ with
$\bS_i$ is $0$ $\mod{4\over N_{\bS_i}}$.
When two GSOPs are in the same class they either work identically
on the gravitino sector or in combination kick out all gravitinos.
Since we desire $N=1$ ST-SUSY, we obviously do not want all
gravitinos removed.
Therefore, once we choose the $K_{{\bV_1},{\bS_i}}$ in $\bV_1$'s
GSOP, $K_{{\bV_2},{\bS_i}}$ is fixed.
For a given reduction set we present one element from each
class of GSOPs. However, the classes of GSOPs for a
given reduction set are not all independent.
For example, when $\bS_1 = (st)S(4,4,1,2)S(2,2,1,2)$
the left-hand sides of the fifth through tenth GSOPs are
linear combinations of those of the first four GSOPs.  (See below.)
Thus, the third general $N=1$ solution for this $\bS_1$ expression
is not independent of the first two.

We define $\bF_0$
as the number operator for periodic $\psi^{\mu}$ and $\bF_{1,2,3}$ for the
(up to) three internal periodic WS fermions.
Recall that $\bF_i\in \{ 0,\,\, -1\}$ for periodic
fermions. In our GSOPs below, we replace
the right-hand side (RHS) of (\puteqn{gso1-a})
with the corresponding set of possible
RHS eigenvalues. For a gravitino state of given chirality and internal
charge vector $\bQ$ to survive a GSOP, the left-hand side (LHS)
of the GSOP equation
must equal one of the possible values of the RHS,
otherwise the gravitino state is projected out of the model.

For any gravitino generator $\bS_i$, the first GSOP
(with all ``+1" $\bF$ coefficients)
is always present, since $\bS_i$ and the
periodic sector $\bV_0$ are both in the class of BVs giving
this projection.
Application of this GSOP to $\bS_i$ and application of the corresponding
all-integer component GSOPs
to the odd multiples of $\bS_i$ always produces $N=4$ ST-SUSY.
This is the starting point for our models below. Hence, we generally do
not discuss the first GSOP of any set.

Several LM BVs that generate GSOPs in our sets
must be paired up with RM BVs of sufficient order
to satisfy (\puteqn{bv-a}-b). Otherwise their GSOPs could not be present.
GSOP equations resulting from BVs
 of this type are denoted by an ``$\ast$'' in front of the
GSOP class name.
Most often such projections result directly in $N=0$ ST-SUSY.
Note, however, that GSOPs from some LM BVs that do not
require specific RM pairing can also automatically eliminate
ST-SUSY. All such GSOPs at odds with $N>0$ are marked with a
``$\rightarrow N=0$" before their class name.

\vfa
\headf
\vfa

\noc XI. \title 6 in $S(6,1,1,1)$

\no \phan{XI.} \gen $\bS_1 $, $\bS_7$

\dotf

\no\poss $\bS_1= (\st)S(6,6,1,2)$:
\vfa

\gayo{+1}{+1}{+1}{0,1}{(1.6.2.1)}

\gayo{+\third}{+1}{-\third}{0,1}{(1.6.2.3,4)}

%\nzero

\ganzo{+0}{+0}{+0}{\pm\half}{(1.6.2.2)}

\ganzo{+\twothird}{+0}{-\twothird}{\pm\half}{(1.6.2.5,6)}

\ganzo{+\fivesixth}{+\half}{+\sixth}{-\fourth,\threefourth}{(1.6.2.7,8)}

\ganzo{+\half}{-\half}{+\half}{\fourth,-\threefourth}{(1.6.2.9,10)}

\ganzo{+\sixth}{+\half}{+\fivesixth}{-\fourth,\threefourth}{(1.6.2.9,10)}

\vfa

\secondz
\vfa

\dotf

\no\poss $m\bS_7 = m\times(\st)S(6,2,1,2)$:

\gcyo{+1}{0,\pm\third,\pm\twothird,1}{(3.6.2.1)}

%\nzero

\gcnzo{+0}{\pm\sixth,\pm\half\pm\fivesixth}{(3.6.2.2)}

\gcnzo{+\half}{-\twelf,\fourth,-\fivetwelf,\seventwelf,-\threefourth,
\eleventwelf}{(3.6.2.3,4)}

\vfa
For $\bS_7$ the gravitinos are divided up between $\bS_7$, $3\bS_7$, and
$5\bS_7$. However, as shown in section 1, states in $\{m\bS_7\}$
are not independent. For every gravitino of
given chirality and charge $\bQ_1$ in $\bS_7$, there is a gravitino of
opposite
chirality and charge $-\bQ_1$ in $5\bS_7$.
Thus, we need only examine the GSOPs
for $\bS_7$ and $3\bS_7$. For $3\bS_7$ we make use of the
$K_{i,3j} = 3K_{i,j}$ identity
and of the equivalence of $3\bS_7$ from $3(\st)S(6,2,1,2)$
with $\bS_1$ from $(\st)S(6,6,1,2)$. (In general,
if $N_i/2$ for $\bS_i$ is odd, then ${N_i\over 2}\bS_i$ equals
$\bS_1$.)
$\bS_7$ can generate $N=1$ SUSY only if all gravitinos are removed
from $3\bS_7$.

The class $(3.6.2.1)$ GSOP for $\bS_7$ originates from the same BVs that
generate classes $(1.6.2.1)$ and $(1.6.2.3,4)$ for $3\bS_7$,
while together the BVs for $(3.6.2.2)$ and $(3.6.2.3,4)$
compose the sets associated with all remaining GSOP classes for $3\bS_7$.
If they are present,
the GSOPs in $(3.6.2.2)$ and $(3.6.2.3,4)$ and the related
in $(1.6.2.2,5-10)$ remove all gravitinos
from $\bS_7$ and $3\bS_7$, respectively, thereby resulting in $N=0$ ST-SUSY.
GSO class $(3.6.2.1)$ either keeps one gravitino of
each chirality (RHS equals 0 or 1)
or no gravitinos at all (RHS equals $\pm\twothird$) in $\bS_7$ .
A RHS from the set $\{0,\, \pm\twothird\}$ in GSOP $(3.6.2.1)$
implies a RHS of $1$ for $(1.6.2.1)$ and a  RHS of $0$ for $(1.6.2.3,4)$;
a RHS from the set $\{1,\, \pm\onethird\}$ in GSOP $(3.6.2.1)$
implies a RHS of $0$ for $(1.6.2.1)$ and a  RHS of $1$ for $(1.6.2.3,4)$.
Thus, when one of each chirality is kept in $\bS_7$, two of each chirality
automatically survive in $3\bS_7$, resulting in $N=4$ ST-SUSY. However, if
no gravitinos are kept in $\bS_7$, either two or zero gravitinos remain in
$3\bS_7$.

The same patterns occurs for the three other sets of BVs in Table 3
that commute with $S(6,1,1,1)$.
Consequently, any model using
non-disjoint cyclical permutations of length six (equivalently, $\Z_{12}$
twists) can never possess $N=1$ ST-SUSY.

\vfa
\headf

\noc X. \title $5\cdot1$ in $S(5,1,1,1)S(1,1,1,1)$

\no \phan{X.} \gen $\bS_1$, $\bS_{10}$

\dotf

\no\poss $\bS_1 = (\st)S(5,5,1,1)S(1,1,1,1)$:
\vfa

\gayo{+1}{+1}{+1}{0,1}{(1.5.1.1)}

%\nzero

\gayzo{+\threefifth}{-\onefifth}{+1}{-\onefifth,\fourfifth}{(1.5.1.3,4)}

\gayzo{+\onefifth}{+\threefifth}{+1}{-\twofifth,\threefifth}{(1.5.1.5,6)}

\ganzo{+0}{+0}{+0}{\pm\half}{(1.5.1.2)}

\ganzo{+\fourfifth}{+\twofifth}{+0}{-\onetenth,\ninetenth}{(1.5.1.7,8)}

\ganzo{+\twofifth}{-\fourfifth}{+0}{-\threetenth,\sevententh}{(1.5.1.9,10)}

\vfa
The initial $N=4$ ST-SUSY can only be broken to $N=0$.

\dotf

\no\poss $m\bS_{10} = m\times(\st)S(5,1,1,1)S(1,1,1,1)$:
\vfa

\gcyo{+1}{0,\pm\onefifth,\pm\twofifth,\pm\threefifth,
\pm\fourfifth,1}{(5.5.1.1)}

%\nzero

\gcnzo{+0}{\pm\onetenth,\pm\threetenth,\pm\fivetenth,
\pm\sevententh,\pm\ninetenth}{(5.5.1.2)}

\vfa
Since $\bS_{10}$ generates gravitinos in each of its odd multiples,
we need to consider the interdependent sets of GSOPs
on $\bS_{10}$, $3\bS_{10}$, and $5\bS_{10}$.
$(5.5.1.1)$ and $(5.5.1.2)$ form the set of GSOPs
for both $\bS_{10}$ and $3\bS_{10}$.
A RHS of 0 (1) in (5.5.1.1) for $\bS_{10}$ leads to
a RHS of 1 (0) for $3\bS_{10}$. This results in
one gravitino of each chirality surviving in both $\bS_{10}$
and $3\bS_{10}$. Any other RHS value for (5.5.1.1) for $\bS_{10}$
removes all gravitinos from both sectors.

No gravitinos can emerge from $5\bS_{10}$.
Since $5\bS_{10}$ equals $\bS_{1}$, GSOP classes (1.5.1.1-10) apply to
$5\bS_{10}$ as well.
Eq.~(\puteqn{const-c}) relates the GSOPs in class (1.5.1.1) coming
from $\bV_0$ and $\bS_{10}$:
$$K_{\bS_{10},5\bS_{10}}= 5K_{\bS_{10},\bS_{10}} = 5K_{\bV_{0},\bS_{10}} +1
\quad\mod{2}\,\, . \eq{gsoout}$$
Hence, $\bS_{10}$ and $\bV_0$ contribute
conflicting versions of  GSOP (1.5.1.1) for $\bS_{10}$
that contain differing RHS's:
$5K_{\bV_0,\bS_{10}} + \threefifth \quad\mod{2}$ and
$5K_{\bV_0,\bS_{10}} + 1 \quad\mod{2}$, respectively.
Therefore, $\{m\bS_{10}\}$ generates either an $N=4$ or $N=0$ ST-SUSY.

\vfa
\headf

\noc IX. \title $4\cdot2$ in $S(4,1,1,1)S(2,1,1,1)$

\no \phan{IX.} \gen $\bS_1$, $\bS_3$, and $\bS_{9}$.

\dotf

\no\poss $\bS_1 = (\st)S(4,4,1,1)S(2,2,1,1)$:
\vfa

\gayw{+1}{+1}{+1}{0,1}{(1.4.1.1)(1.2.1.1)}

\gayw{+1}{+0}{+0}{0,1}{(1.4.1.2)(1.2.1.4)}

%\nzero

\gayzw{+0}{+1}{+1}{\pm\half}{(1.4.1.3)(1.2.1.1)}

\ganzw{+0}{+0}{+0}{\pm\half}{(1.4.1.4)(1.2.1.4)}

\gayzw{+\half}{+1}{+0}{-\fourth,+\threefourth}{(1.4.1.5,6)(1.2.1.2)}

\gayzw{+\half}{+0}{+1}{-\fourth,+\threefourth}{(1.4.1.7,8)(1.2.1.3)}

\vfa
\second
\newpage

\dotf

\no\poss $\bS_1 = (\st)S(4,4,1,2)S(2,2,1,2)$:
\vfa

\gayw{+1}{+1}{+1}{0,1}{(1.4.2.1)(1.2.2.1)}

\gayw{+0}{+0}{+1}{0,1}{(1.4.2.2)(1.2.2.1)}

\gayw{+\threefourth}{+\fourth}{+0}{0,1}{(1.4.2.3,4)(1.2.2.2)}

\gayw{+\threefourth}{+\fourth}{+\half}{-\fourth,+\threefourth}{(1.4.2.w)(1.2.2.w)}

\extra{w=3,\, 4}

\gayw{+\threefourth}{+\fourth}{-\half}{+\fourth,-\threefourth}{(1.4.2.w)(1.2.2.w')}

\extra{(w,w')= (3,4),\, (4,3)}

\gayw{+\fourth}{+\threefourth}{+0}{0,1}{(1.4.2.7,8)(1.2.2.2)}

\gayw{+\fourth}{+\threefourth}{+\half}{-\fourth,+\threefourth}{(1.4.2.w)(1.2.2.w')}

\extra{(w,w')= (7,3),\, (8,4)}

\gayw{+\fourth}{+\threefourth}{-\half}{+\fourth,-\threefourth}{(1.4.2.w)(1.2.2.w')}

\extra{(w,w')= (7,4),\, (8,3)}

\gayw{+\half}{-\half}{+1}{0,1}{(1.4.2.5,6)(1.2.2.1)}

\gayw{+\half}{-\half}{+0}{\pm\half}{(1.4.2.5,6)(1.2.2.2)}

%\nzero

\ganzw{+1}{+1}{+0}{\pm\half}{(1.4.2.1)(1.2.2.2)}

\ganzw{+1}{+1}{+\half}{+\fourth,-\threefourth}{(1.4.2.1)(1.2.2.3,4)}

\ganzw{+0}{+0}{+0}{\pm\half}{(1.4.2.2)(1.2.2.2)}

\ganzw{+0}{+0}{+\half}{\pm\half}{(1.4.2.2)(1.2.2.3,4)}

\ganzw{+\threefourth}{+\fourth}{+1}{\pm\half}{(1.4.2.3,4)(1.2.2.1)}

\ganzw{+\fourth}{+\threefourth}{+1}{\pm\half}{(1.4.2.7,8)(1.2.2.1)}

\ganzw{+\half}{-\half}{+\half}{+\fourth,-\threefourth}{(1.4.2.w)(1.2.2.w')}

\extra{(w,w')= (5,3),\, (6,4)}

\ganzw{+\half}{-\half}{-\half}{-\fourth,+\threefourth}{(1.4.2.w)(1.2.2.w')}

\extra{(w,w')= (5,4),\, (6,3)}

\vfa
There are three general solutions for $N=1$ ST-SUSY.
GSOP class $(1.4.2.1)$$(1.2.2.1)$ can be combined with:
\item{1.} $(1.4.2.3,4)$$(1.2.2.2)$ and/or $(1.4.2.7,8)$$(1.2.2.2)$
plus,
optionally, $(1.4.2.2)(1.2.2.1)$ and
/or $(1.4.2.5,6)$$(1.2.2.1)$,
\item{2.} $(1.4.2.3)$$(1.2.2.3)$/$(1.4.2.4)$$(1.2.2.4)$
 and/or $(1.4.2.7)$$(1.2.2.4)$/$(1.4.2.8)$$(1.2.2.3)$,
plus,
optionally, $(1.4.2.2)(1.2.2.1)$ and/or
$(1.4.2.5,6)$$(1.2.2.2)$, or
\item{3.} $(1.4.2.3)$$(1.2.2.4)$/$(1.4.2.4)$$(1.2.2.3)$ and/or
$(1.4.2.7)$$(1.2.2.3)$/$(1.4.2.8)$$(1.2.2.4)$,
plus,
optionally, $(1.4.2.2)(1.2.2.1)$ and/or $(1.4.2.5,6)$$(1.2.2.2)$.

\noindent For each of these three $N=1$ solution sets,
choice of the RHS value for either of the first two GSOPs fixes those
for all other GSOs used.

\dotf

\no\poss $m\bS_{3} = m\times(\st)S(4,2,1,1)S(2,2,1,1)$:
\vfa

\gbyw{+1}{+1}{0,\pm\half,1}{(2.4.1.1)(1.2.1.1)}

\gbyw{+0}{+0}{0,\pm\half,1}{(2.4.1.4)(1.2.1.4)}

%\nzero

\gbyzw{+1}{+0}{\pm\fourth,\pm\threefourth}{(2.4.1.2)(1.2.1.2)}

\gbyzw{+0}{+1}{\pm\fourth,\pm\threefourth}{(2.4.1.3)(1.2.1.3)}

\vfa
\secondz

\dotf

\no\poss $m\bS_{3} = m\times(\st)S(4,2,1,2)S(2,2,1,2)$:
\vfa

\gbyw{+1}{+1}{0,\pm\half,1}{(2.4.2.1)(1.2.2.1)}

\gbnw{+1}{+0}{0,\pm\half,1}{(2.4.2.1)(1.2.2.2)}

\gbyw{+0}{+1}{0,\pm\half,1}{(2.4.2.2)(1.2.2.1)}

\gbnw{+0}{+0}{0,\pm\half,1}{(2.4.2.2)(1.2.2.2)}

\gbyw{+\half}{+\half}{0,\pm\half,1}{(2.4.2.w)(1.2.2.w),}

\extra{w=3,\, 4}

\gbyw{+\half}{-\half}{0,\pm\half,1}{(2.4.2.w)(1.2.2.w'),}

\extra{(w,w')= (3,4),\, (4,3)}

%\nzero

\gbnzw{+1}{+\half}{\pm\fourth,\pm\threefourth}{(2.4.2.1)(1.2.2.3,4)}

\gbnzw{+0}{+\half}{\pm\fourth,\pm\threefourth}{(2.4.2.2)(1.2.2.3,4)}

\gbyzw{+\half}{+1}{\pm\fourth,\pm\threefourth}{(2.4.2.3,4)(1.2.2.1)}

\gbyzw{+\half}{+0}{\pm\fourth,\pm\threefourth}{(2.4.2.3,4)(1.2.2.2)}

\vfa

$N=1$ ST-SUSY is directly reached through proper choice of RHS values
of (1) the fifth or sixth GSOP, or
(2) any two of the second through fourth GSOPs.
In either case, the RHS values of the
rest of the first six GSOPs also used
are fixed or $N=1$ is broken to $N=0$.
The first four GSOPs imitate $\bS_1$'s NAHE set of projections in I.

\dotf

\no\poss $m\bS_{9} = m\times(\st)S(4,1,1,1)S(2,1,1,1)$:
\vfa

% 1S: 0 -> 3S: 1
\gcyw{+1}{0,\pm\fourth,\pm\half,\pm\threefourth,1}{(4.4.1.w)(2.2.1.w),}

\extra{w=1,\, 2}

% 1S: 0 -> 3S: 0
\gcyw{+1}{0,\pm\fourth,\pm\half,\pm\threefourth,1}{(4.4.1.w)(2.2.1.w'),}

\extra{(w,w')=(1,2),\, (2,1)}

% 1S: 0 -> 3S: 0
\gcyw{+0}{0,\pm\fourth,\pm\half,\pm\threefourth,1}{(4.4.1.w)(2.2.1.w),}

\extra{w=3,\, 4}

% 1S: 0 -> 3S: 1
\gcyw{+0}{0,\pm\fourth,\pm\half,\pm\threefourth,1}{(4.4.1.w)(2.2.1.w'),}

\extra{(w,w')= (3,4),\, (4,3)}

\vfa

For $\bS_{9}$ the gravitinos are divided among $m\bS_{9}$, $m=1,3,5,7$.
$\bS_{9}$ and $3\bS_{9}$ have the same set of projections classes,
with the RHS of  $3\bS_{9}$'s projections dependent upon those for $\bS_{9}$.
%For either $\bS_{9}$ or $3\bS_{9}$, the first (third) and second (fourth)
%projections are eqivalent.
The first GSOP keeps one left-handed
and one right-handed gravitino in $\bS_{9}$ and also in  $3\bS_{9}$.
The internal charge vector $\bQ$ on surviving gravitinos of given chirality
changes sign between $\bS_{9}$ and $3\bS_{9}$ for this GSOP.
However, the second GSOP operator requires
these charge vectors to be equal. Therefore, when both the first and second
GSOPs are applied all gravitinos in $3\bS_{9}$ can be projected out.
This yields $N=2$ ST-SUSY from $\bS_{9}$.
The last two projections eliminate one chirality for
$\bS_{9}$ gravitinos, therefore inducing $N=1$.
%\newpage

\vfa
\headf

\noc VIII. \title $4\cdot1\cdot1$ in $S(4,1,1,1)S(1,1,1,1)^2$

\no \phan{VIII.} \gen $\bS_1$, $\bS_3$

\dotf

\no\poss $\bS_1 = (\st)S(4,4,1,1)S(1,1,1,1)^2$:
\vfa

\gayt{+1}{+1}{+1}{0,1}{(1.4.1.1)(1.1.1.1)(1.1.1.1)}

\gayt{+1}{+0}{+0}{0,1}{(1.4.1.2)(1.1.1.2)(1.1.1.2)}

%\nzero

\gayzt{+0}{+1}{+1}{\pm\half}{(1.4.1.3)(1.1.1.1)(1.1.1.1)}

\ganzt{+0}{+0}{+0}{\pm\half}{(1.4.1.4)(1.1.1.2)(1.1.1.2)}

\gayzt{+\half}{+1}{+0}{-\fourth,+\threefourth}{(1.4.1.5,6)(1.1.1.1)(1.1.1.2)}

\gayzt{+\half}{+0}{+1}{-\fourth,+\threefourth}{(1.4.1.7,8)(1.1.1.2)(1.1.1.1)}

\vfa
These GSOPs are the same as the set for $\bS_1 = (\st)S(4,4,1,1)S(2,2,1,1)$ in
IX.
We replace $(1.2.1.1)$ with $(1.1.1.1)^2$,
$(1.2.1.2)$ with $(1.1.1.1)$$(1.1.1.2)$,
$(1.2.1.3)$ with $(1.1.1.2)$\break
$(1.1.1.1)$, and $(1.2.1.4)$ with $(1.1.1.2)^2$.
\second

\dotf

\no\poss $\bS_1 = (\st)S(4,4,1,2)S(1,1,1,1)^2$:
\vfa

\gayt{+1}{+1}{+1}{0,1}{(1.4.2.1)(1.1.1.1)^2}

\gayt{+0}{+0}{+1}{0,1}{(1.4.2.2)(1.1.1.1)^2}

\gayt{+\threefourth}{+\fourth}{+0}{0,1}{(1.4.2.3,4)(1.1.1.2)^2}

\gayt{+\fourth}{+\threefourth}{+0}{0,1}{(1.4.2.7,8)(1.1.1.2)^2}

\gayt{+\half}{-\half}{+1}{0,1}{(1.4.2.5,6)(1.1.1.1)^2}

\gayt{+\half}{-\half}{+0}{\pm\half}{(1.4.2.5,6)(1.1.1.2)^2}

%\nzero

\ganzt{+1}{+1}{+0}{\pm\half}{(1.4.2.1)(1.1.1.2)^2}

\ganzt{+0}{+0}{+0}{\pm\half}{(1.4.2.2)(1.1.1.2)^2}

\ganzt{+\threefourth}{+\fourth}{+1}{\pm\half}{(1.4.2.3,4)(1.1.1.1)^2}

\ganzt{+\fourth}{+\threefourth}{+1}{\pm\half}{(1.4.2.7,8)(1.1.1.1)^2}

\vfa
These GSOPs are a subset of those for
$\bS_1 = S(4,4,1,2)S(2,2,1,2)$ in IX.
We replace $(2.2.1.1)$ with $(1.1.1.1)^2$,
$(2.2.1.2)$ with $(1.1.1.2)^2$,
and remove the GSOPs involving $(2.2.1.3)$ and $(2.2.1.4)$.
There remains a single general solution
for $N=1$ ST-SUSY. $(1.4.2.1)(1.1.1.1)^2$ can be combined:
\item{1.} $(1.4.2.3,4)(1.1.1.2)^2$ and/or
$(1.4.2.7,8)(1.1.1.2)^2$
plus, optionally, $(1.4.2.2)(1.1.1.1)^2$ and/or
$(1.4.2.5,6)(1.1.1.1)^2$.

Choice of the RHS value for either of the first two GSOPs
in this $N=1$ solution set
fixes those
for all other GSOs used from the set.

\dotf

\no\poss $m\bS_3 = m\times(\st)S(4,2,1,1)S(1,1,1,1)^2$:
\vfa

\gbyt{+1}{+1}{0,\pm\half,1}{(2.4.1.1)(1.1.1.1)^2}

\gbyt{+0}{+0}{0,\pm\half,1}{(2.4.1.4)(1.1.1.2)^2}

%\nzero

\gbyzt{+1}{+0}{\pm\fourth,\pm\threefourth}{(2.4.1.2)(1.1.1.1)(1.1.1.2)}

\gbyzt{+0}{+1}{\pm\fourth,\pm\threefourth}{(2.4.1.3)(1.1.1.2)(1.1.1.1)}

\vfa

These GSOPs are the same as the set for $\bS_3 = (\st)S(4,2,1,1)S(2,2,1,1)$.
We simply replace $(1.2.1.1)$ with $(1.1.1.1)^2$,
$(1.2.1.2)$ with $(1.1.1.1)(1.1.1.2)$,
$(1.2.1.3)$ with $(1.1.1.2)$\break
$(1.1.1.1)$, and $(1.2.1.4)$ with $(1.1.1.2)^2$.
\second

\dotf

\no\poss $m\bS_3 = m\times(\st)S(4,2,1,2)S(1,1,1,1)^2$:
\vfa

\gbyw{+1}{+1}{0,\pm\half,1}{(2.4.2.1)(1.1.1.1)^2}

\gbnw{+1}{+0}{0,\pm\half,1}{(2.4.2.1)(1.1.1.2)^2}

\gbyw{+0}{+1}{0,\pm\half,1}{(2.4.2.2)(1.1.1.1)^2}

\gbnw{+0}{+0}{0,\pm\half,1}{(2.4.2.2)(1.1.1.2)^2}

%\nzero

\gbyzw{+\half}{+1}{\pm\fourth,\pm\threefourth}{(2.4.2.3,4)(1.1.1.1)^2}

\gbyzw{+\half}{+0}{\pm\fourth,\pm\threefourth}{(2.4.2.3,4)(1.1.1.2)^2}

\vfa
These GSOPs are a subset of those for
$\bS_3 = S(4,2,1,1)S(2,2,1,2)$ in IX.
We replace $(1.2.2.1)$ with $(1.1.1.1)^2$ and
$(1.2.2.2)$ with $(1.1.1.2)^2$, and remove the GSOPs involving
$(1.2.2.3)$ and $(1.2.2.4)$.
\FNAHE
However, we cannot consider this as a category VIII solution;
it actually belongs in category III (below). The first four
GSOP classes are not associated with $4\cdot1\cdot1$ cycles,
instead only with cycles of lengths $2\cdot2\cdot1\cdot1$ and
$1\cdot1\cdot1\cdot1\cdot1\cdot1$.

\vfa
\headf

\noc VII. \title $3\cdot3$ in $S(3,1,1,1)S(3,1,1,1)$

\no \phan{VII.} \gen $\bS_1$, $\bS_5$, $\bS_7$

\dotf

\no\poss $\bS_1 = (\st)S(3,3,1,1)^2$:
\vfa

\gayw{+1}{+1}{+1}{0,1}{(1.3.1.1)(1.3.1.1)}

\gayw{+\third}{+\third}{+1}{0,1}{(1.3.1.w)(1.3.1.w)}

\extra{w = 3,\, 4}

\gayw{+\third}{-\third}{+1}{0,1}{(1.3.1.w)(1.3.1.w')}

\extra{(w,w')= (3,4),\, (4,3)}

%\nzero

\gayzw{+1}{+\third}{+1}{+\third,-\twothird}{(1.3.1.1)(1.3.1.3,4)}

\gayzw{+\third}{+1}{+1}{+\third,-\twothird}{(1.3.1.3,4)(1.3.1.1)}

\ganzw{+0}{+0}{+0}{\pm\half}{(1.3.1.2)(1.3.1.2)}

\ganzw{+0}{+\twothird}{+0}{+\sixth,-\fivesixth}{(1.3.1.2)(1.3.1.5,6)}

\ganzw{+\twothird}{+0}{+0}{+\sixth,-\fivesixth}{(1.3.1.5,6)(1.3.1.2)}

\ganzw{+\twothird}{+\twothird}{+0}{-\sixth,+\fivesixth}{(1.3.1.w)(1.3.1.w)}

\extra{w = 5,\, 6}

\ganzw{+\twothird}{-\twothird}{+0}{\pm\half}{(1.3.1.w)(1.3.1.w')}

\extra{(w,w')= (5,6),\, (6,5)}

\vfa
The second and the third GSOPs independently reduce the initial $N=4$
ST-SUSY to either $N=2$ or $N=0$. If both are applied
only $N=0$ results.

\dotf

\no\poss $m\bS_5 = m\times(\st)S(3,1,1,1)S(3,3,1,1)^3$:
\vfa

\gbyw{+1}{+1}{0,\pm\third,\pm\twothird,1}{(3.3.1.1)(1.3.1.1)}

\gbyw{+1}{+\third}{0,\pm\third,\pm\twothird,1}{(3.3.1.1)(1.3.1.3,4)}

\gbnzw{+0}{+0}{\pm\sixth,\pm\half,\pm\fivesixth}{(3.3.1.2)(1.3.1.2)}

\gbnzw{+0}{+\twothird}{\pm\sixth,\pm\half,\pm\fivesixth}{(3.3.1.2)(1.3.1.5,6)}

\vfa
The first GSOP keeps two gravitinos of each chirality in $\bS_5$,
creating $N=4$ from $\bS_5$ and $5\bS_5$.
The second GSOP eliminates from $\bS_5$
one gravitino of each chirality, reducing ST-SUSY to $N=2$.
($3\bS_5$ is equivalent to $\bS_1$, from which
the ever-present GSOP $(1.3.1.1)(1.3.1.3)$
removes all gravitinos.)

\dotf

\no\poss $m\bS_7 = m\times(\st)S(3,1,1,1)^2$:
\vfa
\gcyw{+1}{0,\pm\third,\pm\twothird,1}{(1.3.1.1)(1.3.1.1)}

\gcnzw{+0}{\pm\onesixth,\pm\half,\pm\fivesixth}{(1.3.1.2)(1.3.1.2)}

\vfa

The first GSOP keeps one gravitino of each chirality
in $\bS_7$, giving $N=2$ from $\bS_7$ and $5\bS_7$.
The GSOPs for $3\bS_7$ are the same as those for $\bS_1$;
the BVs giving GSOP class $(1.3.1.1)(1.3.1.1)$
for $\bS_7$ divide up to generate the first five GSOP classes for
$3\bS_7= \bS_1$.
Together GSOP classes $(1.3.1.1)(1.3.1.1)$ and $(1.3.1.3)(1.3.1.3)$
from $\bV_0$ and $\bS_7$, respectivley,
keep two gravitinos of each chirality in $3\bS_7$.
The other possible GSOPs  either keep or eliminate
all of $3\bS_7$'s gravitinos.
Thus, the $\{m\bS_7\}$ set can only produce $N=4$, $2$, or $0$ ST-SUSY.
%\newpage

\vfa
\headf

\noc VI. \title $3\cdot2\cdot1$ in $S(3,1,1,1)S(2,1,1,1)S(1,1,1,1)$

\no \phan{VI.} \gen $\bS_1$, $\bS_5$

\dotf

\no\poss $\bS_1 = (\st)S(3,3,1,1)S(2,2,1,1)S(1,1,1,1)$:
\vfa
\gayt{+1}{+1}{+1}{0,1}{(1.3.1.1)(1.2.1.1)(1.1.1.1)}

\gayt{+0}{+0}{+1}{0,1}{(1.3.1.2)(1.2.1.3)(1.1.1.1)}

%\nzero

\ganzt{+1}{+1}{+0}{\pm\half}{(1.3.1.1)(1.2.1.2)(1.1.1.2)}

\ganzt{+0}{+0}{+0}{\pm\half}{(1.3.1.2)(1.2.1.4)(1.1.1.2)}

\ganzt{+\third}{+1}{+1}{+\third,-\twothird}{(1.3.1.3)(1.2.1.1)(1.1.1.1)}

\ganzt{+\third}{+1}{+0}{-\onesixth,+\fivesixth}{(1.3.1.3)(1.2.1.2)(1.1.1.2)}

\ganzt{+\twothird}{+0}{+1}{-\third,+\twothird}{(1.3.1.4)(1.2.1.3)(1.1.1.1)}

\ganzt{+\twothird}{+0}{+0}{+\onesixth,-\fivesixth}{(1.3.1.4)(1.2.1.4)(1.1.1.2)}

\vfa
\second

\dotf

\no\poss $\bS_1 = (\st)S(3,3,1,1)S(2,2,1,2)S(1,1,1,1)$:
\vfa
\gayt{+1}{+1}{+1}{0,1}{(1.3.1.1)(1.2.2.1)(1.1.1.1)}

\gayt{+0}{+1}{+0}{0,1}{(1.3.1.2)(1.2.2.1)(1.1.1.2)}

%\nzero

\ganzt{+1}{+0}{+1}{\pm\half}{(1.3.1.1)(1.2.2.2)(1.1.1.1)}

\ganzt{+0}{+0}{+0}{\pm\half}{(1.3.1.2)(1.2.2.2)(1.1.1.2)}

\ganzt{+1}{+\half}{+1}{+\onefourth,-\threefourth}{(1.3.1.1)(1.2.2.3,4)(1.1.1.1)}

\ganzt{+0}{+\half}{+0}{+\onefourth,-\threefourth}{(1.3.1.2)(1.2.2.3,4)(1.1.1.2)}

\gayzt{+\onethird}{+1}{+1}{+\third,-\twothird}{(1.3.1.3,4)(1.2.2.1)(1.1.1.1)}

\ganzt{+\onethird}{+0}{+1}{-\onesixth,+\fivesixth}{(1.3.1.3,4)(1.2.2.2)(1.1.1.1)}

\ganzt{+\onethird}{+\half}{+1}{-\fivetwelf,\seventwelf}{(1.3.1.w)(1.2.2.w)(1.1.1.1)}

\extra{w= 3,\, 4}

\ganzt{+\onethird}{-\half}{+1}{+\onetwelf,-\eleventwelf}{(1.3.1.w)(1.2.2.w')(1.1.1.1)}

\extra{(w,w')= (3,4),\, (4,3)}

\gayzt{+\twothird}{+1}{+0}{-\onethird,+\twothird}{(1.3.1.5,6)(1.2.2.1)(1.1.1.2)}

\ganzt{+\twothird}{+0}{+0}{+\onesixth,-\fivesixth}{(1.3.1.5,6)(1.2.2.2)(1.1.1.2)}

\ganzt{+\twothird}{+\half}{+0}{-\onetwelf,\eleventwelf}{(1.3.1.w)(1.2.2.w')(1.1.1.2)}

\extra{(w,w')= (5,3),\, (6,4)}

\ganzt{+\twothird}{-\half}{+0}{+\fivetwelf,-\seventwelf}{(1.3.1.w)(1.2.2.w')(1.1.1.2)}

\extra{(w,w')= (5,4),\, (5,3)}

\vfa
\second

\dotf

\no\poss $m\bS_5 = m\times(\st)S(3,1,1,1)S(2,2,1,1)S(1,1,1,1)$:
\vfa

\gbyt{+1}{+1}{0,\pm\third,\pm\twothird,1}{(3.3.1.1)(1.2.1.1)(1.1.1.1)}

\gbyt{+0}{+1}{0,\pm\third,\pm\twothird,1}{(3.3.1.2)(1.2.1.3)(1.1.1.1.)}

%\nzero

\gbnzt{+1}{+0}{\pm\sixth,\pm\half,\pm\fivesixth}{(3.3.1.1)(1.2.1.2)(1.1.1.2)}

\gbnzt{+0}{+0}{\pm\sixth,\pm\half,\pm\fivesixth}{(3.3.1.2)(1.2.1.4)(1.1.1.2)}

\vfa
\secondz
All gravitinos are removed from $3\bS_5$ for the reasons discussed previously.

\dotf

\no\poss $m\bS_5 = m\times(\st)S(3,1,1,1)S(2,2,1,2)S(1,1,1,1)$:
\vfa
\gbyt{+1}{+1}{0,\pm\third,\pm\twothird,1}{(3.3.1.1)(1.1.1.1)(1.2.2.1)}

\gbyt{+0}{+1}{0,\pm\third,\pm\twothird,1}{(3.3.1.2)(1.1.1.2)(1.2.2.1)}

%\nzero

\gbnzt{+1}{+0}{\pm\sixth,\pm\half,\pm\fivesixth}{(3.3.1.1)(1.1.1.1)(1.2.2.2)}

\gbnzt{+0}{+0}{\pm\sixth,\pm\half,\pm\fivesixth}{(3.3.1.2)(1.1.1.2)(1.2.2.2)}

\gbnztb{+1}{+\half}{-\onetwelf,\threetwelf,-\fivetwelf,\seventwelf,-\ninetwelf,\eleventwelf}{(3.3.1.1)(1.1.1.1)(1.2.2.3,4)}

\gbnztb{+0}{+\half}{-\onetwelf,\threetwelf,-\fivetwelf,\seventwelf,-\ninetwelf,\eleventwelf}{(3.3.1.2)(1.1.1.2)(1.2.2.3,4)}

\vfa
\secondz

\vfa
\headf

\noc V. \title $3\cdot1\cdot1\cdot1$ in $S(3,1,1,1)S(1,1,1,1)^3$

\no \phan{V.} \gen $\bS_1$, $\bS_5$

\dotf

\no\poss $\bS_1 = (\st)S(3,3,1,1)S(1,1,1,1)^3$:
\vfa
\gayt{+1}{+1}{+1}{0,1}{(1.3.1.1)(1.1.1.1)^3}

\gayt{+0}{+0}{+1}{0,1}{(1.3.1.2)(1.1.1.2)(1.1.1.1)^2}

%\nzero

\ganzt{+1}{+1}{+0}{\pm\half}{(1.3.1.1)(1.1.1.1)(1.1.1.2)^2}

\ganzt{+0}{+0}{+0}{\pm\half}{(1.3.1.2)(1.1.1.2)^3}

\ganzt{+\third}{+1}{+1}{+\third,-\twothird}{(1.3.1.3)(1.1.1.1)^3}

\ganzt{+\third}{+1}{+0}{-\onesixth,+\fivesixth}{(1.3.1.3)(1.1.1.1)(1.1.1.2)^2}

\ganzt{+\twothird}{+0}{+1}{-\third,+\twothird}{(1.3.1.4)(1.1.1.2)(1.1.1.1)^2}

\ganzt{+\twothird}{+0}{+0}{+\onesixth,-\fivesixth}{(1.3.1.4)(1.1.1.2)^3}

\vfa
\second
\eject

\dotf

\no\poss $m\bS_5 = m\times(\st)S(3,1,1,1)S(1,1,1,1)^3$:
\vfa
\gbyt{+1}{+1}{0,\pm\third,\pm\twothird,1}{(3.3.1.1)(1.1.1.1)^3}

\gbyt{+0}{+1}{0,\pm\third,\pm\twothird,1}{(3.3.1.2)(1.1.1.2)(1.1.1.1)^2}

%\nzero

\gbnzt{+1}{+0}{\pm\sixth,\pm\half,\pm\fivesixth}{(3.3.1.1)(1.1.1.1)(1.1.1.1)^2}

\gbnzt{+0}{+0}{\pm\sixth,\pm\half,\pm\fivesixth}{(3.3.1.2)(1.1.1.2)^3}

\vfa
\secondz

\vfa
\headf

\noc IV. \title $2\cdot2\cdot2$ in $S(2,1,1,1)^3$

\no \phan{IV.} \gen $\bS_1$, $\bS_3$

\dotf

\no\poss $\bS_1 = (\st)S(2,2,1,1)^2S(2,2,1,2)$:
\vfa

\gayt{+1}{+1}{+1}{0,1}{(1.2.1.1)^2(1.2.2.1)}

\gayt{+1}{+0}{+0}{0,1}{(1.2.1.2)^2(1.2.2.2)}

\gayt{+0}{+1}{+0}{0,1}{(1.2.1.3)^2(1.2.2.2)}

\gayt{+0}{+0}{+1}{0,1}{(1.2.1.4)^2(1.2.2.1)}

%\nzero

\ganzt{+1}{+1}{+0}{\pm\half}{(1.2.1.1)^2(1.2.2.2)}

\ganzt{+1}{+0}{+1}{\pm\half}{(1.2.1.2)^2(1.2.2.1)}

\ganzt{+0}{+1}{+1}{\pm\half}{(1.2.1.3)^2(1.2.2.1)}

\ganzt{+0}{+0}{+0}{\pm\half}{(1.2.1.4)^2(1.2.2.2)}

\ganzt{+1}{+1}{+\half}{+\onefourth,-\threefourth}{(1.2.1.1)^2(1.2.2.3,4)}

\ganzt{+1}{+0}{+\half}{-\onefourth,+\threefourth}{(1.2.1.2)^2(1.2.2.3,4)}

\ganzt{+0}{+1}{+\half}{-\onefourth,+\threefourth}{(1.2.1.3)^2(1.2.2.3,4)}

\ganzt{+0}{+0}{+\half}{+\onefourth,-\threefourth}{(1.2.1.4)^2(1.2.2.3,4)}

\vfa

\NAHE
However, we cannot count this as a category IV solution.
It is in fact a category III solution since the first four
GSOP classes do not come from states with $2\cdot2\cdot2$ cycles,
but from $2\cdot2\cdot1\cdot1$ and $1\cdot1\cdot1\cdot1\cdot1\cdot1$
cycles.
\newpage

\dotf

\no\poss $\bS_1 = (\st)S(2,2,1,2)^3$:
\vfa

\no GROUP 1: GSOPs generated from an embedded $1\cdot1\cdot1\cdot1\cdot1\cdot1$
permutation

\gayt{+1}{+1}{+1}{0,1}{(1.2.2.1)^3}

\gayt{+1}{+0}{+0}{0,1}{(1.2.2.1)(1.2.2.2)^2}

\gayt{+0}{+1}{+0}{0,1}{(1.2.2.2)(1.2.2.1)(1.2.2.2)}

\gayt{+0}{+0}{+1}{0,1}{(1.2.2.2)^2(1.2.2.1)}

\vfa
%**********************************************************************

\no GROUP 2: GSOPs generated from an embedded $2\cdot2\cdot1\cdot1$
permutation

\gayt{+\half}{+\half}{+1}{\pm\half}{(1.2.2.w)^2(1.2.2.1)}

\extra{w=3,\, 4}

\gayt{+\half}{-\half}{+1}{0,1}{(1.2.2.w)(1.2.2.w')(1.2.2.1)}

\extra{(w,w')= (3,4),\, (4,3)}

\gayt{+\half}{+\half}{+0}{0,1}{(1.2.2.w)^2(1.2.2.2)}

\extra{w=3,\, 4}

\gayt{+\half}{-\half}{+0}{\pm\half}{(1.2.2.w)(1.2.2.w')(1.2.2.2)}

\extra{(w,w') = (3,4),\, (4,3)}

\vfa
%**********************************************************************

\no GROUP 3: GSOPs generated from an embedded $2\cdot1\cdot1\cdot2$
permutation

\gayt{+\half}{+1}{+\half}{\pm\half}{(1.2.2.w
)(1.2.2.1)(1.2.2.w)}

\extra{w=3,\, 4}

\gayt{+\half}{+1}{-\half}{0,1}{(1.2.2.w)(1.2.2.1)(1.2.2.w')}

\extra{(w,w') = (3,4),\, (4,3)}

\gayt{+\half}{+0}{+\half}{0,1}{(1.2.2.w)(1.2.2.2)(1.2.2.w)}

\extra{w=3,\, 4}

\gayt{+\half}{+0}{-\half}{\pm\half}{(1.2.2.w)(1.2.2.2)(1.2.2.w')}

\extra{(w,w') = (3,4),\, (4,3)}

\vfa
%**********************************************************************

\no GROUP 4: GSOPs generated from an embedded $1\cdot1\cdot2\cdot2$
permutation

\gayt{+1}{+\half}{+\half}{\pm\half}{(1.2.2.1)(1.2.2.w)^2}

\extra{w=3,\, 4}

\gayt{+1}{+\half}{-\half}{0,1}{(1.2.2.1)(1.2.2.w)(1.2.2.w')}

\extra{(w,w') = (3,4),\, (4,3)}

\gayt{+0}{+\half}{+\half}{0,1}{(1.2.2.2)(1.2.2.w)^2}

\extra{w=3,\, 4}

\gayt{+0}{+\half}{-\half}{\pm\half}{(1.2.2.2)(1.2.2.w)(1.2.2.w')}

\extra{(w,w') = (3,4),\, (4,3)}

\vfa
%**********************************************************************
%\nzero

\ganzt{+1}{+1}{+0}{\pm\half}{(1.2.2.1)^2(1.2.2.2)}

\ganzt{+1}{+0}{+1}{\pm\half}{(1.2.2.1)(1.2.2.2)(1.2.2.1)}

\ganzt{+0}{+1}{+1}{\pm\half}{(1.2.2.2)(1.2.2.1)^2}

\ganzt{+0}{+0}{+0}{\pm\half}{(1.2.2.2)^3}

\ganzt{+1}{+1}{+\half}{+\onefourth,-\threefourth}{(1.2.2.1)^2(1.2.2.3,4)}

\ganzt{+1}{+0}{+\half}{-\onefourth,+\threefourth}{(1.2.2.1)(1.2.2.2)(1.2.2.3,4)}

\ganzt{+0}{+1}{+\half}{-\onefourth,+\threefourth}{(1.2.2.2)(1.2.2.1)(1.2.2.3,4)}

\ganzt{+0}{+0}{+\half}{+\onefourth,-\threefourth}{(1.2.2.2)^2(1.2.2.3,4)}

\ganzt{+1}{+\half}{+1}{+\onefourth,-\threefourth}{(1.2.2.1)(1.2.2.3,4)(1.2.2.1)}

\ganzt{+1}{+\half}{+0}{-\onefourth,+\threefourth}{(1.2.2.1)(1.2.2.3,4)(1.2.2.2)}

\ganzt{+0}{+\half}{+1}{-\onefourth,+\threefourth}{(1.2.2.2)(1.2.2.3,4)(1.2.2.1)}

\ganzt{+0}{+\half}{+0}{+\onefourth,-\threefourth}{(1.2.2.2)(1.2.2.3,4)(1.2.2.2)}

\ganzt{+\half}{+1}{+1}{+\onefourth,-\threefourth}{(1.2.2.3,4)(1.2.2.1)^2}

\ganzt{+\half}{+1}{+0}{-\onefourth,+\threefourth}{(1.2.2.3,4)(1.2.2.1)(1.2.2.2)}

\ganzt{+\half}{+0}{+1}{-\onefourth,+\threefourth}{(1.2.2.3,4)(1.2.2.2)(1.2.2.1)}

\ganzt{+\half}{+0}{+0}{+\onefourth,-\threefourth}{(1.2.2.3,4)(1.2.2.2)^2}

\ganzt{+\half}{+\half}{+\half}{-\onefourth,+\threefourth}{(1.2.2.w)^3}

\extra{w=3,\, 4}

\ganzt{+\half}{-\half}{-\half}{-\onefourth,+\threefourth}{(1.2.2.w)(1.2.2.w')^2}

\extra{(w,w') = (3,4),\, (4,3)}

\ganzt{+\half}{-\half}{+\half}{+\onefourth,-\threefourth}{(1.2.2.w)(1.2.2.w')(1.2.2.w)}

\extra{(w,w') = (3,4),\, (4,3)}

\ganzt{+\half}{+\half}{-\half}{+\onefourth,-\threefourth}{(1.2.2.w)(1.2.2.w)(1.2.2.w')}

\extra{(w,w') = (3,4),\, (4,3)}
%**********************************************************************
\vfa

Within the first sixteen GSOPs
are four isomorphic maximal sets of GSOPs that can render $N=1$ ST-SUSY.
They correspond to various combinations of GSOPs from the three
groups corresponding to embeddings of
$1\cdot1\cdot2\cdot2$, $2\cdot1\cdot1\cdot2$, and $1\cdot1\cdot2\cdot2$.
The first and the fourth GSOPs in one embedding
can appear in the same GSOP set, as may the second and the third.
An $N=1$ set can have either a first/fourth GSOP
or a second/third GSOP contribution from each of these three embedding groups,
as long as the number of groups contributing second/third GSOPs is odd.
The first or second GSOPs in these three groups
give $N=2$ ST-SUSY, whereas the third or four directly give $N=1$.
Whenever two ``first or second" GSOPs
or a single ``third or fourth" GSOP are chosen,
the RHSs must be fixed for all other GSOPs present if $N=1$ ST-SUSY is to
remain.
Each maximal set contains the entire first group, which forms the
actual NAHE projection of category I.
The NAHE set BVs are
contained in the classes of BVs generating this first group.
\newpage

\dotf
\no\poss $m\bS_3 = m\times(\st)S(2,1,1,1)^2 S(2,2,1,1)$:
\vfa

\gbyt{+1}{+1}{0,1}{(2.2.1.1)^2(1.2.1.1)}

\gbyt{+1}{+0}{0,1}{(2.2.1.w)(2.2.1.w')(1.2.1.3)}

\extra{(w,w') = (2,3),\, (1,4)}

\gbyt{+0}{+1}{0,1}{(2.2.1.w)(2.2.1.w')(1.2.1.2)}

\extra{(w,w') = (3,2),\, (4,1)}

\gbyt{+0}{+0}{0,1}{(2.2.1.3)^2(1.2.1.4)}

%\nzero

\gbnzt{+1}{+1}{\pm\half}{(2.2.1.2)^2(1.2.1.1)}

\gbnzt{+1}{+1}{+\fourth,-\threefourth}{(2.2.1.w)(2.2.1.w')(1.2.1.1)}

\extra{(w,w') = (1,2),\, (2,1)}

\gbnzt{+1}{+0}{+\fourth,-\threefourth}{(2.2.1.w)(2.2.1.w')(1.2.1.3)}

\extra{(w,w') = (1,3),\, (2,4)}

\gbnzt{+0}{+1}{+\fourth,-\threefourth}{(2.2.1.w)(2.2.1.w')(1.2.1.2)}

\extra{(w,w') = (3,1),\, (4,2)}

\gbyzt{+0}{+0}{\pm\half}{(2.2.1.4)^2(1.2.1.4)}

\gbnzt{+0}{+0}{+\fourth,-\threefourth}{(2.2.1.w)(2.2.1.w')(1.2.1.4)}

\extra{(w,w') = (3,4),\, (4,3)}

\vfa

\FNAHE

\dotf

\no\poss $m\bS_3 = m\times(\st)S(2,1,1,1)^2 S(2,2,1,2)$:
\vfa

\gbywb{+1}{+1}{0,1}{(2.2.1.1)^2(1.2.2.1)}

\gbywb{+1}{+0}{0,1}{(2.2.1.1)^2(1.2.2.2)}

\gbywb{+0}{+1}{0,1}{(2.2.1.3)^2(1.2.2.1)}

\gbywb{+0}{+0}{0,1}{(2.2.1.4)^2(1.2.2.2)}

%\nzero

\gbyzwb{+1}{+1}{\pm\half}{(2.2.1.2)^2(1.2.2.1)}

\gbyzwb{+1}{+0}{\pm\half}{(2.2.1.2)^2(1.2.2.2)}

\gbyzwb{+0}{+1}{\pm\half}{(2.2.1.4)^2(1.2.2.1)}

\gbyzwb{+0}{+0}{\pm\half}{(2.2.1.3)^2(1.2.2.2)}

\gbnzwb{+1}{+\half}{+\fourth,-\threefourth}{(2.2.1.1,2)^2(1.2.2.3,4)}

\gbnzwb{+0}{+\half}{+\fourth,-\threefourth}{(2.2.1.3,4)^2(1.2.2.3,4)}

\vfa

\FNAHE
%\newpage

\vfa
\headf

\noc III. \title $2\cdot2\cdot1\cdot1$ in $S(2,1,1,1)^2S(1,1,1,1)^2$

\no \phan{III.} \gen $\bS_1$, $\bS_3$

\dotf

\no\poss $\bS_1 = (\st)S(2,2,1,1)^2S(1,1,1,1)^2$:
\vfa

\gayr{+1}{+1}{+1}{0,1}{(1.2.1.1)^2(1.1.1.1)^2}

\gayr{+1}{+0}{+0}{0,1}{(1.2.1.2)^2(1.1.1.2)^2}

\gayr{+0}{+1}{+0}{0,1}{(1.2.1.3)^2(1.1.1.2)^2}

\gayr{+0}{+0}{+1}{0,1}{(1.2.1.4)^2(1.1.1.1)^2}

%\nzero

\ganzr{+1}{+1}{+0}{\pm\half}{(1.2.1.1)^2(1.1.1.2)^2}

\ganzr{+1}{+0}{+1}{\pm\half}{(1.2.1.2)^2(1.1.1.1)^2}

\ganzr{+0}{+1}{+1}{\pm\half}{(1.2.1.3)^2(1.1.1.1)^2}

\ganzr{+0}{+0}{+0}{\pm\half}{(1.2.1.4)^2(1.1.1.2)^2}

\vfa

These GSOPs are a subset of those for
$\bS_1 = (\st)S(2,2,1,1)^2S(2,2,1,2)$ in IV.
We replace $(1.2.2.w)$ with $(1.1.1.w)^2$ for $w=1,2$
and remove the GSOPs corresponding to the $(1.2.2.3,4)$ class.
\NAHE

\vfa

\dotf

\no\poss $\bS_1 = (\st)S(2,2,1,2)^2S(1,1,1,1)^2$:
\vfa

\no GROUP 1: GSOPs generated from an embedded $1\cdot1\cdot1\cdot1\cdot1\cdot1$
permutation

\gayt{+1}{+1}{+1}{0,1}{(1.2.2.1)^2(1.1.1.1)^2}

\gayt{+1}{+0}{+0}{0,1}{(1.2.2.1)(1.2.2.2)(1.1.1.2)^2}

\gayt{+0}{+1}{+0}{0,1}{(1.2.2.2)(1.2.2.1)(1.1.1.2)^2}

\gayt{+0}{+0}{+1}{0,1}{(1.2.2.2)^2(1.1.1.1)^2}

\vfa
%**********************************************************************

\no GROUP 2.

\gayt{+\half}{+\half}{+1}{\pm\half}{(1.2.2.w)^2(1.1.1.1)^2}

\extra{w=3,\, 4}

\gayt{+\half}{-\half}{+1}{0,1}{(1.2.2.w)(1.2.2.w')(1.1.1.1)^2}

\extra{(w,w')= (3,4),\, (4,3)}

\gayt{+\half}{+\half}{+0}{0,1}{(1.2.2.w)^2(1.1.1.2)^2}

\extra{w=3,\, 4}

\gayt{+\half}{-\half}{+0}{\pm\half}{(1.2.2.w)(1.2.2.w')(1.1.1.2)^2}

\extra{(w,w') = (3,4),\, (4,3)}

\vfa
%**********************************************************************

%\nzero

\ganzt{+1}{+1}{+0}{\pm\half}{(1.2.2.1)^2(1.1.1.2)^2}

\ganzt{+1}{+0}{+1}{\pm\half}{(1.2.2.1)(1.2.2.2)(1.1.1.1)^2}

\ganzt{+0}{+1}{+1}{\pm\half}{(1.2.2.2)(1.2.2.1)(1.1.1.1)^2}

\ganzt{+0}{+0}{+0}{\pm\half}{(1.2.2.2)^2(1.1.1.2)^2}

\vfa

\ganzt{+1}{+\half}{+1}{+\onefourth,-\threefourth}{(1.2.2.1)(1.2.2.3,4)(1.1.1.1)^2}

\ganzt{+1}{+\half}{+0}{-\onefourth,+\threefourth}{(1.2.2.1)(1.2.2.3,4)(1.1.1.2)^2}

\ganzt{+0}{+\half}{+1}{-\onefourth,+\threefourth}{(1.2.2.2)(1.2.2.3,4)(1.1.1.1)^2}

\ganzt{+0}{+\half}{+0}{+\onefourth,-\threefourth}{(1.2.2.2)(1.2.2.3,4)(1.1.1.2)^2}

\ganzt{+\half}{+1}{+1}{+\onefourth,-\threefourth}{(1.2.2.3,4)(1.2.2.1)(1.1.1.1)^2}

\ganzt{+\half}{+1}{+0}{-\onefourth,+\threefourth}{(1.2.2.3,4)(1.2.2.1)(1.1.1.2)^2}

\ganzt{+\half}{+0}{+1}{-\onefourth,+\threefourth}{(1.2.2.3,4)(1.2.2.2)(1.1.1.1)^2}

\ganzt{+\half}{+0}{+0}{+\onefourth,-\threefourth}{(1.2.2.3,4)(1.2.2.2)(1.1.1.2)^2}

\vfa

These GSOPs are a subset of those for
$\bS_1 = (\st)S(2,2,1,1)^2S(2,2,1,2)$ in IV.
Of the four groups only the first two survive.
We replace $(1.2.2.w)$ with $(1.1.1.w)^2$ for $w=1,2$
and remove the GSOPs corresponding to the $(1.2.2.3,4)$ class.
Only two maximal sets yielding $N=1$ ST-SUSY remain.
Either the first and fourth GSOPs in the second group
or the second and third GSOPs can be combined with the first group.
The NAHE set BVs are
contained in the classes of BVs generating the first four GSOPs.

\dotf

\no\poss $m\bS_3 = m\times(\st)S(2,1,1,1)^2 S(1,1,1,1)^2$:
\vfa

\gbyr{+1}{+1}{0,1}{(2.2.1.1)^2(1.1.1.1)^2}

\gbyr{+1}{+0}{0,1}{(2.2.1.w)(2.2.1.w')(1.1.1.2)(1.1.1.1)}

\extra{(w,w') = (2,3),\, (1,4)}

\gbyr{+0}{+1}{0,1}{(2.2.1.w)(2.2.1.w')(1.1.1.1)(1.1.1.2)}

\extra{(w,w') = (3,2),\, (4,1)}

\gbyr{+0}{+0}{0,1}{(2.2.1.3)^2(1.1.1.2)^2}

%\nzero

\gbnzr{+1}{+1}{\pm\half}{(2.2.1.2)^2(1.1.1.1)^2}

\gbnzr{+1}{+1}{+\fourth,-\threefourth}{(2.2.1.w)(2.2.1.w')(1.1.1.1)^2}

\extra{(w,w') = (1,2),\, (2,1)}

\gbnzr{+1}{+0}{+\fourth,-\threefourth}{(2.2.1.w)(2.2.1.w')(1.1.1.2)(1.1.1.1)}

\extra{(w,w') = (1,3),\, (2,4)}

\gbnzr{+0}{+1}{+\fourth,-\threefourth}{(2.2.1.w)(2.2.1.w')(1.1.1.1)(1.1.1.2)}

\extra{(w,w') = (3,1),\, (4,2)}

\gbyzr{+0}{+0}{\pm\half}{(2.2.1.4)^2(1.1.1.2)^2}

\gbnzr{+0}{+0}{+\fourth,-\threefourth}{(2.2.1.w)(2.2.1.w')(1.1.1.2)^2}

\extra{(w,w') = (3,4),\, (4,3)}

\vfa

These GSOPs form the same set as those for
$\bS_3 = (\st)S(2,1,1,1)^2 S(2,2,1,1)$ in IV.
We replace $(1.2.1.1)$ with $(1.1.1.1)^2$,
$(1.2.1.2)$ with $(1.1.1.1)(1.1.1.2)$,
$(1.2.1.3)$ with $(1.1.1.2)(1.1.1.1)$, and
$(1.2.1.4)$ with $(1.1.1.2)^2$.
\FNAHE

\dotf

\no\poss $m\bS_3 = m\times(\st)S(2,1,1,1)^2 S(1,1,1,1)^2$:
\vfa

\gbyw{+1}{+1}{0,1}{(2.2.1.1)^2(1.1.1.1)^2}

\gbyw{+1}{+0}{0,1}{(2.2.1.1)^2(1.1.1.2)^2}

\gbyw{+0}{+1}{0,1}{(2.2.1.3)^2(1.1.1.1)^2}

\gbyw{+0}{+0}{0,1}{(2.2.1.4)^2(1.1.1.2)^2}

%\nzero

\gbyzw{+1}{+1}{\pm\half}{(2.2.1.2)^2(1.1.1.1)^2}

\gbyzw{+1}{+0}{\pm\half}{(2.2.1.2)^2(1.1.1.2)^2}

\gbyzw{+0}{+1}{\pm\half}{(2.2.1.4)^2(1.1.1.1)^2}

\gbyzw{+0}{+0}{\pm\half}{(2.2.1.3)^2(1.1.1.2)^2}

\vfa

These GSOPs are a subset of those for
$\bS_3 = (\st)S(2,2,1,1)^2S(2,2,1,2)$ in IV.
We replace $(1.2.2.w)$ with $(1.1.1.w)^2$ for $w=1,2$ and remove
of GSOPs corresponding to the $(1.2.2.3,4)$ class.
\FNAHE

\vfa
\headf

\noc II. \title $2\cdot1\cdot1\cdot1\cdot1$ in $S(2,1,1,1)S(1,1,1,1)^4$

\no \phan{II.} \gen $\bS_1$

\dotf

\no\poss $\bS_1 = (\st)S(2,2,1,1)S(1,1,1,1)^4$:
\vfa

\gayr{+1}{+1}{+1}{0,1}{(1.2.1.1)(1.1.1.1)^4}

\gayr{+1}{+0}{+0}{0,1}{(1.2.1.2)(1.1.1.1)(1.1.1.2)^3}

\gayr{+0}{+1}{+0}{0,1}{(1.2.1.3)(1.1.1.2)(1.1.1.1)(1.1.1.2)^2}

\gayr{+0}{+0}{+1}{0,1}{(1.2.1.4)(1.1.1.2)^2(1.1.1.1)^2}

%\nzero

\ganzr{+1}{+1}{+0}{\pm\half}{(1.2.1.1)(1.1.1.1)^2(1.1.1.2)^2}

\ganzr{+1}{+0}{+1}{\pm\half}{(1.2.1.2)(1.1.1.1)(1.1.1.2)(1.1.1.1)^2}

\ganzr{+0}{+1}{+1}{\pm\half}{(1.2.1.3)(1.1.1.2)(1.1.1.1)^3}

\ganzr{+0}{+0}{+0}{\pm\half}{(1.2.1.4)(1.1.1.2)^4}

\vfa

These GSOPs form the same set as those for
$\bS_1 = (\st)S(2,2,1,1)^2 S(1,1,1,1)$ in III.
We replace the second
$(1.2.1.1)$ with $(1.1.1.1)^2$,
$(1.2.1.2)$ with $(1.1.1.1)(1.1.1.2)$,
$(1.2.1.3)$ with $(1.1.1.2)(1.1.1.1)$, and
$(1.2.1.4)$ with $(1.1.1.2)^2$.
\NAHE

\dotf

\no\poss $\bS_1 = (\st)S(2,2,1,2)S(1,1,1,1)^4$:
\vfa

\gayt{+1}{+1}{+1}{0,1}{(1.2.2.1)(1.1.1.1)^4}

\gayt{+1}{+0}{+0}{0,1}{(1.2.2.1)(1.1.1.2)^4}

\gayt{+0}{+1}{+0}{0,1}{(1.2.2.2)(1.1.1.1)^2(1.1.1.2)^2}

\gayt{+0}{+0}{+1}{0,1}{(1.2.2.2)(1.1.1.2)^2(1.1.1.1)^2}

%\nzero

\ganzt{+1}{+1}{+0}{\pm\half}{(1.2.2.1)(1.1.1.1)^2(1.1.1.2)^2}

\ganzt{+1}{+0}{+1}{\pm\half}{(1.2.2.1)(1.1.1.2)^2(1.1.1.1)^2}

\ganzt{+0}{+1}{+1}{\pm\half}{(1.2.2.2)(1.1.1.1)^4}

\ganzt{+0}{+0}{+0}{\pm\half}{(1.2.2.2)(1.1.1.1)^4}

\ganzt{+\half}{+1}{+1}{+\onefourth,-\threefourth}{(1.2.2.3,4)(1.1.1.1)^4}

\ganzt{+\half}{+1}{+0}{-\onefourth,+\threefourth}{(1.2.2.3,4)(1.1.1.1)^2(1.1.1.2)^2}

\ganzt{+\half}{+0}{+1}{-\onefourth,+\threefourth}{(1.2.2.3,4)(1.1.1.2)^2(1.1.1.1)^2}

\ganzt{+\half}{+0}{+0}{+\onefourth,-\threefourth}{(1.2.2.3,4)(1.1.1.2)^4}

\vfa

These GSOPs are a subset of those for
$\bS_1 = (\st)S(2,2,1,2)^2S(1.1.1.1)^2$ in III.
We replace the second $(1.2.2.w)$ with $(1.1.1.w)^2$ for $w=1,2$
and remove the GSOPs corresponding to the $(1.2.2.3,4)$ class.
However, the first four GSOPs are equivalent to the $N=1$ solution
for $\bS_1 = (\st)S(1.1.1.1)^6$ in I.
The NAHE set BVs are
contained in the classes of BVs generating these four GSOPs.

\vfa
\headf

\noc I. \title $1\cdot1\cdot1\cdot1\cdot1\cdot1$ in $S(1,1,1,1)^6$

\no \phan{I.} \gen $\bS_1$

\dotf

\no\poss $\bS_1 = (\st)S(1,1,1,1)^6$:
\vfa

\gayt{+1}{+1}{+1}{0,1}{(1.1.1.1)^6}

\gayt{+1}{+0}{+0}{0,1}{(1.1.1.1)^2(1.1.1.2)^4}

\gayt{+0}{+1}{+0}{0,1}{(1.1.1.1)^2(1.1.1.2)^2(1.1.1.1)^2}

\gayt{+0}{+0}{+1}{0,1}{(1.1.1.2)^4(1.1.1.1)^2}

%\nzero

\ganzt{+1}{+1}{+0}{\pm\half}{(1.1.1.1)^4(1.1.1.1)^2}

\ganzt{+1}{+0}{+1}{\pm\half}{(1.1.1.1)^2(1.1.1.2)^2(1.1.1.1)^2}

\ganzt{+0}{+1}{+1}{\pm\half}{(1.1.1.2)^2(1.1.1.1)^4}

\ganzt{+0}{+0}{+0}{\pm\half}{(1.1.1.2)^6}

\vfa

The first four GSOPs form the NAHE solution set for $N=1$ ST-SUSY.
This is the set used in all string models to date.
The NAHE set BVs are
contained in the classes of BVs generating these four GSOPs.
The combination of any three of these four GSOPs produce $N=1$ ST-SUSY,
independent of their RHS values.
The remaining GSOP either keeps or breaks $N=1$,
depending upon it RHS value,

\vfa
\headf
\vfa

%We conclude from this exhaustive analysis that only three of the six
%gravitino generators cannot produce $N=1$ models
%for any allowed sets of GSOPs. These three are $\bS_1$,
%$\bS_3$, and $\bS_9$. Whether or not one of these (and its odd multiples)
%generates exactly one gravitino depends on the set of GSOPs applied to it.
%If there are GSOPs from BVs with non-disjoint cycles of length--3, --5, or
%--6, $N=1$ is not possible.

This finishes our classification of allowed sets of GSOPs
acting on a given source of gravitinos in heterotic free fermionic strings.
The findings of our exhaustive search can be summarized as follows:
\item{1.} If sectors involving non-disjoint cycles of length
3, 5, or 6 (corresponding to $\Z_6$, $\Z_{10}$, and
$\Z_{12}$ projections, respectively)
are present in a model, then $N=1$ ST-SUSY is forbidden.
Thus, $\bS_5$, $\bS_7$, and $\bS_{10}$ as gravitino generators
can only result in $N=4$, 2, or 0 ST-SUSY.
\item{2.} $N=1$ ST-SUSY is possible for $\bS_1$, $\bS_3$, and $\bS_9$.
There are six general categories of reduction set solutions
(IX, VIII, IV, III, II and I) for $\bS_1$,
three (IX, IV, III) for $\bS_3$, and
one (IX) for $\bS_9$.

Let us define the effective order of a GSOP for a specific gravitino generator
$\bS_i$ as the order of the components of the BV giving GSOP's $\bF$
coefficients.
By this definition,
each member of the set of GSOPs in category IX rendering $N=1$ ST-SUSY
from $\bS_9$ has an effective order of 2.
For  $\bS_3$ there are $N=1$-producing GSOP sets in the same category
with projections of effective orders 4 and 2,
while the parallel sets in VIII, IV, and III
only contain projections with effective orders of 2.
For  $\bS_1$, the $N=1$-producing GSOP set in IX and in VIII
contains projections with effective orders of 8, 4, and 2.
$\bS_1$'s solution set in IV contains projections with
effective orders of 4 and 2, whereas
the solution sets in III contain projections with
effective orders of either (i) 4 and 2 or (ii) 2.
The sets in classes II and I contain only order 2 projections.

Although we have found nine new solutions for generating $N=1$
ST-SUSY in free fermionic models,
it remains to be shown that these are all
physically unique from the standard NAHE
solution set of GSOPs and BVs (the category I solution).
That is, we must check for instances when an $N=1$ model that does
not use the standard NAHE solution has equivalent phenomenology to an $N=1$
model that does.
Identities relating partition functions for
products of (anti)periodic WS fermions to those for certain products of
complex fermions have been derived.\mpr{bailinx,cleaver1}
These identities will be used to test for
possible physical equivalences.
If physical equivalences do exist,
we suspect they are more likely to
involve our
new solution sets containing only GSOPs
with effective orders of 2.

%\end{ignore}

%\begin{ignore}
{\bigfonts\bf\section{Worldsheet Supercurrent}}

Local $N=1$ ST-SUSY in string theory implies not simply
the local $N=1$ worldsheet supersymmetry (WS-SUSY) generated by
the internal energy momentum tensor $T(z)^{\rm int}$
and  spin-3/2 supercurrent,
$$ T_{\rm F}^{\rm int}(z) = i\sum_{J= 1}^{6}
\chi^{3J}\chi^{3J+1}\chi^{3J+2}\,\, ,
\eqno\eqnlabel{tsc2}$$
but an enlargement of this by
a spin-1 field, $J(z)$, forming a global $N=2$ WS-SUSY.\mpr{lust89}
$J(z)$ is the generator of a U(1) Ka\v c-Moody algebra and
splits $T_{\rm F}^{\rm int}$ into two terms,
$$ T_{\rm F}^{\rm int} = T_{\rm F}^{+1} + T_{\rm F}^{-1}\, ,
\eqno\eqnlabel{tfspli}$$
denoted by their respective $+1$ and $-1$ U(1) charges.
$J(z)$ depends upon the choice of gravitino-generating sector $\bS_i$
and takes
the general form
$$ J(z) = i \bS_i^{\rm int} \cdot \partial_z {\bmit B}\,\, .
\eqno\eqnlabel{defj}$$
$\bmit B$ is a bosonization (vector) of the complexified internal LM
fermionic fields,
$$\exp\left\{ \pm i B_{K,L}\right\}
  = {1\over\sqrt{2}}(\chi^K \pm i\chi^{L})\, ,
\eqno\eqnlabel{bosons}$$
and $\bS_i^{\rm int}$ is the internal part of  $\bS_i$, with
$\mid \bS_i^{\rm int}\mid^2 = 3$.
Selecting ${\bS_1}$\mpr{kalara90,giannakis95a} as the
gravitino generator results in
$$ J_{\bS_1}(z) = i \partial_z (B_{3,6} + B_{9,12} +  B_{15,18})\, .
\eqno\eqnlabel{defj1}$$
This U(1) current divides $T_{\rm F}^{\rm int}$ into
\subequationnumstyle{alphabetic}
$$\eqalignno{
T_{\rm F}^{-1} & = \sum_{J=1,3,5}
                   {i\over \sqrt{2}}\exp{(-iB_{3J,3(J+1)})}
                   (\chi^{3J+1}\chi^{3J+2} + i\chi^{3(J+1)+1}\chi^{3(J+1)+2})
\,\, ,
               &\eqnlabel{tm1-a}\cr
               &{\rm and}\cr
T_{\rm F}^{+1} & = \sum_{J=1,3,5}
                   {i\over \sqrt{2}}\exp{(+iB_{3J,3(J+1)})}
                   (\chi^{3J+1}\chi^{3J+2} - i\chi^{3(J+1)+1}\chi^{3(J+1)+2})
\,\, .
               &\eqnlabel{tm1-b}}
$$
\subequationnumstyle{blank}

The comparable expressions for $T_{\rm F}^{\pm 1}$'s
(written in a diagonalized basis of fields)
become a bit more complicated when either
$\b{S}_3$ or $\b{S}_9$ is the gravitino generator.
Nevertheless,  $\b{S}_1$ is not the only choice of gravitino sector
consistent with the  $N=2$ WS-SUSY.
For example, in the initial non-diagonal $\{ \chi^i\}$ basis, the
${\bS_3}$ automorphism arises from the combination of the non-trivial
inner automorphisms,
%\subequationnumstyle{alphabetic}
$$
 (\chi^{3J},\, \chi^{3J+1},\, \chi^{3J+2})
   \rightarrow (\chi^{3J},\, -\chi^{3J+1},\, -\chi^{3J+2})
      \,\,\,\, {\rm ~for~} J= 1,3,5,6
    \eq{inner3}
$$
% (\chi^J,\, \chi^{J+1},\, \chi^{J+2})
%    & \rightarrow (\chi^J,\, \chi^{J+1},\, \chi^{J+2})
%      \,\,\,\, {\rm ~for~} I= 6,12
%    & \eqnlabel{inner3-b}}
%\subequationnumstyle{blank}
with the outer automorphism,
$$ (\chi^{3J},\, \chi^{3J+1},\, \chi^{3J+2})
    \leftrightarrow (\chi^{3(J+1)},\, \chi^{3(J+1)+1},\, \chi^{3(J+1)+2})
      \,\,\,\, {\rm ~for~} J= 1,3\,\, .
    \eqno\eqnlabel{outer3}
$$
Note that $T_{\rm F}^{\rm int}$  is invariant under these transformations,
as is required by (\puteqn{tsc2}).
The eigenstates of the automorphism (with eigenvalues in LHS parenthesis) are:
\subequationnumstyle{alphabetic}
$$\eqalignno{
  (-1):\,\,\,\, & u^{3,6}
               = {1\over \sqrt{2}}(-\chi^3 + \chi^6)
                 &\eqnlabel{ev3-a}\cr
  (+1):\,\,\,\,  & v^{3,6}
               = {1\over \sqrt{2}}(\phantom{-}\chi^3 + \chi^6)
                 &\eqnlabel{ev3-b}\cr
  (+i):\,\,\,\,  & u^{4,7}
               = {-1\over\sqrt{2}}(\chi^7 - i\chi^4)
                 &\eqnlabel{ev3-c}\cr
  (-i):\,\,\,\,  & u^{5,8}
               = {-1\over\sqrt{2}}(\chi^8 + i\chi^5)
                 &\eqnlabel{ev3-d}\cr
                 &\cr
  (-1):\,\,\,\,  & u^{9,12}
               = {1\over \sqrt{2}}(-\chi^9 + \chi^{12})
                 &\eqnlabel{ev3-e}\cr
  (+1):\,\,\,\,  & v^{9,12}
               = {1\over \sqrt{2}}(\phantom{-}\chi^9 + \chi^{12})
                 &\eqnlabel{ev3-f}\cr
  (+i):\,\,\,\,  & u^{10,13}
               = {-1\over\sqrt{2}}(\chi^{10} - i\chi^{13})
                 &\eqnlabel{ev3-g}\cr
  (-i):\,\,\,\,  & u^{12,14}
               = {-1\over\sqrt{2}}(\chi^{12} + i\chi^{14})
                 &\eqnlabel{ev3-h}\cr
                 &\cr
  (+1):\,\,\,\,  &{1\over\sqrt{2}}(\chi^{15} \pm i\chi^{18})
                 &\eqnlabel{ev3-i}\cr
  (-1):\,\,\,\,  & \chi^{16},\,\,\,\, \chi^{17},\,\,\,\,
                   \chi^{19},\,\,\,\, \chi^{20}\,\, . \cr}
$$
\subequationnumstyle{blank}
and complex conjugates of $u^{4,7}$, $u^{5,8}$, $u^{10,13}$, and $u^{11,14}$.
We can bosonize these fermion eigenstates as follows:
\subequationnumstyle{alphabetic}
$$\eqalignno{
   \exp\left\{(-)i\pi B_{I,I+3} \right\}
                           & = u^{I,I+3\, (*)}
                               {\rm ~for~} I= 4,5,10,12
                           & \eqnlabel{boson3a-a}\cr
   \exp\left\{\pm i\pi B_{3,6(-1)}\right\}
                           & = {1\over\sqrt{2}}(u^{3,6} \pm i u^{9,12})
                           & \eqnlabel{boson3a-b}\cr
   \exp\left\{\pm i\pi B_{3,6(+1)} \right\}
                           & = {1\over\sqrt{2}}(v^{3,6} \pm i v^{9,12})
                           & \eqnlabel{boson3a-c}\cr
   \exp\left\{\pm i\pi B_{15,18} \right\}
                           & = {1\over\sqrt{2}}(\chi^{15} \pm i \chi^{18})
                             \, .
                           & \eqnlabel{boson3a-d}}
$$
\subequationnumstyle{blank}
The U(1) current and $\pm 1$--charged supercurrents then appear as,
%\subequationnumstyle{alphabetic}
\def\htpm{\hbox to 0.1truecm{\hfill}}
$$\eqalignno{
J_{\bS_3} &=  i\partial \left(
             \b{B}_{3,6(+)}  -\half \b{B}_{4,7} +\half \b{B}_{5,8}
            +\b{B}_{9,12(+)} -\half \b{B}_{10,13} +\half \b{B}_{11,14}
            +\b{B}_{15,18}\right) &\eqnlabel{js3-a}\cr
T_{\rm F}^- &= \phantom{+} {i\over {2}}
               \exp\{-i\b{B}_{3,6(-)}\}\left(
               \exp\{+i\b{B}_{4,7}\}  \exp\{-i\b{B}_{5,8}\}
             +i\exp\{+i\b{B}_{10,13}\}\exp\{-i\b{B}_{11,14}\}\right)\cr
            &\htpm\phantom{=} + {i\over {2}}
               \exp\{+i\b{B}_{3,6(-)}\}\left(
               \exp\{+i\b{B}_{4,7}\}  \exp\{-i\b{B}_{5,8}\}
             -i\exp\{+i\b{B}_{10,13}\}\exp\{-i\b{B}_{11,14}\}\right)\cr
            &\htpm\phantom{=} + {i\over {2}}
               \exp\{-i\b{B}_{3,6(+)}\}\left(
               \exp\{-i\b{B}_{4,7}\}  \exp\{-i\b{B}_{5,8}\}
             +i\exp\{-i\b{B}_{10,13}\}\exp\{-i\b{B}_{11,14}\}\right)\cr
            &\htpm\phantom{=} + {i\over {2}}
               \exp\{-i\b{B}_{3,6(+)}\}\left(
               \exp\{+i\b{B}_{4,7}\}  \exp\{+i\b{B}_{5,8}\}
             +i\exp\{+i\b{B}_{10,13}\}\exp\{+i\b{B}_{11,14}\}\right)\cr
            &\htpm\phantom{=} + {i\over {2}} \exp\{-i\b{B}_{18,21}\}
               \left( \chi^{16}\chi^{17} + i \chi^{19}\chi^{20} \right)\,\, .
            &\eqnlabel{js3-b}\cr
T_{\rm F}^+ &=  -(T_{\rm F}^+)^{\ast}\,\, .
            &\eqnlabel{js3-c}\cr
}$$

{\bigfonts\bf\section{Comments}}

%This means that any complete $N=1$ ST-SUSY model must have embedded in
%it one of our $N=1$ solutions for left-movers.

We have completely
classified the ways by which the number of spacetime supersymmetries
in heterotic free fermionic strings
may be reduced from $N=4$ to the phenomenologically preferred $N=1$.
This means that the set of LM boundary vectors
in any free fermionic model  with $N=1$ ST-SUSY
must be reproducible from a combination of one of the three
gravitino sectors, $\bS_1$, $\bS_3$, or $\bS_9$, with one of our
accompanying  $N=1$ reduction sets.
The only variations from our
LM BV sets that true $N=1$ models could have are
(1) trivial reordering of the BV worldsheet fermions, or (2)
trivial phase changes by minus signs.
Neither variation leads to physically distinct models. The
latter variation corresponds to either
(1) using one of the odd multiples of $\bS_i$ (of the same order as $\bS_i$),
rather than $\bS_i$ itself, to generate the surviving gravitino, or
(2) (if applicable) using
boundary vector components that commute with $S(4,1,2,1)$
rather than with $S(4,1,1,1)$.

To this date, only the gravitino generator $\bS_1$
has been used in actual $N=1$ free fermionic models.
Reduction to $N=1$ ST-SUSY has been accomplished through
use of the LM NAHE set discussed previously.
Thus, our new results should be especially useful for
model building
when the NAHE set may be inconsistent with other properties
specifically desired of a model.
This appears to be the situation with regard to
current searches for consistent three generation
SO(10) level-2 models. Initial results of this search
were discussed in refs.~\pr{cleaver94a} and \pr{chaudhuri94a}.
Initial attempts to simultaneously produce
$N=1$ SUSY and a three generation SO(10) level-2 grand
unified theory were thought to be successful
using $\bS_1$, the LM NAHE set, and one or two
additional BVs containing some non-integer components.
However, later it was discovered that
the extra non-integer BVs required did not correspond to proper
SU(2)$^6$ automorphisms and, therefore, produced
additional sectors containing tachyonic spacetime fermions, making
the models inconsistent.

The next step in classification of free fermionic models with
potentially good phenomenology is to test for physical equivalences
between our new $N=1$ ST-SUSY solutions and the standard NAHE solution.
Following this, we will
investigate which of our physically
unique LM $N=1$ ST-SUSY solutions may be consistent with
three generations.
While the research presented herein was underway,
a paper exploring this issue appeared.
Ref.~\pr{giannakis95a} shows that the number of generations in
an $N=1$ ST-SUSY model is related to the
index of the underlying $N=2$ internal superconformal field theory.
Determination of the index requires knowledge of both
the LM and RM components of the BVs in a model.
Thus, this technique cannot be applied directly to our $N=1$ solutions
since they have only LM components.
However, slightly modifying this techinique, we will investigate
in upcoming papers\mpr{cleaver95x}
which
(if any) of our sets of left-movers have the {\it potential},
when paired with appropriate right-movers,
to yield exactly three chiral generations for a given gauge group,
in particular for SO(10) level-2.

\sectionnumstyle{blank}
{\bigfonts\bf\section{Acknowledgements}}

Our sincere
thanks to Henry Tye, Jorge Lopez, and Kajia Yuan for helpful discussions.
We also note that unpublished work on classification of free fermionic
spacetime supersymmetries has been done by the authors of ref.~\pr{dreiner89b}.
The contents of Tables 1 and 2 and some of the boundary vector components
in Table 3 were first presented in \pr{dreiner89b}.\newpage

%******************************************************************************
%\input table1.tex

\no TABLE 1. The $\alpha$ in $-\exp\{ i\pi \alpha\}$ eigenvalues
of length-$n$ cyclic permutations combined with
degree of freedom $\theta$ from SU(2) inner automorphism.

\vskip .5truecm
\no
{\hbox to .5truecm{\hfill n \hfill}
\hskip .5truecm
\hbox to 5cm{\hfill $\alpha$'s for $J_{3}$ \hfill}
\hskip .5truecm
\hbox to 8.5truecm{\hfill $\alpha$'s for  $J_+$, $J_-$
(for $J_1$, $J_2$ if $\theta= 0,1$)
                    \hfill}}

\no
\underline{\hbox to .5truecm{\hfill $\phantom{*}$ \hfill}}
\hskip .5truecm
\underline{\hbox to 5cm{\hfill $\phantom{*}$ \hfill}}
\hskip .5truecm
\underline{\hbox to 8.5truecm{\hfill $\phantom{*}$ \hfill}}

\no
\hbox to .5truecm{\hfill 1 \hfill}
\hskip .5truecm
\hbox to 5cm{\hfill 1 \hfill}
\hskip .5truecm
\hbox to 8.5truecm{\hfill $1-\theta$ \hfill}
\vskip .5truecm

\no
\hbox to .5truecm{\hfill 2 \hfill}
\hskip .5truecm
\hbox to 5cm{\hfill 0, 1 \hfill}
\hskip .5truecm
\hbox to 8.5truecm{\hfill $-\thetaa$, $1-\thetaa$ \hfill}
\vskip .5truecm

\no
\hbox to .5truecm{\hfill 3 \hfill}
\hskip .5truecm
\hbox to 5cm{\hfill $\pm\third$, 1\hfill}
\hskip .5truecm
\hbox to 8.5truecm{\hfill $\pm\third -\thetab$, $1-\thetab$ \hfill}
\vskip .5truecm

\no
\hbox to .5truecm{\hfill 4 \hfill}
\hskip .5truecm
\hbox to 5cm{\hfill 0, $\pm\half$, 1\hfill}
\hskip .5truecm
\hbox to 8.5truecm{\hfill $ -\thetac$, $\pm\half -\thetac$,
                    $1- \thetac$ \hfill}
\vskip .5truecm

\no
\hbox to .5truecm{\hfill 5 \hfill}
\hskip .5truecm
\hbox to 5cm{\hfill $\pm\fifth$, $\pm\threefifth$, 1\hfill}
\hskip .5truecm
\hbox to 8.5truecm{\hfill $\pm\fifth -\thetad$, $\pm\threefifth -\thetad$,
                        $ 1 - \thetad$ \hfill}
\vskip .5cm

\no
\hbox to .5truecm{\hfill 6 \hfill}
\hskip .5truecm
\hbox to 5cm{\hfill 0, $\pm\third$, $\pm\twothird$, 1\hfill}
\hskip .5truecm
\hbox to 8.5truecm{\hfill $- \thetae$, $\pm\third - \thetae$,
                    $\pm\twothird - \thetae$, $1 - \thetae$ \hfill}
\newpage

%******************************************************************************
%\input table2.tex

\no TABLE 2. Representatives, $\bS_i$, of the eleven distinct classes of
left-moving boundary vectors that can yield massless fermions.

Classes
are defined by lengths of non-disjoint cycles in the SU(2)$^6$ permutation
upon which a boundary vector is based.
Only $\bS_1$, $\bS_3$, $\bS_5$, $\bS_7$, $\bS_9$, and
$\bS_{10}$ have an even number of perioidic components (1's).
They are the only potential gravitino generators.

The first two components of each $\bS_i$ are the transverse
spacetime components $\psi^{1,2}$ for lightcone gauge.
The boundary vector components for each non-disjoint cycle of length--$n_k$
in $\bS_i$ are enclosed in parentheses.
The components on the left-hand side of the semi-colons
correspond to the $n_k$ eigenvalues for the $J_3$ fermions in each group.
The components in the first (second) half of the right-hand side of the
semicolon correspond to the eigenvalues for the $J_1$ $(J_2)$ fermions.
A ``$\hat{\phantom{*}}$" denotes an boundary vector component
for a complex fermion.
For each complex fermion eigenstate with a complex eigenvalue
(non-integer boundary component), there is
also a complex congugate eigenstate.

\def\va{\vskip .3truecm}

\vskip .5truecm
\no
\hbox to 2.4truecm{\bf Class\hfill}
\hskip .5truecm
\hbox to 13.4truecm{\hfill \bf Representative Massless Boundary Vector \hfill}

\no
\underline{\hbox to 2.4truecm{\hfill $\phantom{*}$ \hfill}}
\hskip .5truecm
\underline{\hbox to 13.4truecm{\hfill $\phantom{*}$\hfill}}

\no
\hbox to 2.4truecm{$1\cdot1\cdot1\cdot1\cdot1\cdot1$\hfill}
\hskip .5truecm
\hbox to 1.1truecm{\hfill $\bS_1$ \hfill $=$}
\hskip .5truecm
\hbox to 11.8truecm{$\{ 1,1\quad (1; 0, 0)\quad (1; 0, 0)\quad
                                      (1; 0, 0)\quad (1; 0, 0)\quad
                                      (1; 0, 0)\quad (1; 0, 0)\}$ \hfill}\hfill

\va
\no
\hbox to 2.4truecm{$2\cdot1\cdot1\cdot1\cdot1$\hfill}
\hskip .5truecm
\hbox to 1.1truecm{\hfill $\bS_2$ \hfill $=$}
\hskip .5truecm
\hbox to 11.8truecm{$\{ 1,1\quad (0,1; -\hat{\half},\, \hat{\half})\quad
                                      (1; 0, 0)\quad (1; 0, 0)\quad
                                      (1; 0, 0)\quad (1; 0, 0)\}$ \hfill}\hfill

\va
\no
\hbox to 2.4truecm{$2\cdot2\cdot1\cdot1$\hfill}
\hskip .5truecm
\hbox to 1.1truecm{\hfill $\bS_3$ \hfill $=$}
\hskip .5truecm
\hbox to 11.8truecm{$\{ 1,1\quad (0,1; -\hat{\half}, \hat{\half})\quad
                                      (0,1; -\hat{\half}, \hat{\half})\quad
                                      (1; 0, 0)\quad (1; 0, 0)\}$ \hfill}\hfill

\va
\no
\hbox to 2.4truecm{$2\cdot2\cdot2$\hfill}
\hskip .5truecm
\hbox to 1.1truecm{\hfill $\bS_4$ \hfill $=$}
\hskip .5truecm
\hbox to 11.8truecm{$\{ 1,1\quad (0,1; -\hat{\half}, \hat{\half})\quad
                                      (0,1; -\hat{\half}, \hat{\half})\quad
                                      (0,1; -\hat{\half}, \hat{\half}
                                                              \}$ \hfill}\hfill

\va
\no
\hbox to 2.4truecm{$3\cdot1\cdot1\cdot1$\hfill}
\hskip .5truecm
\hbox to 1.1truecm{\hfill $\bS_5$ \hfill $=$}
\hskip .5truecm
\hbox to 11.8truecm{$\{ 1,1\quad
                 (\hat{\third},1; -\hat{\twothird}, 0,0,\hat{\twothird})\quad
                                      (1; 0, 0)\quad (1; 0, 0)\quad
                                      (1; 0, 0)\}$ \hfill}\hfill

\va
\no
\hbox to 2.4truecm{$3\cdot2\cdot1$\hfill}
\hskip .5truecm
\hbox to 1.1truecm{\hfill $\bS_6$ \hfill $=$}
\hskip .5truecm
\hbox to 11.8truecm{$\{ 1,1\quad
                 (\hat{\third},1; -\hat{\twothird}, 0,0,\hat{\twothird})\quad
                                      (0,1; -\hat{\half}, \hat{\half})\quad
                                      (1; 0, 0)\}$ \hfill}\hfill

\va
\no
\hbox to 2.4truecm{$3\cdot3$\hfill}
\hskip .5truecm
\hbox to 1.1truecm{\hfill $\bS_7$ \hfill $=$}
\hskip .5truecm
\hbox to 11.8truecm{$\{ 1,1\quad
                 (\hat{\third},1; -\hat{\twothird}, 0,0,\hat{\twothird})\quad
                 (\hat{\third},1; -\hat{\twothird}, 0,0,\hat{\twothird})
                                                              \}$ \hfill}\hfill

\va
\no
\hbox to 2.4truecm{$4\cdot1\cdot1$\hfill}
\hskip .5truecm
\hbox to 1.1truecm{\hfill $\bS_8$ \hfill $=$}
\hskip .5truecm
\hbox to 11.8truecm{$\{ 1,1\quad
   (0,\hat\half, 1; -\hat{\threefourth}, -\hat{\fourth},
                     \hat{\fourth},  \hat{\threefourth})\quad
                                      (1; 0, 0)\quad
                                      (1; 0, 0)\}$ \hfill}\hfill

\va
\no
\hbox to 2.4truecm{$4\cdot2$\hfill}
\hskip .5truecm
\hbox to 1.1truecm{\hfill $\bS_9$ \hfill $=$}
\hskip .5truecm
\hbox to 11.8truecm{$\{ 1,1\quad
   (0,\hat\half, 1; -\hat{\threefourth}, -\hat{\fourth},
                     \hat{\fourth},  \hat{\threefourth})\quad
                                      (0,1; -\hat{\half}, \hat{\half})
                                                              \}$ \hfill}\hfill

\va
\no
\hbox to 2.4truecm{$5\cdot1$\hfill}
\hskip .5truecm
\hbox to 1.1truecm{\hfill $\bS_{10}$ \hfill $=$}
\hskip .5truecm
\hbox to 11.8truecm{$\{ 1,1\quad
   (\hat{\fifth},\hat{\threefifth},1; -\hat{\fourfifth},-\hat{\twofifth},0,0,
                                      \hat{\twofifth},  \hat{\fourfifth})\quad
                                      (1; 0, 0)\}$ \hfill}\hfill

\va
\no
\hbox to 2.4truecm{$6$\hfill}
\hskip .5truecm
\hbox to 1.1truecm{\hfill $\bS_{11}$ \hfill $=$}
\hskip .5truecm
\hbox to 11.8truecm{$\{ 1,1\quad
(0,\hat{\third},\hat{\twothird},1;
                                  -\hat{\fivesixth},-\hat{\half},-\hat{\sixth},
                                   \hat{\sixth},\hat{\half},\hat{\fivesixth})
                                                \}$\hfill}\hfill

\va
\no Order $N_{\bS_i} = 2$, 4, 4, 4, 6, 12, 6, 8, 8, 10, 12
for $i = 1$ to 11, respectively.
\newpage

%******************************************************************************
%\input table3.tex

\def\hm{\hbox to 0.8cm{\hfill}}
\def\hn{\hbox to 0.4cm{\hfill}}
\def\hp{\hbox to 0.3cm{\hfill}}

\def\tailer{\vskip 0.2truecm

\hrule height0.1truecm}

\def\tailerb{\hrule height0.1truecm}

\def\headings{
%\vskip .5truecm
\hrule height0.1truecm

\vskip .5truecm
\no
\hbox to 5.9truecm{\hbox to 3.0truecm{\hfill {\bf Cycles of}\hfill}\hn\hbox to
1.4truecm{\hfill {\bf Order}\hfill}\hm\hfill}\hbox to 2.8truecm{\hfill}\hp\hbox
to 7.412truecm{\hfill {\bf Boundary Vector}\hfill}

\no
\hbox to 5.9truecm{\hbox to 3.0truecm{\hfill {\bf Lengths} ${\bmit
n}_i$\hfill}\hn\hbox to 1.4truecm{\hfill $\bmit N$\hfill}\hm\hfill}\hbox to
2.8truecm{\hfill {\bf Designation}\hfill}\hp\hbox to 7.412truecm{\hfill {\bf
Components}\hfill}

\no
\hbox to 5.9truecm{\underline{\hbox to
3.0truecm{$\phantom{*}$\hfill}}\hn\underline{\hbox to
1.4truecm{$\phantom{*}$\hfill}}\hm\hfill}\underline{\hbox to
2.8truecm{$\phantom{*}$\hfill}}\hp\underline{\hbox to
7.412truecm{$\phantom{*}$\hfill}}}

\def\ra{\hfill$\rightarrow$\hbox to .3truecm{\hfill}}

\def\fch#1#2{\no\hbox to 5.9truecm{\hbox to 3.0truecm{#1\hfill}\hn\hbox to
1.4truecm{\hfill #2\hfill}\hm}\hfill}

\def\fcha#1#2{\no\hbox to 5.9truecm{\hbox to 3.0truecm{#1\hfill}\hn\hbox to
1.4truecm{\hfill #2\hfill}\ra}}

\def\fchn{\vskipa
\no\hbox to 5.9truecm{\hbox to 3.0truecm{\hfill }\hn\hbox to
1.4truecm{\hfill}\hm\hfill}}

\def\fchna{\vskipa
\no\hbox to 5.9truecm{\hbox to 3.0truecm{\hfill}\hn\hbox to
1.4truecm{\hfill}\ra}}

\def\hbs#1{\hbox to 2.8truecm{#1\hfill=}\hp}
\def\hbv#1{\hbox to 7.412truecm{#1\hfill}}

\def\cont#1#2{
\no
\underline{\hbox to 16.5truecm{GSO projection classes
for #1 contribution to gravitino sectors #2\hfill}}
\vskip .2truecm}

\def\conto#1#2{
\no
\underline{\hbox to 16.5truecm{GSO projection classes
for #1 contribution to gravitino sector #2\hfill}}
\vskip .2truecm}

%%%%%%%%%%%%%%%%%%%%%%%%%%%%%%%%%%%%%%%%%%%%%%%%
\def\sone{$1\cdot1\cdot1\cdot1\cdot1\cdot1$}

\def\sseven{$3\cdot3$}

\def\class#1{\no
\hbox to 3.0truecm{class #1:\hfill}\hbox{ $\bF$ coef = }}

\def\hbc#1{\hbox to 0.6truecm{{#1}\hfill}}

\def\hdp#1#2{
\hbox{ and $\half$dp = }
\hbox to 1cm{$#1$\hfill}
\hbox to 2cm{$\mod{#2}$\hfill}}

%start table

\no TABLE 3. Sets of commuting boundary vectors and classes of GSO projections
for components of gravitino-generating basis vectors.
\vskip .2truecm

Listed below are all possible sets of mutually commuting boundary vectors
for the worldsheet supersymmetric sector of heterotic free fermionic
strings. Each element of length $3\sum n_k = 3n_{\ast}$
(with complex components counting double)
in a set of commuting boundary vectors is derived from a
cyclic permutation that is a
power of the non-disjoint cycle of length $n_{\ast}$
from  the set's first boundary vector.
In the notation below, the boundary vectors designated as $S(n_{\ast},y,z,w)$
are classified by:
(1) the total length of their permutations,  $n_{\ast}$,
(2) the power of the cyclic permutations used, $y$,
(3) the various ways of dividing their eigenvalues among the $J_1$'s
and $J_2$'s, numbered by $z$,
and
(4) the choice of inner automorphism, specified by $w$.

In each commuting set, the distinct length-$3n_{\ast}$ boundary vectors
that could potentially appear in one (or more) of the six gravitino
sectors are
marked with a ``$\rightarrow$''. All boundary vectors are classified according
to their contributions to the GSO projections operatings on these
``arrowed'' boundary vectors.
The important properties defining a boundary vector's contributions to such
GSO operators are:
(i) the boundary vector's components that correspond to the ``arrowed"
vector's periodic components, referred to as $\bF$ coefficients,
and
(ii) half of the dot product of the boundary vector with the ``arrowed''
vector.
Since we can shown that varying the division of the components among the
$J_1$'s and $J_2$'s does not alter the set of boundary vectors
composing a class, GSO data is given only for $z=1$.
\vskip 0.5truecm

%***************************************************************************
\headings

\fcha{1}{2}\hbox to 10.57truecm{\hbs{$S(1,1,1,1)$}\hbv{$(1; 0, 0)$}}

\fchn\hbox to 10.57truecm{\hbs{$S(1,1,1,2)$}\hbv{$(0; 1, 0)$}}

\fchn\hbox to 10.57truecm{\hbs{$S(1,1,1,3)$}\hbv{$(0; 0, 1)$}}

\fchn\hbox to 10.57truecm{\hbs{$S(1,1,1,4)$}\hbv{$(1; 1, 1)$}}

\vskipa
\cont{$S(1,1,1,1)$}{$\bS_1$, $\bS_3$, $\bS_5$, $\bS_7$}
\class{$(1.1.1.1)$}
\hbc{$1$}
\hdp{\fourth}{1}

\no\hskip 2.2truecm{$\{S(1,1,1,1),\,\, S(1,1,1,4)\}$}

\vskipa
\class{$(1.1.1.2)$}
\hbc{$0$}
\hdp{0}{1}

\no\hskip 2.2truecm{$\{S(1,1,1,2),\,\, S(1,1,1,3)\}$}

\tailer

\newpage

%*****************
\headings

\fcha{2}{4}\hbox to 10.57truecm{\hbs{$S(2,1,1,1)$}\hbv{$(0,1; -\hat{\half},
\phm\hat{\half})$}}

\fchn\hbox to 10.57truecm{\hbs{$S(2,1,1,2)$}\hbv{$(1,0; \phm\hat{\half},
\phm\hat{\half})$}}

\fchn\hbox to 10.57truecm{\hbs{$S(2,1,1,3)$}\hbv{$(1,0; -\hat{\half},
-\hat{\half})$}}

\fchn\hbox to 10.57truecm{\hbs{$S(2,1,1,4)$}\hbv{$(0,1; \phm\hat{\half},
-\hat{\half})$}}

\vskipa
\fcha{$1\cdot1$}{2}\hbox to 10.57truecm{\hbs{$S(2,2,1,1)$}\hbv{$(1,1;
\hj\phm\hat{0}, \hj\phm\hat{0})$}}

\fchna\hbox to 10.57truecm{\hbs{$S(2,2,1,2)$}\hbv{$(0,0; \hj\phm\hat{1},
\hj\phm\hat{0})$}}

\fchn
\hbs{$S(2,2,1,3)$}\hbv{$(0,0; \hj\phm\hat{0}, \hj\phm\hat{1})$}

\fchn\hbs{$S(2,2,1,4)$}\hbv{$(1,1; \hj\phm\hat{1}, \hj\phm\hat{1})$}

%***************
\vskipa
\cont{$S(2,1,1,1)$}{$\bS_3$, $\bS_9$}
\class{$(2.2.1.1)$}
\hbc{$\phm1$}
\hdp{\phm\half}{\half}

\no\hskip 2.2truecm{$\{S(2,1,1,1),\,\, S(2,1,1,4)\}$}

\vskipa
\class{$(2.2.1.2)$}
\hbc{$\phm1$}
\hdp{\phm\fourth}{\half}

\no\hskip 2.2truecm{$\{S(2,2,1,1),\,\, S(2,2,1,4)\}$}

\vskipa
\class{$(2.2.1.3)$}
\hbc{$\phm0$}
\hdp{\phm0}{\half}

\no\hskip 2.2truecm{$\{S(2,1,1,2),\,\, S(2,1,1,3)\}$}

\vskipa
\class{$(2.2.1.4)$}
\hbc{$\phm0$}
\hdp{-\fourth}{\half}

\no\hskip 2.2truecm{$\{S(2,2,1,2),\,\, S(2,2,1,3)\}$}

%***************
\vskipa
\cont{$S(2,2,1,1)$}{$\bS_1$, $\bS_3$, $\bS_5$}
\class{$(1.2.1.1)$}
\hbc{$\phm1$},
\hbc{$\phm1$}
\hdp{\phm\half}{1}

\no\hskip 2.2truecm{$\{S(2,2,1,1),\,\, S(2,2,1,4)\}$}

\vskipa
\class{$(1.2.1.2)$}
\hbc{$\phm1$},
\hbc{$\phm0$}
\hdp{\phm\fourth}{1}

\no\hskip 2.2truecm{$\{S(2,1,1,2),\,\, S(2,1,1,3)\}$}

\vskipa
\class{$(1.2.1.3)$}
\hbc{$\phm0$},
\hbc{$\phm1$}
\hdp{\phm\fourth}{1}

\no\hskip 2.2truecm{$\{S(2,1,1,1),\,\, S(2,1,1,4)\}$}

\vskipa
\class{$(1.2.1.4)$}
\hbc{$\phm0$},
\hbc{$\phm0$}
\hdp{\phm0}{1}

\no\hskip 2.2truecm{$\{S(2,2,1,2),\,\, S(2,2,1,3)\}$}

%**************
\vskipa
\cont{$S(2,2,1,2)$}{$\bS_1$, $\bS_3$, $\bS_5$}
\class{$(1.2.2.1)$}
\hbc{$\phm\hat{1}$}
\hdp{\phm\half}{1}

\no\hskip 2.2truecm{$\{S(2,2,1,2),\,\, S(2,2,1,4)\}$}

\vskipa
\class{$(1.2.2.2)$}
\hbc{$\phm\hat{0}$}
\hdp{\phm0}{1}

\no\hskip 2.2truecm{$\{S(2,2,1,1),\,\, S(2,2,1,3)\}$}

\vskipa
\class{$(1.2.2.3)$}
\hbc{$\phm\hat{\half}$}
\hdp{\phm\onefourth}{1}

\no\hskip 2.2truecm{$\{S(2,1,1,2),\,\, S(2,1,1,4)\}$}

\vskipa
\class{$(1.2.2.4)$}
\hbc{$-\hat{\half}$}
\hdp{-\onefourth}{1}

\no\hskip 2.2truecm{$\{S(2,1,1,1),\,\, S(2,1,1,3)\}$}

\tailer

\newpage

%\begin{ignore}
%****************
\headings

\fcha{3}{6}\hbs{$S(3,1,1,1)$}\hbv{$(\phm\hat{\third}, 1; -\hat{\twothird},0,
                                   0,\phm\hat{\twothird})$}

\fchn\hbs{$S(3,1,1,2)$}\hbv{$(-\hat{\twothird}, 0; \phm\hat{\third},1,
                            0,\phm\hat{\twothird})$}

\fchn\hbs{$S(3,1,1,3)$}\hbv{$(-\hat{\twothird}, 0; -\hat{\twothird},0,
                                       1,-\hat{\third})$}

\fchn\hbs{$S(3,1,1,4)$}\hbv{$(\phm\hat{\third}, 1; \phm\hat{\third},1,
                                  1,-\hat{\third})$}

\fchn\hbs{$S(3,2,1,1)$}\hbv{$(-\hat{\third}, 1; -\hat{\third},1,
                                  1,\phm\hat{\third})$}

\fchn\hbs{$S(3,2,1,2)$}\hbv{$(\phm\hat{\twothird}, 0; \phm\hat{\twothird},0,
                                  1,\phm\hat{\third})$}

\fchn\hbs{$S(3,2,1,3)$}\hbv{$(\phm\hat{\twothird}, 0; -\hat{\third},1,
                                  0,-\hat{\twothird})$}

\fchn\hbs{$S(3,2,1,4)$}\hbv{$(-\hat{\third}, 1; \phm\hat{\twothird},0,
                                  0,-\hat{\twothird})$}

\vskipa\fcha{$1\cdot1\cdot1$}{2}\hbs{$S(3,3,1,1)$}\hbv{$(\phm\hat{1}, 1;
                                  \hj\phm\hat{0},0,
                                  0, \hj\phm\hat{0})$}

\fchn\hbs{$S(3,3,1,2)$}\hbv{$(\phm\hat{0}, 0; \hj\phm\hat{1},1,
                                  0, \hj\phm\hat{0})$}

\fchn\hbs{$S(3,3,1,3)$}\hbv{$(\phm\hat{0}, 0; \hj\phm\hat{0},0,
                                  1, \hj\phm\hat{1})$}

\fchn\hbs{$S(3,3,1,4)$}\hbv{$(\phm\hat{1}, 1; \hj\phm\hat{1},1,
                                  1, \hj\phm\hat{1})$}

\vskipa
\cont{$S(3,1,1,1)$}{$\bS_5$, $\bS_7$}
\class{$(3.3.1.1)$}
\hbc{$\phm1$}
\hdp{\phm\threefourth}{\third}

\no\hskip 2.2truecm{$\{S(3,1,1,1),\,\, S(3,1,1,4),\,\,
                       S(3,2,1,1),\,\, S(3,2,1,4),\,\,
                       S(3,3,1,1),\,\, S(3,3,1,4)\}$}

\vskipa
\class{$(3.3.1.2)$}
\hbc{$\phm0$}
\hdp{\phm0}{\third}

\no\hskip 2.2truecm{$\{S(3,1,1,2),\,\, S(3,1,1,3),\,\,
                       S(3,2,1,2),\,\, S(3,2,1,3),\,\,
                       S(3,3,1,2),\,\, S(3,3,1,3)\}$}

%******************
\vskipa
\cont{$S(3,3,1,1)$}{$\bS_1$, $\bS_5$}
\class{$(1.3.1.1)$}
\hbc{$\phm\hat{1}$},
\hbc{$\phm 1$}
\hdp{\phm\threefourth}{1}

\no\hskip 2.2truecm{$\{S(3,3,1,1),\,\, S(3,3,1,4)\}$}

\vskipa
\class{$(1.3.1.2)$}
\hbc{$\phm\hat{0}$},
\hbc{$\phm 0$}
\hdp{\phm0}{1}

\no\hskip 2.2truecm{$\{S(3,3,1,2),\,\, S(3,3,1,3)\}$}

\vskipa
\class{$(1.3.1.3)$}
\hbc{$\phm\hat{\onethird}$},
\hbc{$\phm 1$}
\hdp{\phm\fivetwelf}{1}

\no\hskip 2.2truecm{$\{S(3,1,1,1),\,\, S(3,1,1,4)\}$}

\vskipa
\class{$(1.3.1.4)$}
\hbc{$-\hat{\onethird}$},
\hbc{$\phm 1$}
\hdp{\phm\onetwelf}{1}

\no\hskip 2.2truecm{$\{S(3,2,1,1),\,\, S(3,2,1,4)\}$}

\vskipa
\class{$(1.3.1.5)$}
\hbc{$\phm\hat{\twothird}$},
\hbc{$\phm0$}
\hdp{\phm\onethird}{1}

\no\hskip 2.2truecm{$\{S(3,2,1,2),\,\, S(3,2,1,3)\}$}

\vskipa
\class{$(1.3.1.6)$}
\hbc{$-\hat{\twothird}$},
\hbc{$\phm0$}
\hdp{-\onethird}{1}

\no\hskip 2.2truecm{$\{S(3,1,1,2),\,\, S(3,1,1,3)\}$}

\tailer

\newpage

%***************
\headings

\fcha{4}{8}\hbs{$S(4,1,1,1)$}\hbv{$(\phm\hat{\half},0,1;
                                   -\hat{\threefourth},
                                   -\hat{\fourth}, \phm\hat{\fourth},
                                    \phm\hat{\threefourth})$}

\fchn\hbs{$S(4,1,1,2)$}\hbv{$(-\hat{\half},1,0; \phm\hat{\fourth},
                                       \phm\hat{\threefourth},
                                       \phm\hat{\fourth},
                                       \phm\hat{\threefourth})$}

\fchn\hbs{$S(4,1,1,3)$}\hbv{$(-\hat{\half},1,0; -\hat{\threefourth},
                                       -\hat{\fourth},
                                       -\hat{\threefourth},-\hat{\fourth})$}

\fchn\hbs{$S(4,1,1,4)$}\hbv{$(\phm\hat{\half},0,1; \phm\hat{\fourth},
                                          \phm\hat{\threefourth},
                                         -\hat{\threefourth},-\hat{\fourth})$}

\fchn\hbs{$S(4,3,1,1)$}\hbv{$(-\hat{\half},0,1; -\hat{\fourth},
                                      -\hat{\threefourth},
                                       \phm\hat{\threefourth},
                                       \phm\hat{\fourth})$}
\fchn\hbs{$S(4,3,1,2)$}\hbv{$(\phm\hat{\half},1,0; \phm\hat{\threefourth},
                                      \phm\hat{\fourth},
                                      \phm\hat{\threefourth},
                                      \phm\hat{\fourth})$}

\fchn\hbs{$S(4,3,1,3)$}\hbv{$(\phm\hat{\half},1,0; -\hat{\fourth},
                                       -\hat{\threefourth},
                                      -\hat{\fourth},-\hat{\threefourth})$}

\fchn\hbs{$S(4,3,1,4)$}\hbv{$(-\hat{\half},0,1; \phm\hat{\threefourth},
                                       \phm\hat{\fourth},
                                      -\hat{\fourth},-\hat{\threefourth})$}

\vskipa\fcha{$2\cdot2$}{4}\hbs{$S(4,2,1,1)$}\hbv{$(\hj\phm\hat{0},1,1;
                                   -\hat{\half},\phm\hat{\half},
                                   -\hat{\half},\phm\hat{\half})$}

\fchna\hbs{$S(4,2,1,2)$}\hbv{$(\hj\phm\hat{1},0,0; \phm\hat{\half},
                                  -\hat{\half},
                                  -\hat{\half},\phm\hat{\half})$}

\fchn\hbs{$S(4,2,1,3)$}\hbv{$(\phm\hat{1},0,0; -\hat{\half},\phm\hat{\half},
                                   \phm\hat{\half},-\hat{\half})$}

\fchn\hbs{$S(4,2,1,4)$}\hbv{$(\phm\hat{0},1,1;
                                       \phm\hat{\half},-\hat{\half},
                                       \phm\hat{\half},-\hat{\half})$}

\vskipa
\fcha{$1\cdot1\cdot1\cdot1$}{2}\hbs{$S(4,4,1,1)$}\hbv{$(\hj\phm\hat{1},1,1;
                                          \hj\phm\hat{0},\hj\phm\hat{0},
                                          \hj\phm\hat{0},\hj\phm\hat{0})$}

\fchna\hbs{$S(4,4,1,2)$}\hbv{$(\hj\phm\hat{0},0,0;
                                         \hj\phm\hat{1},\hj\phm\hat{1},
                                         \hj\phm\hat{0},\hj\phm\hat{0})$}

\fchn\hbs{$S(4,4,1,3)$}\hbv{$(\hj\phm\hat{0},0,0;
                                          \hj\phm\hat{0},\hj\phm\hat{0},
                                          \hj\phm\hat{1},\hj\phm\hat{1})$}

\fchn\hbs{$S(4,4,1,4)$}\hbv{$(\hj\phm\hat{1},1,1;  \hj\phm\hat{1},
                                          \hj\phm\hat{1},
                                          \hj\phm\hat{1},\hj\phm\hat{1})$}

\vskipa
\conto{$S(4,1,1,1)$}{$\bS_9$}
\class{$(4.4.1.1)$}
\hbc{$\phm1$}
\hdp{\phm1}{\fourth}

\no\hskip 2.2truecm{$\{S(4,1,1,1),\,\, S(4,1,1,4),\,\,
                       S(4,3,1,1),\,\, S(4,3,1,4)\}$}

\vskipa
\class{$(4.4.1.2)$}
\hbc{$\phm1$}
\hdp{\phm\half}{\fourth}

\no\hskip 2.2truecm{$\{S(4,2,1,1),\,\, S(4,2,1,4),\,\,
                       S(4,4,1,1),\,\, S(4,4,1,4)\}$}

\vskipa
\class{$(4.4.1.3)$}
\hbc{$\phm0$}
\hdp{\phm0}{\fourth}

\no\hskip 2.2truecm{$\{S(4,1,1,2),\,\, S(4,1,1,3),\,\,
                       S(4,3,1,2),\,\, S(4,3,1,3)\}$}

\vskipa
\class{$(4.4.1.4)$}
\hbc{$\phm0$}
\hdp{\phm\fourth}{\fourth}

\no\hskip 2.2truecm{$\{S(4,2,1,2),\,\, S(4,2,1,3),\,\,
                       S(4,4,1,2),\,\, S(4,4,1,3)\}$}

\vskipa
\conto{$S(4,2,1,1)$}{$\bS_3$}
\class{$(2.4.1.1)$}
\hbc{$\phm1$},
\hbc{$\phm1$}
\hdp{\phm1}{\half}

\no\hskip 2.2truecm{$\{S(4,2,1,1),\,\, S(4,2,1,4),\,\,
                       S(4,4,1,1),\,\, S(4,4,1,4)\}$}

\vskipa
\class{$(2.4.1.2)$}
\hbc{$\phm1$},
\hbc{$\phm0$}
\hdp{\phm\half}{\half}

\no\hskip 2.2truecm{$\{S(4,1,1,2),\,\, S(4,1,1,3),\,\,
                       S(4,3,1,2),\,\, S(4,3,1,3)\}$}

\vskipa
\class{$(2.4.1.3)$}
\hbc{$\phm0$},
\hbc{$\phm1$}
\hdp{\phm\half}{\half}

\no\hskip 2.2truecm{$\{S(4,1,1,1),\,\, S(4,1,1,4),\,\,
                       S(4,3,1,1),\,\, S(4,3,1,4)\}$}

\vskipa
\class{$(2.4.1.4)$}
\hbc{$\phm0$},
\hbc{$\phm0$}
\hdp{\phm0}{\half}

\no\hskip 2.2truecm{$\{S(4,2,1,2),\,\, S(4,2,1,3),\,\,
                       S(4,4,1,2),\,\, S(4,4,1,3)\}$}

\vskipa
\conto{$S(4,2,1,2)$}{$\bS_3$}
\class{$(2.4.2.1)$}
\hbc{$\phm\hat{1}$}
\hdp{\phm1}{\half}

\no\hskip 2.2truecm{$\{S(4,2,1,2),\,\, S(4,2,1,3),\,\,
                       S(4,4,1,1),\,\, S(4,4,1,4)\}$}

\vskipa
\class{$(2.4.2.2)$}
\hbc{$\phm\hat{0}$}
\hdp{\phm0}{\half}

\no\hskip 2.2truecm{$\{S(4,2,1,1),\,\, S(4,2,1,4),\,\,
                       S(4,4,1,2),\,\, S(4,4,1,3)\}$}

\vskipa
\class{$(2.4.2.3)$}
\hbc{$\phm\hat{\half}$}
\hdp{\phm\onefourth}{\half}

\no\hskip 2.2truecm{$\{S(4,1,1,1),\,\, S(4,1,1,4),\,\,
                       S(4,3,1,2),\,\, S(4,3,1,3)\}$}

\vskipa
\class{$(2.4.2.4)$}
\hbc{$-\hat{\half}$}
\hdp{-\onefourth}{\half}

\no\hskip 2.2truecm{$\{S(4,1,1,2),\,\, S(4,1,1,3),\,\,
                       S(4,3,1,1),\,\, S(4,3,1,4)\}$}

\vskipa
\conto{$S(4,4,1,1)$}{$\bS_1$}
\class{$(1.4.1.1)$}
\hbc{$\phm\hat{1}$},
\hbc{$\phm1$},
\hbc{$\phm1$}
\hdp{\phm1}{1}

\no\hskip 2.2truecm{$\{S(4,4,1,1),\,\, S(4,4,1,4)\}$}

\vskipa
\class{$(1.4.1.2)$}
\hbc{$\phm\hat{1}$},
\hbc{$\phm0$},
\hbc{$\phm0$}
\hdp{\phm\half}{1}

\no\hskip 2.2truecm{$\{S(4,2,1,2),\,\, S(4,2,1,3)\}$}

\vskipa
\class{$(1.4.1.3)$}
\hbc{$\phm\hat{0}$},
\hbc{$\phm1$},
\hbc{$\phm1$}
\hdp{\phm\half}{1}

\no\hskip 2.2truecm{$\{S(4,2,1,1),\,\, S(4,2,1,4)\}$}

\vskipa
\class{$(1.4.1.4)$}
\hbc{$\phm\hat{0}$},
\hbc{$\phm0$},
\hbc{$\phm0$}
\hdp{\phm0}{1}

\no\hskip 2.2truecm{$\{S(4,4,1,2),\,\, S(4,4,1,3)\}$}

\vskipa
\class{$(1.4.1.5)$}
\hbc{$\phm\hat{\half}$},
\hbc{$\phm1$},
\hbc{$\phm0$}
\hdp{\phm\half}{1}

\no\hskip 2.2truecm{$\{S(4,3,1,2),\,\, S(4,3,1,3)\}$}

\vskipa
\class{$(1.4.1.6)$}
\hbc{$-\hat{\half}$},
\hbc{$\phm1$},
\hbc{$\phm0$}
\hdp{\phm0}{1}

\no\hskip 2.2truecm{$\{S(4,1,1,2),\,\, S(4,1,1,3)\}$}

\vskipa
\class{$(1.4.1.7)$}
\hbc{$\phm\hat{\half}$},
\hbc{$\phm0$},
\hbc{$\phm1$}
\hdp{\phm\half}{1}

\no\hskip 2.2truecm{$\{S(4,1,1,1),\,\, S(4,1,1,4)\}$}

\vskipa
\class{$(1.4.1.8)$}
\hbc{$-\hat{\half}$}
\hbc{$\phm0$},
\hbc{$\phm1$}
\hdp{\phm0}{1}

\no\hskip 2.2truecm{$\{S(4,3,1,1),\,\, S(4,3,1,4)\}$}

\vskipa
\conto{$S(4,4,1,2)$}{$\bS_1$}
\class{$(1.4.2.1)$}
\hbc{$\phm\hat{1}$},
\hbc{$\phm\hat{1}$}
\hdp{\phm1}{1}

\no\hskip 2.2truecm{$\{S(4,4,1,2),\,\, S(4,4,1,4)\}$}

\vskipa
\class{$(1.4.2.2)$}
\hbc{$\phm\hat{0}$},
\hbc{$\phm\hat{0}$}
\hdp{\phm0}{1}

\no\hskip 2.2truecm{$\{S(4,4,1,1),\,\, S(4,4,1,3)\}$}

\vskipa
\class{$(1.4.2.3)$}
\hbc{$\phm\hat{\threefourth}$},
\hbc{$\phm\hat{\onefourth}$}
\hdp{\phm\half}{1}

\no\hskip 2.2truecm{$\{S(4,3,1,2),\,\, S(4,3,1,4)\}$}

\vskipa
\class{$(1.4.2.4)$}
\hbc{$-\hat{\threefourth}$},
\hbc{$-\hat{\onefourth}$}
\hdp{-\half}{1}

\no\hskip 2.2truecm{$\{S(4,1,1,1),\,\, S(4,1,1,3)\}$}

\vskipa
\class{$(1.4.2.5)$}
\hbc{$\phm\hat{\half}$},
\hbc{$-\hat{\half}$}
\hdp{\phm0}{1}

\no\hskip 2.2truecm{$\{S(4,2,1,2),\,\, S(4,2,1,4)\}$}

\vskipa
\class{$(1.4.2.6)$}
\hbc{$-\hat{\half}$},
\hbc{$\phm\hat{\half}$}
\hdp{\phm0}{1}

\no\hskip 2.2truecm{$\{S(4,2,1,1),\,\, S(4,2,1,3)\}$}

\vskipa
\class{$(1.4.2.7)$}
\hbc{$\phm\hat{\onefourth}$},
\hbc{$\phm\hat{\threefourth}$}
\hdp{\phm\half}{1}

\no\hskip 2.2truecm{$\{S(4,1,1,2),\,\, S(4,1,1,4)\}$}

\vskipa
\class{$(1.4.2.8)$}
\hbc{$-\hat{\fourth}$},
\hbc{$-\hat{\threefourth}$}
\hdp{-\half}{1}

\no\hskip 2.2truecm{$\{S(4,3,1,1),\,\, S(4,3,1,3)\}$}

\tailer

\newpage

%-------------------------------
\headings

\fch{4}{8}\hbs{$S(4,1,2,1)$}\hbv{$(\phm\hat{\half},0,1; -\hat{\threefourth},
                                       \phm\hat{\fourth},
                                          -\hat{\fourth},
                                       \phm\hat{\threefourth})$}

\fchn\hbs{$S(4,1,2,2)$}\hbv{$(-\hat{\half},1,0; \phm\hat{\fourth},
                                          -\hat{\threefourth},
                                          -\hat{\fourth},
                                       \phm\hat{\threefourth})$}

\fchn\hbs{$S(4,1,2,3)$}\hbv{$(-\hat{\half},1,0; -\hat{\threefourth},
                                       \phm\hat{\fourth},
                                       \phm\hat{\threefourth},-\hat{\fourth})$}

\fchn\hbs{$S(4,1,2,4)$}\hbv{$(\phm\hat{\half},0,1; \phm\hat{\fourth},
                                             -\hat{\threefourth},
                                          \phm\hat{\threefourth},
                                             -\hat{\fourth})$}

\fchn\hbs{$S(4,3,2,1)$}\hbv{$(-\hat{\half},0,1;
         -\hat{\fourth},
         \phm\hat{\threefourth},
         -\hat{\threefourth},
          \phm\hat{\fourth})$}

\fchn\hbs{$S(4,3,2,2)$}\hbv{$(\phm\hat{\half},1,0; \phm\hat{\threefourth},
                            -\hat{\fourth},
                            -\hat{\threefourth},
                            \phm\hat{\fourth})$}

\fchn\hbs{$S(4,3,2,3)$}\hbv{$(\phm\hat{\half},1,0; -\hat{\fourth},
                            \phm\hat{\threefourth},
                            \phm\hat{\fourth},
                            -\hat{\threefourth})$}

\fchn\hbs{$S(4,3,2,4)$}\hbv{$(-\hat{\half},0,1; \phm\hat{\threefourth},
                            -\hat{\fourth},
                         \phm\hat{\fourth},
                         -\hat{\threefourth})$}

\vskipa
\fch{$2\cdot2$}{4}\hbs{$S(4,2,2,1)$}\hbv{$(\hj\phm\hat{0},1,1; -\hat{\half},
                           -\hat{\half},
                           \phm\hat{\half},
                           \phm\hat{\half})$}

\fchn\hbs{$S(4,2,2,2)$}\hbv{$(\hj\phm\hat{1},0,0; \phm\hat{\half},
                           \phm\hat{\half},
                           \phm\hat{\half},\phm\hat{\half})$}

\fchn\hbs{$S(4,2,2,3)$}\hbv{$(\hj\phm\hat{1},0,0; -\hat{\half},-\hat{\half},
                           -\hat{\half},-\hat{\half})$}

\fchn\hbs{$S(4,2,2,4)$}\hbv{$(\hj\phm\hat{0},1,1;
                                \phm\hat{\half},\phm\hat{\half},
                               -\hat{\half},-\hat{\half})$}

\vskipa
\fch{$1\cdot1\cdot1\cdot1$}{2}\hbs{$S(4,4,2,1)$}\hbv{$(\hj\phm\hat{1},1,1;
                                          \hj\phm\hat{0},\hj\phm\hat{0},
                                          \hj\phm\hat{0},\hj\phm\hat{0})$}

\fchn\hbs{$S(4,4,2,2)$}\hbv{$(\hj\phm\hat{0},0,0;
                                          \hj\phm\hat{1},\hj\phm\hat{1},
                                          \hj\phm\hat{0},\hj\phm\hat{0})$}

\fchn\hbs{$S(4,4,2,3)$}\hbv{$(\hj\phm\hat{0},0,0;
                                          \hj\phm\hat{0},\hj\phm\hat{0},
                                          \hj\phm\hat{1},\hj\phm\hat{1})$}

\fchn\hbs{$S(4,4,2,4)$}\hbv{$(\hj\phm\hat{1},1,1;
                                          \hj\phm\hat{1},\hj\phm\hat{1},
                                          \hj\phm\hat{1},\hj\phm\hat{1})$}

\tailer

\newpage

%***********
\headings

\fcha{5}{10}\hbs{$S(5,1,1,1)$}\hbv{$(\phm\hat{\fifth},\phm\hat{\threefifth},1;
                     -\hat{\fourfifth},-\hat{\twofifth}, 0,0,
                     \phm\hat{\twofifth}, \phm\hat{\fourfifth})$}

\fchn\hbs{$S(5,1,1,2)$}\hbv{$(-\hat{\fourfifth},-\hat{\twofifth},0;
                     \phm\hat{\onefifth},\phm\hat{\threefifth}, 1,0,
                     \phm\hat{\twofifth}, \phm\hat{\fourfifth})$}

\fchn\hbs{$S(5,1,1,3)$}\hbv{$(-\hat{\fourfifth},-\hat{\twofifth},0;
                     -\hat{\fourfifth},-\hat{\twofifth},0 ,1,
                     -\hat{\threefifth},-\hat{\onefifth})$}

\fchn\hbs{$S(5,1,1,4)$}\hbv{$(\phm\hat{\onefifth},\phm\hat{\threefifth},1;
                     \phm\hat{\onefifth},\phm\hat{\threefifth},1,1,
                     -\hat{\threefifth},\phm\hat{\onefifth})$}

\fchn\hbs{$S(5,2,1,1)$}\hbv{$(-\hat{\threefifth},\phm\hat{\onefifth},1;
                     -\hat{\threefifth},\phm\hat{\onefifth},1,1,
                     -\hat{\onefifth},\phm\hat{\threefifth})$}

\fchn\hbs{$S(5,2,1,2)$}\hbv{$(\phm\hat{\twofifth},-\hat{\fourfifth},0;
                     \phm\hat{\twofifth},-\hat{\fourfifth},0,1,
                     -\hat{\onefifth},\phm\hat{\threefifth})$}

\fchn\hbs{$S(5,2,1,3)$}\hbv{$(\phm\hat{\twofifth},-\hat{\fourfifth},0;
                     -\hat{\threefifth},\phm\hat{\onefifth},1,0,
                     \phm\hat{\fourfifth},-\hat{\twofifth})$}

\fchn\hbs{$S(5,2,1,4)$}\hbv{$(-\hat{\threefifth},\phm\hat{\onefifth},1;
                     \phm\hat{\twofifth},-\hat{\fourfifth},0,0,
                     \phm\hat{\fourfifth},-\hat{\twofifth})$}

\fchn\hbs{$S(5,3,1,1)$}\hbv{$(\phm\hat{\threefifth},-\hat{\onefifth},1;
                     -\hat{\twofifth},\phm\hat{\fourfifth},0,0,
                     -\hat{\fourfifth},\phm\hat{\twofifth})$}

\fchn\hbs{$S(5,3,1,2)$}\hbv{$(-\hat{\twofifth},\phm\hat{\fourfifth},0;
                     \phm\hat{\threefifth},-\hat{\onefifth},1,0,
                     -\hat{\fourfifth},\phm\hat{\twofifth})$}

\fchn\hbs{$S(5,3,1,3)$}\hbv{$(-\hat{\twofifth},\phm\hat{\fourfifth},0;
                     -\hat{\twofifth},\phm\hat{\fourfifth},0,1,
                     \phm\hat{\onefifth},-\hat{\threefifth})$}

\fchn\hbs{$S(5,3,1,4)$}\hbv{$(\phm\hat{\threefifth},-\hat{\onefifth},1;
                     \phm\hat{\threefifth},-\hat{\onefifth},1,1,
                     \phm\hat{\onefifth},-\hat{\threefifth})$}

\fchn\hbs{$S(5,4,1,1)$}\hbv{$(-\hat{\onefifth},-\hat{\threefifth},1;
                     -\hat{\onefifth},-\hat{\threefifth},1,1,
                     \phm\hat{\threefifth},\phm\hat{\onefifth})$}

\fchn\hbs{$S(5,4,1,2)$}\hbv{$(\phm\hat{\fourfifth},\phm\hat{\twofifth},0;
                     \phm\hat{\fourfifth},\phm\hat{\twofifth},0,1,
                     \phm\hat{\threefifth},\phm\hat{\onefifth})$}

\fchn\hbs{$S(5,4,1,3)$}\hbv{$(\phm\hat{\fourfifth},\phm\hat{\twofifth},0;
                     -\hat{\onefifth},-\hat{\threefifth},1,0,
                     -\hat{\twofifth},-\hat{\fourfifth})$}

\fchn\hbs{$S(5,4,1,4)$}\hbv{$(-\hat{\onefifth},-\hat{\threefifth},1;
                     \phm\hat{\fourfifth},\phm\hat{\twofifth},0,0,
                     -\hat{\twofifth},-\hat{\fourfifth})$}

\vskipa
\fcha{$1\cdot1\cdot1\cdot1\cdot1$}{2}\hbs{$S(5,5,1,1)$}\hbv{$(\hj\phm\hat{1},
                     \hj\phm\hat{1},1;
                     \hj\phm\hat{0},\hj\phm\hat{0},0,0,
                     \hj\phm\hat{0},\hj\phm\hat{0})$}

\fchn\hbs{$S(5,5,1,2)$}\hbv{$(\hj\phm\hat{0},\hj\phm\hat{0},0;
                     \hj\phm\hat{1},\hj\phm\hat{1},1,0,
                     \hj\phm\hat{0},\hj\phm\hat{0})$}

\fchn\hbs{$S(5,5,1,3)$}\hbv{$(\hj\phm\hat{0},\hj\phm\hat{0},0;
                     \hj\phm\hat{0},\hj\phm\hat{0},0,1,
                     \hj\phm\hat{1},\hj\phm\hat{1})$}

\fchn\hbs{$S(5,5,1,4)$}\hbv{$(\hj\phm\hat{0},\hj\phm\hat{0},0;
                     \hj\phm\hat{0},\hj\phm\hat{0},0,1,
                     \hj\phm\hat{1},\hj\phm\hat{1})$}

\vskipa
\conto{$S(5,1,1,1)$}{$\bS_{10}$}
\class{$(5.5.1.1)$}
\hbc{$\phm1$}
\hdp{-\threefourth}{\onefifth}

\no\hskip 2.0truecm{$\{S(5,1,1,1),\,\, S(5,1,1,4),\,\,
                       S(5,2,1,1),\,\, S(5,2,1,4),\,\,
                       S(5,3,1,1),\,\, S(5,3,1,4),$}

\no\hskip 2.0truecm{$\phantom{\{}  S(5,4,1,1),\,\, S(5,4,1,4),\,\,
                       S(5,5,1,1),\,\, S(5,5,1,4)\}$}

\class{$(5.5.1.2)$}
\hbc{$\phm0$}
\hdp{\phm0}{\onefifth}

\no\hskip 2.0truecm{$\{S(5,1,1,2),\,\, S(5,1,1,3),\,\,
                       S(5,2,1,2),\,\, S(5,2,1,3),\,\,
                       S(5,3,1,2),\,\, S(5,3,1,3),$}

\no\hskip 2.0truecm{$\phantom{\{} S(5,4,1,2),\,\, S(5,4,1,3),\,\,
                       S(5,5,1,2),\,\, S(5,5,1,3)\}$}

\vskipa
\conto{$S(5,3,1,1)$}{$3\bS_{10}$}

\no Same $\bF$ coef and $\half$dp Classes as $3\bS_{10}$ above.

\vskipa
\conto{$S(5,5,1,1)$}{$\bS_1$}
\class{$(1.5.1.1)$}
\hbc{$\phm\hat{1}$},
\hbc{$\phm\hat{1}$},
\hbc{$\phm1$}
\hdp{-1\onefourth}{1}

\no\hskip 2.0truecm{$\{S(5,5,1,1),\,\, S(5,5,1,4)\}$}

\class{$(1.5.1.2)$}
\hbc{$\phm\hat{0}$},
\hbc{$\phm\hat{0}$},
\hbc{$\phm0$}
\hdp{\hskip0.08truecm\phm0}{1}

\no\hskip 2.0truecm{$\{S(5,5,1,2),\,\, S(5,5,1,3)\}$}

\class{$(1.5.1.3)$}
\hbc{$\phm\hat{\threefifth}$},
\hbc{$-\hat{\onefifth}$},
\hbc{$\phm1$}
\hdp{\phm\fract{9}{20}}{1}

\no\hskip 2.0truecm{$\{S(5,3,1,1),\,\, S(5,3,1,4)\}$}

\class{$(1.5.1.4)$}
\hbc{$-\hat{\threefifth}$},
\hbc{$\phm\hat{\onefifth}$},
\hbc{$\phm1$}
\hdp{\phm\fract{1}{20}}{1}

\no\hskip 2.0truecm{$\{S(5,2,1,1),\,\, S(5,2,1,4)\}$}

\class{$(1.5.1.5)$}
\hbc{$\phm\hat{\onefifth}$},
\hbc{$\phm\hat{\threefifth}$},
\hbc{$\phm1$}
\hdp{\phm\fract{13}{20}}{1}

\no\hskip 2.0truecm{$\{S(5,1,1,1),\,\, S(5,1,1,4)\}$}

\class{$(1.5.1.6)$}
\hbc{$-\hat{\onefifth}$},
\hbc{$-\hat{\threefifth}$},
\hbc{$\phm1$}
\hdp{-\fract{3}{20}}{1}

\no\hskip 2.0truecm{$\{S(5,4,1,1),\,\, S(5,4,1,4)\}$}

\class{$(1.5.1.7)$}
\hbc{$\phm\hat{\fourfifth}$},
\hbc{$\phm\hat{\twofifth}$},
\hbc{$\phm0$}
\hdp{\phm\threefifth}{1}

\no\hskip 2.0truecm{$\{S(5,4,1,2),\,\, S(5,4,1,3)\}$}

\class{$(1.5.1.8)$}
\hbc{$-\hat{\fourfifth}$},
\hbc{$-\hat{\twofifth}$},
\hbc{$\phm0$}
\hdp{-\threefifth}{1}

\no\hskip 2.0truecm{$\{S(5,1,1,2),\,\, S(5,1,1,3)\}$}

\class{$(1.5.1.9)$}
\hbc{$\phm\hat{\twofifth}$},
\hbc{$-\hat{\fourfifth}$},
\hbc{$\phm0$}
\hdp{-\onefifth}{1}

\no\hskip 2.0truecm{$\{S(5,2,1,2),\,\, S(5,2,1,3)\}$}

\class{$(1.5.1.10)$}
\hbc{$-\hat{\twofifth}$},
\hbc{$\phm\hat{\fourfifth}$},
\hbc{$\phm0$}
\hdp{\phm\onefifth}{1}

\no\hskip 2.0truecm{$\{S(5,3,1,2),\,\, S(5,3,1,3)\}$}

\tailer

\newpage

%------------
\headings

\fch{5}{10}\hbs{$S(5,1,2,1)$}\hbv{$(\phm\hat{\fifth},\phm\hat{\threefifth},1;
                     -\hat{\fourfifth},\phm\hat{\twofifth}, 0,0,
                     -\hat{\twofifth}, \phm\hat{\fourfifth})$}

\fchn\hbs{$S(5,1,2,2)$}\hbv{$(-\hat{\fourfifth},-\hat{\twofifth},0;
                     \phm\hat{\onefifth},-\hat{\threefifth}, 1,0,
                     -\hat{\twofifth}, \phm\hat{\fourfifth})$}

\fchn\hbs{$S(5,1,2,3)$}\hbv{$(-\hat{\fourfifth},-\hat{\twofifth},0;
                     -\hat{\fourfifth},\phm\hat{\twofifth},0 ,1,
                     \phm\hat{\threefifth},-\hat{\onefifth})$}

\fchn\hbs{$S(5,1,2,4)$}\hbv{$(\phm\hat{\onefifth},\phm\hat{\threefifth},1;
                     \phm\hat{\onefifth},-\hat{\threefifth},1,1,
                     \phm\hat{\threefifth},\phm\hat{\onefifth})$}

\fchn\hbs{$S(5,2,2,1)$}\hbv{$(-\hat{\threefifth},\phm\hat{\onefifth},1;
                     -\hat{\threefifth},-\hat{\onefifth},1,1,
                     \phm\hat{\onefifth},\phm\hat{\threefifth})$}

\fchn\hbs{$S(5,2,2,2)$}\hbv{$(\phm\hat{\twofifth},-\hat{\fourfifth},0;
                     \phm\hat{\twofifth},\phm\hat{\fourfifth},0,1,
                     \phm\hat{\onefifth},\phm\hat{\threefifth})$}

\fchn\hbs{$S(5,2,2,3)$}\hbv{$(\phm\hat{\twofifth},-\hat{\fourfifth},0;
                     -\hat{\threefifth},-\hat{\onefifth},1,0,
                     -\hat{\fourfifth},-\hat{\twofifth})$}

\fchn\hbs{$S(5,2,2,4)$}\hbv{$(-\hat{\threefifth},\phm\hat{\onefifth},1;
                     \phm\hat{\twofifth},\phm\hat{\fourfifth},0,0,
                     -\hat{\fourfifth},-\hat{\twofifth})$}

\fchn\hbs{$S(5,3,2,1)$}\hbv{$(\phm\hat{\threefifth},-\hat{\onefifth},1;
                     -\hat{\twofifth},-\hat{\fourfifth},0,0,
                     \phm\hat{\fourfifth},\phm\hat{\twofifth})$}

\fchn\hbs{$S(5,3,2,2)$}\hbv{$(-\hat{\twofifth},\phm\hat{\fourfifth},0;
                     \phm\hat{\threefifth},\phm\hat{\onefifth},1,0,
                     \phm\hat{\fourfifth},\phm\hat{\twofifth})$}

\fchn\hbs{$S(5,3,2,3)$}\hbv{$(-\hat{\twofifth},\phm\hat{\fourfifth},0;
                     -\hat{\twofifth},-\hat{\fourfifth},0,0,
                     -\hat{\onefifth},-\hat{\threefifth})$}

\fchn\hbs{$S(5,3,2,4)$}\hbv{$(\phm\hat{\threefifth},-\hat{\onefifth},1;
                     \phm\hat{\threefifth},\phm\hat{\onefifth},1,1,
                     -\hat{\onefifth},-\hat{\threefifth})$}

\fchn\hbs{$S(5,4,2,1)$}\hbv{$(-\hat{\onefifth},-\hat{\threefifth},1;
                     -\hat{\onefifth},\phm\hat{\threefifth},1,1,
                     -\hat{\threefifth},\phm\hat{\onefifth})$}

\fchn\hbs{$S(5,4,2,2)$}\hbv{$(\phm\hat{\fourfifth},\phm\hat{\twofifth},0;
                     \phm\hat{\fourfifth},-\hat{\twofifth},0,1,
                     -\hat{\threefifth},\phm\hat{\onefifth})$}

\fchn\hbs{$S(5,4,2,3)$}\hbv{$(\phm\hat{\fourfifth},\phm\hat{\twofifth},0;
                     -\hat{\onefifth},\phm\hat{\threefifth},1,0,
                     \phm\hat{\twofifth},-\hat{\fourfifth})$}

\fchn\hbs{$S(5,4,2,4)$}\hbv{$(-\hat{\onefifth},-\hat{\threefifth},1;
                     \phm\hat{\fourfifth},-\hat{\twofifth},0,0,
                     \phm\hat{\twofifth},-\hat{\fourfifth})$}

\vskipa
\fch{$1\cdot1\cdot1\cdot1\cdot1$}{2}\hbs{$S(5,5,2,1)$}\hbv{$(\hj\phm\hat{1},
                     \hj\phm\hat{1},1;
                     \hj\phm\hat{0},\hj\phm\hat{0},0,0,
                     \hj\phm\hat{0},\hj\phm\hat{0})$}

\fchn\hbs{$S(5,5,2,2)$}\hbv{$(\hj\phm\hat{0},\hj\phm\hat{0},0;
                     \hj\phm\hat{1},\hj\phm\hat{1},1,0,
                     \hj\phm\hat{0},\hj\phm\hat{0})$}

\fchn\hbs{$S(5,5,2,3)$}\hbv{$(\hj\phm\hat{0},\hj\phm\hat{0},0;
                     \hj\phm\hat{0},\hj\phm\hat{0},0,1,
                     \hj\phm\hat{1},\hj\phm\hat{1})$}

\fchn\hbs{$S(5,5,2,4)$}\hbv{$(\hj\phm\hat{0},\hj\phm\hat{0},0;
                     \hj\phm\hat{0},\hj\phm\hat{0},0,1,
                     \hj\phm\hat{1},\hj\phm\hat{1})$}

\tailer

\newpage

%*************
\headings

\fch{6}{12}\hbs{$S(6,1,1,1)$}\hbv{$(\phm\hat{\onethird},
                     \phm\hat{\twothird},0,1;
                     -\hat{\fivesixth},-\hat{\half},-\hat{\onesixth},
                     \phm\hat{\onesixth},
                     \phm\hat{\half},
                     \phm\hat{\fivesixth}
)$}

\fchn\hbs{$S(6,1,1,2)$}\hbv{$(-\hat{\twothird},-\hat{\onethird},1,0;
                     \phm\hat{\onesixth},
                     \phm\hat{\half},
                     \phm\hat{\fivesixth},
                     \phm\hat{\onesixth},
                     \phm\hat{\half},
                     \phm\hat{\fivesixth}
)$}

\fchn\hbs{$S(6,1,1,3)$}\hbv{$(-\hat{\twothird},-\hat{\onethird},1,0;
                     -\hat{\fivesixth},
                     -\hat{\half},
                     -\hat{\onesixth},
                     -\hat{\fivesixth},
                     -\hat{\half},
                     -\hat{\onesixth}
)$}

\fchn\hbs{$S(6,1,1,4)$}\hbv{$(\phm\hat{\onethird},\phm\hat{\twothird},0,1;
                     \phm\hat{\onesixth},
                     \phm\hat{\half},
                     \phm\hat{\fivesixth},
                     -\hat{\fivesixth},
                     -\hat{\half},
                     -\hat{\onesixth}
)$}

\fchn\hbs{$S(6,5,1,1)$}\hbv{$(-\hat{\onethird},-\hat{\twothird},0,1;
                     -\hat{\onesixth},
                     -\hat{\half},
                     -\hat{\fivesixth},
                     \phm\hat{\fivesixth},
                     \phm\hat{\half},
                     \phm\hat{\onesixth}
)$}

\fchn\hbs{$S(6,5,1,2)$}\hbv{$(\phm\hat{\twothird},\phm\hat{\onethird},1,0;
                      \phm\hat{\fivesixth},
                      \phm\hat{\half},
                      \phm\hat{\onesixth},
                      \phm\hat{\fivesixth},
                      \phm\hat{\half},
                      \phm\hat{\onesixth}
)$}

\fchn\hbs{$S(6,5,1,3)$}\hbv{$(\phm\hat{\twothird},\phm\hat{\onethird},1,0;
                     -\hat{\onesixth},
                     -\hat{\half},
                     -\hat{\fivesixth},
                     -\hat{\onesixth},
                     -\hat{\half},
                     -\hat{\fivesixth}
)$}

\fchn\hbs{$S(6,5,1,4)$}\hbv{$(-\hat{\onethird},-\hat{\twothird},0,1;
                     \phm\hat{\fivesixth},
                     \phm\hat{\half},
                     \phm\hat{\onesixth},
                     -\hat{\onesixth},
                     -\hat{\half},
                     -\hat{\fivesixth}
)$}

\vskipa\fch{\sseven}{6}\hbs{$S(6,2,1,1)$}\hbv{$(-\hat{\onethird},
                     \phm\hat{\onethird},1,1;
                     -\hat{\twothird},
                     \hj\phm\hat{0},
                     \phm\hat{\twothird},
                     -\hat{\twothird},
                     \hj\phm\hat{0},
                     \phm\hat{\twothird}
)$}

\fchna\hbs{$S(6,2,1,2)$}\hbv{$(\phm\hat{\twothird},-\hat{\twothird},0,0;
                     \phm\hat{\onethird},
                     \hj\phm\hat{1},
                     -\hat{\onethird},
                     -\hat{\twothird},
                     \hj\phm\hat{0},
                     \phm\hat{\twothird}
)$}

\fchn\hbs{$S(6,2,1,3)$}\hbv{$(\phm\hat{\twothird},-\hat{\twothird},0,0;
                     -\hat{\twothird},
                     \hj\phm\hat{0},
                     \phm\hat{\twothird},
                     \phm\hat{\onethird},
                     \hj\phm\hat{1},
                     -\hat{\onethird}
)$}

\fchn\hbs{$S(6,2,1,4)$}\hbv{$(-\hat{\onethird},\phm\hat{\onethird},1,1;
                     \phm\hat{\onethird},
                     \hj\phm\hat{1},
                    -\hat{\onethird},
                     \phm\hat{\onethird},
                     \hj\phm\hat{1},
                     -\hat{\onethird}
)$}

\fchn\hbs{$S(6,4,1,1)$}\hbv{$(\phm\hat{\onethird},-\hat{\onethird},1,1;
                     -\hat{\onethird},
                     \hj\phm\hat{1},
                     \phm\hat{\onethird},
                     -\hat{\onethird},
                     \hj\phm\hat{1},
                     \phm\hat{\onethird}
)$}

\fchn\hbs{$S(6,4,1,2)$}\hbv{$(-\hat{\twothird},\phm\hat{\twothird},0,0;
                     \phm\hat{\twothird},
                     \hj\phm\hat{0},
                        -\hat{\twothird},
                     -\hat{\onethird},
                     \hj\phm\hat{1},
                     \phm\hat{\onethird}
)$}

\fchn\hbs{$S(6,4,1,3)$}\hbv{$(-\hat{\twothird},\phm\hat{\twothird},0,0;
                        -\hat{\onethird},
                     \hj\phm\hat{1},
                     \phm\hat{\onethird},
                     \phm\hat{\twothird},
                     \hj\phm\hat{0},
                        -\hat{\twothird}
)$}

\fchn\hbs{$S(6,4,1,4)$}\hbv{$(\phm\hat{\onethird},-\hat{\onethird},1,1;
                     \phm\hat{\twothird},
                     \hj\phm\hat{0},
                    -\hat{\twothird},
                     \phm\hat{\twothird},
                     \hj\phm\hat{0},
                    -\hat{\twothird}
)$}

\vskipa\fch{$2\cdot2\cdot2$}{4}\hbs{$S(6,3,1,1)$}\hbv{$(\hj\phm\hat{1},
                     \hj\phm\hat{0},0,1;
                        -\hat{\onehalf},
                     \phm\hat{\onehalf},
                        -\hat{\onehalf},
                     \phm\hat{\onehalf},
                        -\hat{\onehalf},
                     \phm\hat{\onehalf}
)$}

\fchn\hbs{$S(6,3,1,2)$}\hbv{$(\hj\phm\hat{0},\hj\phm\hat{1},1,0;
                     \phm\hat{\onehalf},
                        -\hat{\onehalf},
                     \phm\hat{\onehalf},
                     \phm\hat{\onehalf},
                        -\hat{\onehalf},
                     \phm\hat{\onehalf}
)$}

\fchn\hbs{$S(6,3,1,3)$}\hbv{$(\hj\phm\hat{0},\hj\phm\hat{1},1,0;
                        -\hat{\onehalf},
                     \phm\hat{\onehalf},
                        -\hat{\onehalf},
                        -\hat{\onehalf},
                     \phm\hat{\onehalf},
                        -\hat{\onehalf}
)$}

\fchn\hbs{$S(6,3,1,4)$}\hbv{$(\hj\phm\hat{1},\hj\phm\hat{0},0,1;
                     \phm\hat{\onehalf},
                        -\hat{\onehalf},
                     \phm\hat{\onehalf},
                        -\hat{\onehalf},
                     \phm\hat{\onehalf},
                        -\hat{\onehalf}
)$}

\vskipa
\fch{\sone}{2}\hbs{$S(6,6,1,1)$}\hbv{$(\hj\phm\hat{1},\hj\phm\hat{1},1,1;
                     \hj\phm\hat{0},
                     \hj\phm\hat{0},
                     \hj\phm\hat{0},
                     \hj\phm\hat{0},
                     \hj\phm\hat{0},
                     \hj\phm\hat{0}
)$}

\fchna\hbs{$S(6,6,1,2)$}\hbv{$(\hj\phm\hat{0},\hj\phm\hat{0},0,0;
                     \hj\phm\hat{1},
                     \hj\phm\hat{1},
                     \hj\phm\hat{1},
                     \hj\phm\hat{0},
                     \hj\phm\hat{0},
                     \hj\phm\hat{0}
)$}

\fchn\hbs{$S(6,6,1,3)$}\hbv{$(\hj\phm\hat{0},\hj\phm\hat{0},0,0;
                     \hj\phm\hat{0},
                     \hj\phm\hat{0},
                     \hj\phm\hat{0},
                     \hj\phm\hat{1},
                     \hj\phm\hat{1},
                     \hj\phm\hat{1}
)$}

\fchn\hbs{$S(6,6,1,4)$}\hbv{$(\hj\phm\hat{1},\hj\phm\hat{1},1,1;
                     \hj\phm\hat{1},
                     \hj\phm\hat{1},
                     \hj\phm\hat{1},
                     \hj\phm\hat{1},
                     \hj\phm\hat{1},
                     \hj\phm\hat{1}
)$}

\vskipa
\conto{$S(6,2,1,2)$}{$\bS_7$}
\class{$(3.6.2.1)$}
\hbc{$\phm\hat{1}$}
\hdp{-\half}{\third}

\no\hskip 2.2truecm{$\{S(6,2,1,2),\,\, S(6,2,1,4),\,\,
                       S(6,4,1,1),\,\, S(6,4,1,3),\,\,
                       S(6,6,1,2),\,\, S(6,6,1,4)\}$}

\vskipa
\class{$(3.6.2.2)$}
\hbc{$\phm\hat{0}$}
\hdp{\phm0}{\third}

\no\hskip 2.2truecm{$\{S(6,2,1,1),\,\, S(6,2,1,3),\,\,
                       S(6,4,1,2),\,\, S(6,4,1,4),\,\,
                       S(6,6,1,1),\,\, S(6,6,1,3)\}$}

\vskipa
\class{$(3.6.2.3)$}
\hbc{$\phm\hat{\half}$}
\hdp{\phm\onefourth}{\third}

\no\hskip 2.2truecm{$\{S(6,1,1,1),\,\, S(6,1,1,3),\,\,
                       S(6,3,1,2),\,\, S(6,3,1,4),\,\,
                       S(6,5,1,1),\,\, S(6,5,1,3)\}$}

\vskipa
\class{$(3.6.2.4)$}
\hbc{$-\hat{\half}$}
\hdp{-\onefourth}{\third}

\no\hskip 2.2truecm{$\{S(6,1,1,2),\,\, S(6,1,1,4),\,\,
                       S(6,3,1,1),\,\, S(6,3,1,3),\,\,
                       S(6,5,1,2),\,\, S(6,6,1,4)\}$}

\vskipa
\conto{$S(6,6,1,2)$}{$\bS_1$}
\class{$(1.6.2.1)$}
\hbc{$\phm\hat{1}$},
\hbc{$\phm\hat{1}$},
\hbc{$\phm\hat{1}$}
\hdp{-\half}{1}

\no\hskip 2.2truecm{$\{S(6,6,1,2),\,\, S(6,6,1,4)\}$}

\vskipa
\class{$(1.6.2.2)$}
\hbc{$\phm\hat{0}$},
\hbc{$\phm\hat{0}$},
\hbc{$\phm\hat{0}$}
\hdp{\phm0}{1}

\no\hskip 2.2truecm{$\{S(6,6,1,1),\,\, S(6,5,1,3)\}$}

\vskipa
\class{$(1.6.2.3)$}
\hbc{$\phm\hat{\third}$},
\hbc{$\phm\hat{1}$},
\hbc{$-\hat{\third}$}
\hdp{\phm\half}{1}

\no\hskip 2.2truecm{$\{S(6,2,1,2),\,\, S(6,2,1,4)\}$}

\vskipa
\class{$(1.6.2.4)$}
\hbc{$-\hat{\third}$},
\hbc{$\phm\hat{1}$},
\hbc{$\phm\hat{\third}$}
\hdp{\phm\half}{1}

\no\hskip 2.2truecm{$\{S(6,4,1,1),\,\, S(6,4,1,4)\}$}

\vskipa
\class{$(1.6.2.5)$}
\hbc{$\phm\hat{\twothird}$},
\hbc{$\phm\hat{0}$},
\hbc{$-\hat{\twothird}$}
\hdp{\phm0}{1}

\no\hskip 2.2truecm{$\{S(6,4,1,2),\,\, S(6,4,1,4)\}$}

\vskipa
\class{$(1.6.2.6)$}
\hbc{$-\hat{\twothird}$},
\hbc{$\phm\hat{0}$},
\hbc{$\phm\hat{\twothird}$}
\hdp{\phm0}{1}

\no\hskip 2.2truecm{$\{S(6,2,1,1),\,\, S(6,2,1,3)\}$}

\vskipa
\class{$(1.6.2.7)$}
\hbc{$\phm\hat{\fivesixth}$},
\hbc{$\phm\hat{\half}$},
\hbc{$\phm\hat{\onesixth}$}
\hdp{\phm\threefourth}{1}

\no\hskip 2.2truecm{$\{S(6,5,1,2),\,\, S(6,5,1,4)\}$}

\vskipa
\class{$(1.6.2.8)$}
\hbc{$-\hat{\fivesixth}$},
\hbc{$-\hat{\half}$},
\hbc{$-\hat{\onesixth}$}
\hdp{-\threefourth}{1}

\no\hskip 2.2truecm{$\{S(6,1,1,1),\,\, S(6,1,1,3)\}$}

\vskipa
\class{$(1.6.2.9)$}
\hbc{$\phm\hat{\half}$},
\hbc{$-\hat{\half}$},
\hbc{$\phm\hat{\half}$}
\hdp{\phm\fourth}{1}

\no\hskip 2.2truecm{$\{S(6,3,1,2),\,\, S(6,3,1,4)\}$}

\vskipa
\class{$(1.6.2.10)$}
\hbc{$-\hat{\half}$},
\hbc{$\phm\hat{\half}$},
\hbc{$-\hat{\half}$}
\hdp{-\fourth}{1}

\no\hskip 2.2truecm{$\{S(6,3,1,1),\,\, S(6,3,1,4)\}$}

\vskipa
\class{$(1.6.2.11)$}
\hbc{$\phm\hat{\onesixth}$},
\hbc{$\phm\hat{\half}$},
\hbc{$\phm\hat{\fivesixth}$}
\hdp{\phm\threefourth}{1}

\no\hskip 2.2truecm{$\{S(6,1,1,2),\,\, S(6,1,1,4)\}$}

\vskipa
\class{$(1.6.2.12)$}
\hbc{$-\hat{\onesixth}$},
\hbc{$-\hat{\half}$},
\hbc{$-\hat{\fivesixth}$}
\hdp{-\threefourth}{1}

\no\hskip 2.2truecm{$\{S(6,5,1,1),\,\, S(6,5,1,3)\}$}

\tailer

\newpage

%*******************
\headings

\fch{6}{12}\hbs{$S(6,1,2,1)$}\hbv{$(\phm\hat{\onethird},
                     \phm\hat{\twothird},0,1;
                     -\hat{\fivesixth},\phm\hat{\half},-\hat{\onesixth},
                     \phm\hat{\onesixth},
                     -\hat{\half},
                     \phm\hat{\fivesixth}
)$}

\fchn\hbs{$S(6,1,2,2)$}\hbv{$(-\hat{\twothird},-\hat{\onethird},1,0;
                     \phm\hat{\onesixth},
                     -\hat{\half},
                     \phm\hat{\fivesixth},
                     \phm\hat{\onesixth},
                     -\hat{\half},
                     \phm\hat{\fivesixth}
)$}

\fchn\hbs{$S(6,1,2,3)$}\hbv{$(-\hat{\twothird},-\hat{\onethird},1,0;
                     -\hat{\fivesixth},
                     \phm\hat{\half},
                     -\hat{\onesixth},
                     -\hat{\fivesixth},
                     \phm\hat{\half},
                     -\hat{\onesixth}
)$}

\fchn\hbs{$S(6,1,2,4)$}\hbv{$(\phm\hat{\onethird},\phm\hat{\twothird},0,1;
                     \phm\hat{\onesixth},
                     -\hat{\half},
                     \phm\hat{\fivesixth},
                     -\hat{\fivesixth},
                     \phm\hat{\half},
                     -\hat{\onesixth}
)$}

\fchn\hbs{$S(6,5,2,1)$}\hbv{$(-\hat{\onethird},-\hat{\twothird},0,1;
                     -\hat{\onesixth},
                     \phm\hat{\half},
                     -\hat{\fivesixth},
                     \phm\hat{\fivesixth},
                     -\hat{\half},
                     \phm\hat{\onesixth}
)$}

\fchn\hbs{$S(6,5,2,2)$}\hbv{$(\phm\hat{\twothird},\phm\hat{\onethird},1,0;
                      \phm\hat{\fivesixth},
                      -\hat{\half},
                      \phm\hat{\onesixth},
                      \phm\hat{\fivesixth},
                      -\hat{\half},
                      \phm\hat{\onesixth}
)$}

\fchn\hbs{$S(6,5,2,3)$}\hbv{$(\phm\hat{\twothird},\phm\hat{\onethird},1,0;
                     -\hat{\onesixth},
                     \phm\hat{\half},
                     -\hat{\fivesixth},
                     -\hat{\onesixth},
                     \phm\hat{\half},
                     -\hat{\fivesixth}
)$}

\fchn\hbs{$S(6,5,2,4)$}\hbv{$(-\hat{\onethird},-\hat{\twothird},0,1;
                     \phm\hat{\fivesixth},
                     -\hat{\half},
                     \phm\hat{\onesixth},
                     -\hat{\onesixth},
                     \phm\hat{\half},
                     -\hat{\fivesixth}
)$}

\vskipa
\fch{\sseven}{6}\hbs{$S(6,2,2,1)$}\hbv{$(-\hat{\onethird},
                     \phm\hat{\onethird},1,1;
                     -\hat{\twothird},
                     \hj\phm\hat{0},
                     \phm\hat{\twothird},
                     -\hat{\twothird},
                     \hj\phm\hat{0},
                     \phm\hat{\twothird}
)$}

\fchn\hbs{$S(6,2,2,2)$}\hbv{$(\phm\hat{\twothird},-\hat{\twothird},0,0;
                     \phm\hat{\onethird},
                     \hj\phm\hat{1},
                     -\hat{\onethird},
                     -\hat{\twothird},
                     \hj\phm\hat{0},
                     \phm\hat{\twothird}
)$}

\fchn\hbs{$S(6,2,2,3)$}\hbv{$(\phm\hat{\twothird},-\hat{\twothird},0,0;
                     -\hat{\twothird},
                     \hj\phm\hat{0},
                     \phm\hat{\twothird},
                     \phm\hat{\onethird},
                     \hj\phm\hat{1},
                     -\hat{\onethird}
)$}

\fchn\hbs{$S(6,2,2,4)$}\hbv{$(-\hat{\onethird},\phm\hat{\onethird},1,1;
                     \phm\hat{\onethird},
                     \hj\phm\hat{1},
                    -\hat{\onethird},
                     \phm\hat{\onethird},
                     \hj\phm\hat{1},
                     -\hat{\onethird}
)$}

\fchn\hbs{$S(6,4,2,1)$}\hbv{$(\phm\hat{\onethird},-\hat{\onethird},1,1;
                     -\hat{\onethird},
                     \hj\phm\hat{1},
                     \phm\hat{\onethird},
                     -\hat{\onethird},
                     \hj\phm\hat{1},
                     \phm\hat{\onethird}
)$}

\fchn\hbs{$S(6,4,2,2)$}\hbv{$(-\hat{\twothird},\phm\hat{\twothird},0,0;
                     \phm\hat{\twothird},
                     \hj\phm\hat{0},
                        -\hat{\twothird},
                        -\hat{\onethird},
                     \hj\phm\hat{1},
                     \phm\hat{\onethird}
)$}

\fchn\hbs{$S(6,4,2,3)$}\hbv{$(-\hat{\twothird},\phm\hat{\twothird},0,0;
                        -\hat{\onethird},
                     \hj\phm\hat{1},
                     \phm\hat{\onethird},
                     \phm\hat{\twothird},
                     \hj\phm\hat{0},
                     -\hat{\twothird}
)$}

\fchn\hbs{$S(6,4,2,4)$}\hbv{$(\phm\hat{\onethird},-\hat{\onethird},1,1;
                     \phm\hat{\twothird},
                     \hj\phm\hat{0},
                    -\hat{\twothird},
                    \phm\hat{\twothird},
                     \hj\phm\hat{0},
                    -\hat{\twothird}
)$}

\vskipa
\fch{$2\cdot2\cdot2$}{4}\hbs{$S(6,3,2,1)$}\hbv{$(\hj\phm\hat{1},
                        \hj\phm\hat{0},0,1;
                        -\hat{\onehalf},
                        -\hat{\onehalf},
                        -\hat{\onehalf},
                     \phm\hat{\onehalf},
                     \phm\hat{\onehalf},
                     \phm\hat{\onehalf}
)$}

\fchn\hbs{$S(6,3,2,2)$}\hbv{$(\hj\phm\hat{0},\hj\phm\hat{1},1,0;
                     \phm\hat{\onehalf},
                     \phm\hat{\onehalf},
                     \phm\hat{\onehalf},
                     \phm\hat{\onehalf},
                     \phm\hat{\onehalf},
                     \phm\hat{\onehalf}
)$}

\fchn\hbs{$S(6,3,2,3)$}\hbv{$(\hj\phm\hat{0},\hj\phm\hat{1},1,0;
                        -\hat{\onehalf},
                        -\hat{\onehalf},
                        -\hat{\onehalf},
                        -\hat{\onehalf},
                        - \hat{\onehalf},
                        -\hat{\onehalf}
)$}

\fchn\hbs{$S(6,3,2,4)$}\hbv{$(\hj\phm\hat{1},\hj\phm\hat{0},0,1;
                     \phm\hat{\onehalf},
                     \phm\hat{\onehalf},
                     \phm\hat{\onehalf},
                        -\hat{\onehalf},
                        -\hat{\onehalf},
                        -\hat{\onehalf}
)$}

\vskipa\fch{\sone}{2}\hbs{$S(6,6,2,1)$}\hbv{$(\hj\phm\hat{1},
                     \hj\phm\hat{1},1,1;
                     \hj\phm\hat{0},
                     \hj\phm\hat{0},
                     \hj\phm\hat{0},
                     \hj\phm\hat{0},
                     \hj\phm\hat{0},
                     \hj\phm\hat{0}
)$}

\fchn\hbs{$S(6,6,2,2)$}\hbv{$(\hj\phm\hat{0},\hj\phm\hat{0},0,0;
                     \hj\phm\hat{1},
                     \hj\phm\hat{1},
                     \hj\phm\hat{1},
                     \hj\phm\hat{0},
                     \hj\phm\hat{0},
                     \hj\phm\hat{0}
)$}

\fchn\hbs{$S(6,6,2,3)$}\hbv{$(\hj\phm\hat{0},\hj\phm\hat{0},0,0;
                     \hj\phm\hat{0},
                     \hj\phm\hat{0},
                     \hj\phm\hat{0},
                     \hj\phm\hat{1},
                     \hj\phm\hat{1},
                     \hj\phm\hat{1}
)$}

\fchn\hbs{$S(6,6,2,4)$}\hbv{$(\hj\phm\hat{1},\hj\phm\hat{1},1,1;
                     \hj\phm\hat{1},
                     \hj\phm\hat{1},
                     \hj\phm\hat{1},
                     \hj\phm\hat{1},
                     \hj\phm\hat{1},
                     \hj\phm\hat{1}
)$}

\tailerb
\newpage

%*************
\headings

\fch{6}{12}\hbs{$S(6,1,3,1)$}\hbv{$(\phm\hat{\onethird},
                     \phm\hat{\twothird},0,1;
                     -\hat{\fivesixth},
                     -\hat{\half},
                     \phm\hat{\onesixth},
                     -\hat{\onesixth},
                     \phm\hat{\half},
                     \phm\hat{\fivesixth}
)$}

\fchn\hbs{$S(6,1,3,2)$}\hbv{$(-\hat{\twothird},-\hat{\onethird},1,0;
                     \phm\hat{\onesixth},
                     \phm\hat{\half},
                     -\hat{\fivesixth},
                     -\hat{\onesixth},
                     \phm\hat{\half},
                     \phm\hat{\fivesixth}
)$}

\fchn\hbs{$S(6,1,3,3)$}\hbv{$(-\hat{\twothird},-\hat{\onethird},1,0;
                     -\hat{\fivesixth},
                     -\hat{\half},
                     \phm\hat{\onesixth},
                     \phm\hat{\fivesixth},
                     -\hat{\half},
                     -\hat{\onesixth}
)$}

\fchn\hbs{$S(6,1,3,4)$}\hbv{$(\phm\hat{\onethird},\phm\hat{\twothird},0,1;
                     \phm\hat{\onesixth},
                     \phm\hat{\half},
                     -\hat{\fivesixth},
                     \phm\hat{\fivesixth},
                     -\hat{\half},
                     -\hat{\onesixth}
)$}

\fchn\hbs{$S(6,5,3,1)$}\hbv{$(-\hat{\onethird},-\hat{\twothird},0,1;
                     -\hat{\onesixth},
                     -\hat{\half},
                     \phm\hat{\fivesixth},
                     -\hat{\fivesixth},
                     \phm\hat{\half},
                     \phm\hat{\onesixth}
)$}

\fchn\hbs{$S(6,5,3,2)$}\hbv{$(\phm\hat{\twothird},\phm\hat{\onethird},1,0;
                      \phm\hat{\fivesixth},
                      \phm\hat{\half},
                      -\hat{\onesixth},
                      -\hat{\fivesixth},
                      \phm\hat{\half},
                      \phm\hat{\onesixth}
)$}

\fchn\hbs{$S(6,5,3,3)$}\hbv{$(\phm\hat{\twothird},\phm\hat{\onethird},1,0;
                     -\hat{\onesixth},
                     -\hat{\half},
                     \phm\hat{\fivesixth},
                     \phm\hat{\onesixth},
                     -\hat{\half},
                     -\hat{\fivesixth}
)$}

\fchn\hbs{$S(6,5,3,4)$}\hbv{$(-\hat{\onethird},-\hat{\twothird},0,1;
                     \phm\hat{\fivesixth},
                     \phm\hat{\half},
                     -\hat{\onesixth},
                     \phm\hat{\onesixth},
                     -\hat{\half},
                     -\hat{\fivesixth}
)$}

\vskipa
\fch{\sseven}{6}\hbs{$S(6,2,3,1)$}\hbv{$(-\hat{\onethird},
                     \phm\hat{\onethird},1,1;
                     -\hat{\twothird},
                     \hj\phm\hat{0},
                     -\hat{\twothird},
                     \phm\hat{\twothird},
                     \hj\phm\hat{0},
                     \phm\hat{\twothird}
)$}

\fchn\hbs{$S(6,2,3,2)$}\hbv{$(\phm\hat{\twothird},-\hat{\twothird},0,0;
                     \phm\hat{\onethird},
                     \hj\phm\hat{1},
                     \phm\hat{\onethird},
                     \phm\hat{\twothird},
                     \hj\phm\hat{0},
                     \phm\hat{\twothird}
)$}

\fchn\hbs{$S(6,2,3,3)$}\hbv{$(\phm\hat{\twothird},-\hat{\twothird},0,0;
                     -\hat{\twothird},
                     \hj\phm\hat{0},
                     -\hat{\twothird},
                     -\hat{\onethird},
                     \hj\phm\hat{1},
                     -\hat{\onethird}
)$}

\fchn\hbs{$S(6,2,3,4)$}\hbv{$(-\hat{\onethird},\phm\hat{\onethird},1,1;
                     \phm\hat{\onethird},
                     \hj\phm\hat{1},
                     \phm\hat{\onethird},
                     -\hat{\onethird},
                     \hj\phm\hat{1},
                     -\hat{\onethird}
)$}

\fchn\hbs{$S(6,4,3,1)$}\hbv{$(\phm\hat{\onethird},-\hat{\onethird},1,1;
                     -\hat{\onethird},
                     \hj\phm\hat{1},
                     -\hat{\onethird},
                     \phm\hat{\onethird},
                     \hj\phm\hat{1},
                     \phm\hat{\onethird}
)$}

\fchn\hbs{$S(6,4,3,2)$}\hbv{$(-\hat{\twothird},\phm\hat{\twothird},0,0;
                     \phm\hat{\twothird},
                     \hj\phm\hat{0},
                     \phm\hat{\twothird},
                     \phm\hat{\onethird},
                     \hj\phm\hat{1},
                     \phm\hat{\onethird}
)$}

\fchn\hbs{$S(6,4,3,3)$}\hbv{$(-\hat{\twothird},\phm\hat{\twothird},0,0;
                        -\hat{\onethird},
                     \hj\phm\hat{1},
                        -\hat{\onethird},
                     -\hat{\twothird},
                     \hj\phm\hat{0},
                     -\hat{\twothird}
)$}

\fchn\hbs{$S(6,4,3,4)$}\hbv{$(\phm\hat{\onethird},-\hat{\onethird},1,1;
                     \phm\hat{\twothird},
                     \hj\phm\hat{0},
                     \phm\hat{\twothird},
                     -\hat{\twothird},
                     \hj\phm\hat{0},
                     -\hat{\twothird}
)$}

\vskipa
\fch{$2\cdot2\cdot2$}{4}\hbs{$S(6,3,3,1)$}\hbv{$(\hj\phm\hat{1},
                      \hj\phm\hat{0},0,1;
                        -\hat{\onehalf},
                     \phm\hat{\onehalf},
                     \phm\hat{\onehalf},
                     -\hat{\onehalf},
                        -\hat{\onehalf},
                     \phm\hat{\onehalf}
)$}

\fchn\hbs{$S(6,3,3,2)$}\hbv{$(\hj\phm\hat{0},\hj\phm\hat{1},1,0;
                     \phm\hat{\onehalf},
                        -\hat{\onehalf},
                        -\hat{\onehalf},
                        -\hat{\onehalf},
                        -\hat{\onehalf},
                     \phm\hat{\onehalf}
)$}

\fchn\hbs{$S(6,3,3,3)$}\hbv{$(\hj\phm\hat{0},\hj\phm\hat{1},1,0;
                        -\hat{\onehalf},
                     \phm\hat{\onehalf},
                     \phm\hat{\onehalf},
                     \phm\hat{\onehalf},
                     \phm\hat{\onehalf},
                        -\hat{\onehalf}
)$}

\fchn\hbs{$S(6,3,3,4)$}\hbv{$(\hj\phm\hat{1},\hj\phm\hat{0},0,1;
                     \phm\hat{\onehalf},
                        -\hat{\onehalf},
                        -\hat{\onehalf},
                     \phm\hat{\onehalf},
                     \phm\hat{\onehalf},
                        -\hat{\onehalf}
)$}

\vskipa\fch{\sone}{2}\hbs{$S(6,6,3,1)$}\hbv{$(\hj\phm\hat{1},
                     \hj\phm\hat{1},1,1;
                     \hj\phm\hat{0},
                     \hj\phm\hat{0},
                     \hj\phm\hat{0},
                     \hj\phm\hat{0},
                     \hj\phm\hat{0},
                     \hj\phm\hat{0}
)$}

\fchn\hbs{$S(6,6,3,2)$}\hbv{$(\hj\phm\hat{0},\hj\phm\hat{0},0,0;
                     \hj\phm\hat{1},
                     \hj\phm\hat{1},
                     \hj\phm\hat{1},
                     \hj\phm\hat{0},
                     \hj\phm\hat{0},
                     \hj\phm\hat{0}
)$}

\fchn\hbs{$S(6,6,3,3)$}\hbv{$(\hj\phm\hat{0},\hj\phm\hat{0},0,0;
                     \hj\phm\hat{0},
                     \hj\phm\hat{0},
                     \hj\phm\hat{0},
                     \hj\phm\hat{1},
                     \hj\phm\hat{1},
                     \hj\phm\hat{1}
)$}

\fchn\hbs{$S(6,6,3,4)$}\hbv{$(\hj\phm\hat{1},\hj\phm\hat{1},1,1;
                     \hj\phm\hat{1},
                     \hj\phm\hat{1},
                     \hj\phm\hat{1},
                     \hj\phm\hat{1},
                     \hj\phm\hat{1},
                     \hj\phm\hat{1}
)$}

\tailerb
\newpage

%*************
\headings

\fch{6}{12}\hbs{$S(6,1,4,1)$}\hbv{$(\phm\hat{\onethird},
                      \phm\hat{\twothird},0,1;
                     -\hat{\fivesixth},+\hat{\half},+\hat{\onesixth},
                     -\hat{\onesixth},
                     -\hat{\half},
                     \phm\hat{\fivesixth}
)$}

\fchn\hbs{$S(6,1,4,2)$}\hbv{$(-\hat{\twothird},-\hat{\onethird},1,0;
                     \phm\hat{\onesixth},
                     -\hat{\half},
                     -\hat{\fivesixth},
                     -\hat{\onesixth},
                     -\hat{\half},
                     \phm\hat{\fivesixth}
)$}

\fchn\hbs{$S(6,1,4,3)$}\hbv{$(-\hat{\twothird},-\hat{\onethird},1,0;
                     -\hat{\fivesixth},
                     \phm\hat{\half},
                     \phm\hat{\onesixth},
                     \phm\hat{\fivesixth},
                     \phm\hat{\half},
                     -\hat{\onesixth}
)$}

\fchn\hbs{$S(6,1,4,4)$}\hbv{$(\phm\hat{\onethird},\phm\hat{\twothird},0,1;
                     \phm\hat{\onesixth},
                     -\hat{\half},
                     -\hat{\fivesixth},
                     \phm\hat{\fivesixth},
                     \phm\hat{\half},
                     -\hat{\onesixth}
)$}

\fchn\hbs{$S(6,5,4,1)$}\hbv{$(-\hat{\onethird},-\hat{\twothird},0,1;
                     -\hat{\onesixth},
                     \phm\hat{\half},
                     \phm\hat{\fivesixth},
                     -\hat{\fivesixth},
                     \phm\hat{\half},
                     \phm\hat{\onesixth}
)$}

\fchn\hbs{$S(6,5,4,2)$}\hbv{$(\phm\hat{\twothird},\phm\hat{\onethird},1,0;
                      \phm\hat{\fivesixth},
                      -\hat{\half},
                      -\hat{\onesixth},
                      -\hat{\fivesixth},
                      -\hat{\half},
                      \phm\hat{\onesixth}
)$}

\fchn\hbs{$S(6,5,4,3)$}\hbv{$(\phm\hat{\twothird},\phm\hat{\onethird},1,0;
                     -\hat{\onesixth},
                     \phm\hat{\half},
                     \phm\hat{\fivesixth},
                     \phm\hat{\onesixth},
                     \phm\hat{\half},
                     -\hat{\fivesixth}
)$}

\fchn\hbs{$S(6,5,4,4)$}\hbv{$(-\hat{\onethird},-\hat{\twothird},0,1;
                     \phm\hat{\fivesixth},
                     -\hat{\half},
                     -\hat{\onesixth},
                     \phm\hat{\onesixth},
                     \phm\hat{\half},
                     -\hat{\fivesixth}
)$}

\vskipa
\fch{\sseven}{6}\hbs{$S(6,2,4,1)$}\hbv{$(-\hat{\onethird},
                      \phm\hat{\onethird},1,1;
                     -\hat{\twothird},
                     \hj\phm\hat{0},
                     -\hat{\twothird},
                     \phm\hat{\twothird},
                     \hj\phm\hat{0},
                     \phm\hat{\twothird}
)$}

\fchn\hbs{$S(6,2,4,2)$}\hbv{$(\phm\hat{\twothird},-\hat{\twothird},0,0;
                     \phm\hat{\onethird},
                     \hj\phm\hat{1},
                     \phm\hat{\onethird},
                     \phm\hat{\twothird},
                     \hj\phm\hat{0},
                     \phm\hat{\twothird}
)$}

\fchn\hbs{$S(6,2,4,3)$}\hbv{$(\phm\hat{\twothird},-\hat{\twothird},0,0;
                     -\hat{\twothird},
                     \hj\phm\hat{0},
                     -\hat{\twothird},
                     -\hat{\onethird},
                     \hj\phm\hat{1},
                     -\hat{\onethird}
)$}

\fchn\hbs{$S(6,2,4,4)$}\hbv{$(-\hat{\onethird},\phm\hat{\onethird},1,1;
                     \phm\hat{\onethird},
                     \hj\phm\hat{1},
                     \phm\hat{\onethird},
                     -\hat{\onethird},
                     \hj\phm\hat{1},
                     -\hat{\onethird}
)$}

\fchn\hbs{$S(6,4,4,1)$}\hbv{$(\phm\hat{\onethird},-\hat{\onethird},1,1;
                     -\hat{\onethird},
                     \hj\phm\hat{1},
                     -\hat{\onethird},
                     \phm\hat{\onethird},
                     \hj\phm\hat{1},
                     \phm\hat{\onethird}
)$}

\fchn\hbs{$S(6,4,4,2)$}\hbv{$(-\hat{\twothird},\phm\hat{\twothird},0,0;
                     \phm\hat{\twothird},
                     \hj\phm\hat{0},
                     \phm\hat{\twothird},
                     \phm\hat{\onethird},
                     \hj\phm\hat{1},
                     \phm\hat{\onethird}
)$}

\fchn\hbs{$S(6,4,4,3)$}\hbv{$(-\hat{\twothird},\phm\hat{\twothird},0,0;
                        -\hat{\onethird},
                     \hj\phm\hat{1},
                        -\hat{\onethird},
                        -\hat{\twothird},
                     \hj\phm\hat{0},
                        -\hat{\twothird}
)$}

\fchn\hbs{$S(6,4,4,4)$}\hbv{$(\phm\hat{\onethird},-\hat{\onethird},1,1;
                     \phm\hat{\twothird},
                     \hj\phm\hat{0},
                     \phm\hat{\twothird},
                    -\hat{\twothird},
                     \hj\phm\hat{0},
                     -\hat{\twothird}
)$}

\vskipa\fch{$2\cdot2\cdot2$}{4}\hbs{$S(6,3,4,1)$}\hbv{$(\hj\phm\hat{1},
                     \hj\phm\hat{0},0,1;
                        -\hat{\onehalf},
                        -\hat{\onehalf},
                        \phm\hat{\onehalf},
                        -\hat{\onehalf},
                     \phm\hat{\onehalf},
                     \phm\hat{\onehalf}
)$}

\fchn\hbs{$S(6,3,4,2)$}\hbv{$(\hj\phm\hat{0},\hj\phm\hat{1},1,0;
                     \phm\hat{\onehalf},
                     \phm\hat{\onehalf},
                        -\hat{\onehalf},
                        -\hat{\onehalf},
                     \phm\hat{\onehalf},
                     \phm\hat{\onehalf}
)$}

\fchn\hbs{$S(6,3,4,3)$}\hbv{$(\hj\phm\hat{0},\hj\phm\hat{1},1,0;
                        -\hat{\onehalf},
                        -\hat{\onehalf},
                     \phm\hat{\onehalf},
                     \phm\hat{\onehalf},
                        -\hat{\onehalf},
                        -\hat{\onehalf}
)$}

\fchn\hbs{$S(6,3,4,4)$}\hbv{$(\hj\phm\hat{1},\hj\phm\hat{0},0,1;
                     \phm\hat{\onehalf},
                     \phm\hat{\onehalf},
                        -\hat{\onehalf},
                     \phm\hat{\onehalf},
                        -\hat{\onehalf},
                        -\hat{\onehalf}
)$}

\vskipa
\fch{\sone}{2}\hbs{$S(6,6,4,1)$}\hbv{$(\hj\phm\hat{1},\hj\phm\hat{1},1,1;
                     \hj\phm\hat{0},
                     \hj\phm\hat{0},
                     \hj\phm\hat{0},
                     \hj\phm\hat{0},
                     \hj\phm\hat{0},
                     \hj\phm\hat{0}
)$}

\fchn\hbs{$S(6,6,4,2)$}\hbv{$(\hj\phm\hat{0},\hj\phm\hat{0},0,0;
                      \hj\phm\hat{1},
                      \hj\phm\hat{1},
                      \hj\phm\hat{1},
                      \hj\phm\hat{0},
                      \hj\phm\hat{0},
                      \hj\phm\hat{0}
)$}

\fchn\hbs{$S(6,6,4,3)$}\hbv{$(\hj\phm\hat{0},\hj\phm\hat{0},0,0;
                      \hj\phm\hat{0},
                      \hj\phm\hat{0},
                      \hj\phm\hat{0},
                      \hj\phm\hat{1},
                      \hj\phm\hat{1},
                      \hj\phm\hat{1}
)$}

\fchn\hbs{$S(6,6,4,4)$}\hbv{$(\hj\phm\hat{1},\hj\phm\hat{1},1,1;
                      \hj\phm\hat{1},
                      \hj\phm\hat{1},
                      \hj\phm\hat{1},
                      \hj\phm\hat{1},
                      \hj\phm\hat{1},
                      \hj\phm\hat{1}
)$}

\tailerb
\newpage

%\end{ignore}

%******************************************************************************
%\input refs.tex

%\vskip .2cm
\no{\bf References}
\sectionnumstyle{arabic}

\def\am#1{    {\it Adv.~Math.~}{\bf #1}  }
\def\cmp#1{ {\it Comm.~Math.~Phys.~}{\bf #1}  }
\def\ijmp#1{ {\it Int.~Jour.~Mod.~Phys.~}{\bf A#1}  }
\def\jetp#1{ {\it Sov.~Phys.~}JETP {\bf #1}  }
\def\mpl#1{ {\it Mod.~Phys.~Lett.} {\bf #1}  }
\def\np#1{  {\it Nucl.~Phys.~}{\bf B#1}  }
\def\pl#1{  {\it Phys.~Lett.~}{\bf B#1}  }
\def\pr#1{  {\it Phys.~Rev.~}{\bf D#1}   }
\def\prl#1{ {\it Phys.~Rev.~Lett.~}{\bf #1}  }
\def\ptp#1{ {\it Prog.~Theo.~Phys.~}{\bf #1}  }
\def\tmp#1{ {\it Teor.~Mat.~Phys.~}{\bf #1}  }

\def\pp{ preprint  }
\def\pps{ preprints  }
\def\ny{ New York}

\begin{putreferences}

%******************************************************************************
% My Publications:

\ref{cleaver1}{G.~Cleaver, Unpublished research.}
\ref{cleaver92a}{G.~Cleaver. {\it ``Comments on Fractional Superstrings,''}
                 In the Proceedings of the International Workshop
                 on String Theory, Quantum Gravity and the Unification of
                 Fundamental Interactions, Rome, 1992.}
\ref{cleaver93a}{G.~Cleaver and D.~Lewellen, \pr{300} (1993) 354.}
% Aspects ...
\ref{cleaver93b}{G.~Cleaver and P.~Rosenthal, \cmp{167} (1995) 155.}
% The Dimension of Spacetime
\ref{cleaver93c}{G.~Cleaver and P.~Rosenthal, \pp CALT 68/1878.}

\ref{cleaver94a}{G.~ Cleaver and K.~Dienes,
                {\it ``Internal Projection Operators for Fractional
                 Superstrings,''}
                 OHSTPY-HEP-T-93-022, McGill/93-43.}
\ref{cleaver94a}{G.~Cleaver, {\it ``GUT's With Adjoint Higgs Fields
                 From Superstrings,''} OHSTPY-HEP-T-94-007. In the
                 Proceedings of PASCOS '94, Syracuse, \ny, 1994.}

\ref{cleaver95a}{G.~Cleaver, {\it ``Supersymmetries in Free
                 Fermionic Strings,''}
                 OHSTPY-HEP-T-95-004; DOE/ER/01545-643;
                 hep-th/9505080;\\
                 {\it ``What's New in Stringy}\/ SO(10) {\it SUSY-GUTs},''
                 OHSTPY-HEP-T-95-003. To appear in the Proceedings of
                 Strings '95, Los Angeles, CA, 1995.}

\ref{cleaver95b}{G.~Cleaver, {\it ``Free Fermionic Solutions
                 To N=1 SUSY and Three Generations.''} To appear.}
\ref{cleaver95c}{G.~Cleaver, {\it ``Unpairable Real Fermions
                 in Free Fermionic Strings,''} OHSTPY-HEP-T-95-XXX.
                 In preparation}
\ref{cleaver95d}{G.~Cleaver, {\it ``The Search For Stringy
                 SO(10) SUSY-GUTS.''} To appear.}
\ref{cleaver95x}{G.~Cleaver, {\it ``Free Fermionic Solutions
                 To N=1 SUSY and Three Generations'';}
                 {\it ``The Search For Stringy
                 SO(10) SUSY-GUTS.''} Research in Progress.}

%******************************************************************************

% Other Publications:

%---> Freq. Referenced Groups:
%******************************************************************************

\ref{antoniadis86} {I.~Antoniadis and C.~Bachas, \np{278} (1986) 343;\\
     M.~Hama, M.~Sawamura, and H.~Suzuki, RUP-92-1.}
\ref{li88} {K.~Li and N.~Warner, \pl{211} (1988) 101;\\
     A.~Bilal, \pl{226} (1989) 272;\\
     G.~Delius, \pp ITP-SB-89-12.}

\ref{antoniadis87}{I.~Antoniadis, C.~Bachas, C.~Kounnas, and P.~Windey,
     \pl{171} (1986) 51;\\
     I.~Antoniadis, C.~Bachas, and C.~Kounnas, \np{289} (1987) 87;\\
     I.~Antoniadis and C.~Bachas, \np{298} (1988) 586.}

\ref{antoniadis87b}{I.~Antoniadis, J.~Ellis, J.~Hagelin, and D.V.~Nanopoulos,
     \pl{194} (1987) 231.}
\ref{antoniadis88}{I.~Antoniadis and C.~Bachas, \np{298} (1988) 586.}

%The Flipped SU(5)xU(1) Model Revamped.
\ref{antoniadis89a}{I.~Antoniadis, J.~Ellis, J.~Hagelin, and D.V.~Nanopoulos,
    \pl{231} (1989) 65.}

\ref{dreiner89a}{H.~Dreiner, J.~Lopez, D.V.~Nanopoulos, and D.~Reiss,
     \pps MAD/TH/89-2; CTP-TAMU-06.}
\ref{dreiner89b}{H.~Dreiner, J.~Lopez, D.V.~Nanopoulos, and D.~Reiss,
     \np{320} (1989) 401.}
\ref{dreiner89c}{H.~Dreiner, J.~Lopez, D.V.~Nanopoulos, and D.~Reiss,
     \pl{216} (1989) 283.}

\ref{faraggi89a}{A.~Faraggi, D.V.~Nanopoulos, and K.~Yuan, \np{335} (1990)
     347.}
\ref{faraggi92a}{A.~Faraggi, \np{387} (1992) 239.}
\ref{faraggi93a}{A.~Faraggi and D.V.~Nanopoulos, \pr{48} (1993)
     3288.}

\ref{faraggix}{A.~Faraggi, D.V.~Nanopoulos, and K.~Yuan, \np{335} (1990) 347.\\
     A.~Faraggi, \np{387} (1992) 239.\\
     A.~Faraggi and D.V.~Nanopoulos, \pr{48} (1993) 3288.\\
     I.~Antoniadis, J.~Ellis, J.~Hagelin, and D.V.~Nanopoulos,
     \pl{194} (1987) 231.}
%******************************************************************************

\ref{argyres91a}{P.~Argyres, A.~LeClair, and S.-H.~Tye, \pl{235} (1991).}
\ref{argyres91b}{P.~Argyres and S.~-H.~Tye, \prl{67} (1991) 3339.}
\ref{argyres91c}{P.~Argyres, J.~Grochocinski, and S.-H.~Tye, \pp
     CLNS 91/1126.}
\ref{argyres91d}{P.~Argyres, K.~Dienes and S.-H.~Tye, \pps CLNS 91/1113;
     McGill-91-37.}
\ref{argyres91e} {P.~Argyres, E.~Lyman, and S.-H.~Tye, \pp CLNS 91/1121.}
\ref{argyres91f}{P.~Argyres, J.~Grochocinski, and S.-H.~Tye, \np{367} (1991)
     217.}
\ref{argyres93}{P.~Argyres and K.~Dienes, \prl{71} (1993) 819.}

\ref{dienes92b}{K.~Dienes and S.~-H.~Tye, \np{376} (1992) 297.}
\ref{dienes93a}{K.~Dienes, \np{413} (1994) 103 (hep-th/9305094).}
\ref{dienes93b}{K.~Dienes, McGill \pp McGill/93-18. To appear in \np{}.}

%******************************************************************************

\ref{kawai87a} {H.~Kawai, D.~ Lewellen, and S.-H.~Tye, \np{288} (1987) 1;\\
     H.~Kawai, D.~Lewellen, J.A.~Schwartz, and S.-H.~Tye, \np{299} (1988) 431.}
\ref{kawai87b} {H.~Kawai, D.~Lewellen, J.A.~Schwartz, and S.-H.~Tye, \np{299}
     (1988) 431.}

\ref{lewellen87}{H.~Kawai, D.~Lewellen, and S.-H.`Tye, \np{288} (1987) 1.}
\ref{lewellen}{D.~C.~Lewellen, \np{337} (1990) 61.}

%******************************************************************************
\ref{chaudhuri94a}{S.~Chaudhuri, S.~Chung, and J.~Lykken,
     {\it ``Fermion Masses from Superstring Models with Adjoint Scalars,''}
     Fermilab-PUB-94/137-T; NSF-ITP-94-50.}

%******************************************************************************

%---> Not So Freq. Referenced Groups:

\ref{ardalan74}{F.~Ardalan and F.~Mansouri, \pr{9} (1974) 3341;
     \prl{56} (1986) 2456;
     \pl{176} (1986) 99.}

\ref{abbott84}{L.F.~Abbott and M.B.~Wise, \np{244} (1984) 541.}

\ref{alvarez86}{L.~Alvarez-Gaum\'e, G.~Moore, and C.~Vafa, \cmp{106} (1986) 1.}

\ref{athanasiu88}{G.~Athanasiu and J.~Atick, \pp IASSNS/HEP-88/46.}

%\ref{atick88}{J.~Atick and E.~Witten, \np{2} (1988) .}

\ref{axenides88}{M.~Axenides, S.~Ellis, and C.~Kounnas, \pr{37} (1988) 2964.}

\ref{bailin89a}{D.~Bailin, \mpl{4} (1989) 2339.}
\ref{bailin90a}{D.~Bailin, D.~Dunbar, and A.~Love, \np{330} (1990) 124;
     \ijmp{5} (1990) 939.}
\ref{bailin92}{D.~Bailin and A.~Love, \pl{292} (1992) 315.}

\ref{bailinx}{D.~Bailin, \mpl{4} (1989) 2339.\\
     D.~Bailin, D.~Dunbar, and A.~Love, \np{330} (1990) 124;
     {\it Int.~Jour.~Mod.} {\it Phys.~}{\bf A5} (1990) 939.\\
     D.~Bailin and A.~Love, \pl{292} (1992) 315.}

\ref{barnsley88}{M.~Barnsley, {\underbar{Fractals Everywhere}} (Academic Press,
     Boston, 1988).}

\ref{bouwknegt87}{P.~Bouwknegt and W.~Nahm, \pl{184} (1987) 359;\\
     F.~Bais and P.~Bouwknegt, \np{279} (1987) 561;\\
     P.~Bouwknegt, Ph.D.~Thesis.}

\ref{bowick89}{M.~Bowick and S.~Giddings, \np{325} (1989) 631.}
\ref{bowick92}{M.~Bowick, SUHEP-4241-522 (1992).}
\ref{bowick93}{M.~Bowick, Private communications.}

\ref{brustein92}{R.~Brustein and P.~Steinhardt, \pp UPR-541T.}

\ref{capelli87} {A.~Cappelli, C.~Itzykson, and J.~Zuber, \np{280 [FS 18]}
     (1987) 445; \cmp{113} (1987) 1.}

\ref{carlitz}{R.~Carlitz, \pr{5} (1972) 3231.}

\ref{candelas85a}{P.~Candelas, G.~Horowitz, A.~Strominger, and E.~Witten,
     \np{258} (1985) 46.}
\ref{candelas88a}{P.~Candelas, A.M.~Dale, C.A.~Lutken, and R.~Schimmrigk,
     \np{298} (1988) 493.}

\ref{cateau92}{H.~Cateau and K.~Sumiyoshi, \pr{46} (1992) 2366.}

\ref{christe87}{P.~Christe, \pl{188} (1987) 219; \pl{198} (1987) 215;
     Ph.D.~thesis (1986).}

\ref{clavelli90}{L.~Clavelli {\it et al.}, \ijmp{5}
     (1990) 175.}

\ref{clgan}{For recent progress in classification of asymmetric modular
     invariants see\\
     T.~Gannon, Carleton \pp 92-0407; hep-th 9408119, 9407055, 9404185,
     9402074,\\
     G.~Cleaver and D.~Lewellen, \pl{300} (1993) 354.}

\ref{cornwell89}{J.~F.~Cornwell, {\underbar{Group Theory in Physics}},
     {\bf Vol. III}, (Academic Press, London, 1989).}

\ref{deo89a}{N.~Deo, S.~Jain, and C.~Tan, \pl{220} (1989) 125.}
\ref{deo89b}{N.~Deo, S.~Jain, and C.~Tan, \pr{40} (1989) 2626.}
\ref{deo92}{N.~Deo, S.~Jain, and C.~Tan, \pr{45} (1992) 3641.}
\ref{deo90a}{N.~Deo, S.~Jain, and C.-I.~Tan, in
     {\underbar{Modern Quantum Field Theory}}, (World Scientific, Bombay,
     S.~Das {\it et al.} editors, 1990).}

\ref{distler90}{J.~Distler, Z.~Hlousek, and H.~Kawai, \ijmp{5} (1990) 1093.}
\ref{distler93}{J.~Distler, private communication.}

\ref{dixon85}{L.~Dixon, J.~Harvey, C.~Vafa and E.~Witten, \np{261} (1985) 651;
    {\bf B274} (1986) 285.}
\ref{dixon87}{L.~Dixon, V.~Kaplunovsky, and C.~Vafa,
\np{294} (1987) 443.}

\ref{drees90}{W.~Drees, {\underbar{Beyond the Big Bang},} (Open Court,
     La Salle, 1990).}

\ref{ellis90}{J.~Ellis, J.~Lopez, and D.V.~Nanopoulos, \pl{245} (1990) 375.}
%Constraints on Grand Unified Superstring Theories

\ref{fernandez92}{R.~Fern\' andez, J.~Fr\" ohlich, and A.~Sokal,
     {\underbar{Random Walks, Critical Phenomena, and Triviality in}}
     {\underbar{Quantum Mechanics}}, (Springer-Verlag, 1992).}

\ref{font90}{A.~Font, L.~Ib\'a\~ nez, and F.~Quevedo, \np{345} (1990) 389.}

\ref{frampton88}{P.~Frampton and M.~Ubriaco, {\bf D38} (1988) 1341.}
\ref{francesco87}{P.~di Francesco, H.~Saleur, and J.B.~Zuber, \np{28 [FS19]}
     (1987) 454.}

\ref{frautschi71}{S.~Frautschi, \pr{3} (1971) 2821.}

\ref{gannon92}{T.~Gannon, Carleton \pp 92-0407.}

\ref{gasperini91}{M.~Gasperini, N.~S\'anchez, and G.~Veneziano, \ijmp{6}
     (1991) 3853; \np{364} (1991) 365.}

\ref{gepner87}{D.~Gepner and Z.~Qiu, \np{285} (1987) 423.}
\ref{gepner87b}{D.~Gepner, \pl{199} (1987) 380.}
\ref{gepner88a}{D.~Gepner, \np{296} (1988) 757.}

\ref{giannakis95a}{I.~Giannakis, D.V.~Nanopoulos, and K.~Yuan,
%     ``{\it Number of Generations in Free\break Fermionic String Models,}"
     \pr{52} (1995) 1026.}

\ref{ginsparg88}{P.~Ginsparg, \np{295 [FS211]} (1988) 153.}
\ref{ginsparg89}{P.~Ginsparg, in \underbar{Fields, Strings and Critical
     Phenomena}, (Elsevier Science Publishers, E.~Br\' ezin and
     J.~Zinn-Justin editors, 1989).}

\ref{gross84}{D.~Gross, \pl{138} (1984) 185.}

\ref{green53} {H.~S.~Green, {\it Phys.~Rev.~}{\bf 90} (1953) 270.}

\ref{hagedorn68}{R.~Hagedorn, {\it Nuovo Cim.~}{\bf A56} (1968) 1027.}

\ref{kac80}{V.~Ka\v c, {\it Adv.~Math.~}{\bf 35} (1980) 264;\\
     V.~Ka\v c and D.~Peterson, {\it Bull.~AMS} {\bf 3} (1980) 1057;
     \am{53} (1984) 125.}
\ref{kac83}{V.~Ka\v c, {\underbar{Infinite Dimensional Lie Algebras}},
     (Birkh\" auser, Boston, 1983);\\
      V.~Ka\v c editor, {\underbar{Infinite Dimensional Lie Algebras
      and Groups}}, (World Scientific, Singapore, 1989).}

\ref{kaku91}{M.~Kaku, \underbar{Strings, Conformal Fields and Topology},
     (Springer-Verlag, \ny, 1991).}

\ref{kalara90}{S.~Kalara, J.~Lopez, and D.V.~Nanopoulos, \pl{245} (1990) 421.}

\ref{kazakov85}{V.~Kazakov, I.~Kostov, and A.~Migdal, \pl{157} (1985) 295.}

\ref{khuri92}{R.~Khuri, CTP/TAMU-80/1992; CTP/TAMU-10/1993.}

\ref{kikkawa84}{K.~Kikkawa and M.~Yamasaki, \pl{149}(1984) 357.}

\ref{kiritsis88}{E.B.~Kiritsis, \pl{217} (1988) 427.}

\ref{langacker92}{P.~Langacker, \pr{47} (1993) 4028.}

\ref{leblanc88}{Y.~Leblanc, \pr{38} (1988) 38.}

\ref{lerche87}{W.~Lerche, D.~L\" ust, and A.N.~Schellekins, \np{287} (1987)
     477.}

\ref{lizzi90}{F.~Lizzi and I.~Senda, \pl{244} (1990) 27.}
\ref{lizzi91}{F.~Lizzi and I.~Senda, \np{359} (1991) 441.}

\ref{lust89}{D.~L\" ust and S.~Theisen, {\underbar{Lectures on String Theory,}}
     (Springer-Verlag, Berlin, 1989).}

\ref{maggiore93}{M.~Maggiore, \pp IFUP-TH 3/93.}

\ref{mansouri87} {F.~Mansouri and X.~Wu, \mpl{2} (1987) 215; \pl{203} (1988)
     417;
     {\it J.~Math.~Phys.~}{\bf 30} (1989) 892;\\
     A. Bhattacharyya {\it et al.,} \mpl{4} (1989) 1121; \pl{224} (1989) 384.}

\ref{narain86} {K.~S.~Narain, \pl{169} (1986) 41.}
\ref{narain87} {K.~S.~Narain, M.H.~Sarmadi, and C.~Vafa, \np{288} (1987) 551.}

\ref{obrien87}{K.~O'Brien and C.~Tan, \pr{36} (1987) 1184.}

\ref{parisi79}{G.~Parisi, \pl{81} (1979) 357.}

\ref{polchinski88}{J.~Polchinski, \pl{209} (1988) 252.}
\ref{polchinski93}{J.~Polchinski, Private communications.}

\ref{pope92}{C.~Pope, \pp CTP TAMU-30/92  (1992).}

\ref{raby}{G.~Anderson, S.~Raby, S.~Dimopoulos, L.J.~Hall, and G.D.~Starkman,
    \pr{49} (1994) 3660;\\
     S.~Raby, talk presented at IFT Workshop on Yukawa Couplings, Gainsville,
     Florida, February 1993, hep-th 9406333.}

\ref{raiten91}{E.~Raiten, Thesis, (1991).}

\ref{roberts92}{P.~Roberts and H.~Terao, \ijmp{7} (1992) 2207;\\
      P.~Roberts, \pl{244} (1990) 429.}

\ref{sakai86}{N.~Sakaii and I.~Senda, \ptp{75} (1986) 692.}

\ref{salomonson86}{P.~Salomonson and B.-S.~Skagerstam, \np{268} (1986) 349.}

\ref{schellekens89} {A.~N.~Schellekens and S.~Yankielowicz, \np{327} (1989) 3;
     \\
     A.~N.~Schellekens, {\it Phys.~Lett.~}{\bf 244} (1990) 255;\\
     B.~Gato-Rivera and A.~N.~Schellekens, \np{353} (1991) 519; \cmp{145}
     (1992) 85.}
\ref{schellekens89b}{B.~Schellekens, ed. \underbar{Superstring Construction},
     (North-Holland Physics, Amsterdam, 1989).}
\ref{schellekens89c}{B.~Schellekens, CERN-TH-5515/89.}

\ref{schwarz87}{M.~Green, J.~Schwarz, and E.~Witten,
     \underbar{Superstring Theory}, {\bf Vols. I \& II},
     (Cambridge University Press, \ny, 1987).}

\ref{turok87a}{D.~Mitchell and N.~Turok, \np{294} (1987) 1138.}
\ref{turok87b}{N.~Turok, Fermilab 87/215-A (1987).}

\ref{verlinde88}{E.~Verlinde, \np{300} (1988) 360.}

\ref{warner90}{N.~Warner, \cmp{130} (1990) 205.}

\ref{wilczek90} {F.~Wilczek, ed. \underbar {Fractional Statistics and Anyon
     Superconductivity}, (World Scientific, Singaore, 1990) 11-16.}

\ref{witten92}{E.~Witten, \pp IASSNS-HEP-93-3.}

\ref{vafa1}{R.~Brandenberger and C.~Vafa, \np{316} (1989) 391.}
\ref{vafa2}{A.A.~Tseytlin and C.~Vafa, \np{372} (1992) 443.}

\ref{zamol87}{A.~Zamolodchikov and V.~Fateev, \jetp{62} (1985)  215;
     \tmp{71} (1987) 163.}

\end{putreferences}
\newpage
\bye